\documentclass[aps,prb,showpacs,twocolumn]{revtex4}
\usepackage{graphicx}
\usepackage{eufrak}
\newcommand{\goth}[1]{\mathfrak{#1}}
\begin{document}

\title
{


\begin{minipage}[t]{7.0in}
\scriptsize
\begin{quote}
\leftline{{\it Phys. Rev. B}, {\bf in press}
}
\raggedleft {\rm arXiv:1911.00923}
\end{quote}
\end{minipage}
\medskip

An On-Site Density Matrix Description of the Extended 
Falicov--Kimball Model at Finite Temperatures}

\author{D. I. Golosov}
\email{Denis.Golosov@biu.ac.il}
\affiliation{Department of Physics and the Resnick Institute, Bar-Ilan 
University, Ramat-Gan 52900, Israel.}

\date{\today}

\begin{abstract}
We propose a single-site mean-field description, an analogue of Weiss mean 
field theory, suitable for narrow-band 
systems with correlation-induced hybridisation at finite temperatures. 
Presently this approach, based on the 
notion of a fluctuating on-site density matrix (OSDM), is developed for
the case of extended Falicov-Kimball model (EFKM).  
In an EFKM, an excitonic insulator phase can be
stabilised at zero temperature. With increasing temperature, the excitonic
order parameter (interaction-induced hybridisation on-site, characterised by the
absolute value and phase) eventually becomes disordered, which involves 
fluctuations of both its phase and (at higher $T$) its absolute value. In order 
to build an adequate finite-temperature description, it is important to 
clarify the nature of degrees of freedom associated with the phase and 
absolute value of the induced hybridisation, and correctly account for the
corresponding phase-space volume. We show that the OSDM-based treatment of
the local fluctuations indeed provides an intuitive and concise description 
(including the phase-space integration measure). This allows to 
describe both the lower-temperature regime where phase fluctuations destroy 
the long-range order, and the higher temperature crossover corresponding to a 
decrease of the absolute value of  hybridisation. In spite of the 
rapid progress
in the studies of  excitonic insulators, a unified picture of this kind has
not been available to date. We briefly discuss recent
experiments on  ${\rm Ta_2 Ni Se_5}$ and also address the 
amplitude mode of collective excitations in relation to 
the measurements reported for ${\rm 1}T-{\rm TiSe}_2$. 
Both the overall scenario and the theoretical framework are also
expected to be relevant in 
other contexts, including the Kondo lattice model. 

\typeout{polish abstract}
\end{abstract}
\maketitle

\section{INTRODUCTION}
\label{sec:intro}

Interaction-induced pairing commonly occurs in many different 
contexts including excitonic and Kondo insulators and
superconductivity. This can involve either particle-hole 
or particle-particle pairs, and gives rise to an induced hybridisation or to
a superconducting pairing amplitude, both of which can be viewed as scalar 
products between formerly orthogonal many-body states, {\it i.e.}, as 
off-diagonal elements of some density matrix.  The corresponding systems 
are characterised by
the ratio of the induced spectral gap (or pair binding energy) to the bandwidth 
energy scales. The case of small binding energy (weak interaction) corresponds
to the well-known BCS picture, where the crucial r\^{o}le is played by restructuring of
the quasiparticle spectra in the vicinity of the Fermi level only. Broadly 
speaking, this 
case is amenable to a long-wavelength perturbative treatment, leading to the 
familiar results. The opposite limiting case, which is commonly referred 
to as that of BEC (Bose--Einstein condensation), is typically realised in the
narrow-band systems and continues to command much attention from  
experimental and theoretical standpoints. It has been suggested that this BEC 
physics might be relevant for
Kondo lattices and heavy-fermion compounds\cite{Piers,Lonzarich}, for 
high-temperature superconductors 
(``pre-formed pairs'' scenario\cite{NSR,Randeria,Kathy}), as well 
as for various aspects of 
excitonic-insulating behaviour in narrow-band 
systems\cite{SmTm,1TTiSe2,Kogar2017} 
(including ``electronic ferroelectricity''\cite{Portengen}). One may also 
note a rather direct 
connexion with much discussed ``Higgs bosons'' in correlated electron 
systems\cite{Higgs}, due to the difference in the energy cost of phase 
and amplitude fluctuations of, {\it e.g.}, induced hybridisation.

In the BEC regime, there are  two distinct energy scales, corresponding
to the energy of strongly-bound excitons or pairs and to their interaction 
with each other. This gives rise to a peculiar evolution of the system with
increasing temperature, as will be further discussed below. Importantly, the 
BEC pairing is {\it not} a phenomenon which concerns only the carriers in
the vicinity of the Fermi level, and new theoretical tools are needed (and 
were indeed suggested, see, {\it e.g.,} Refs. 
\onlinecite{Piers,Randeria,Kathy,Apinyan}) in order
to study the behaviour of a system in this regime. Owing to a small spatial 
size of an exciton or a pair, it appears highly desirable to construct a
simplified {\it local} mean-field description of a single-site type, an 
analogue of an elementary Weiss mean field approach familiar from the theory 
of magnets. Hitherto, this important benchmark appears to be missing, and our 
present objective is to begin filling this gap.

Arguably, the simplest situation where this BEC regime arises is that of the
excitonic insulating state in an extended  spinless Falicov-Kimball model 
(EFKM). 
In this paper, we develop a single-site mean-field description for 
this case, while adaptation of the method and of the insights to other systems 
is relegated to  future work. It should be noted that Falicov--Kimball 
model throughout its history attracted a massive research effort\cite{Zlatic}, 
owing to
its simplicity, peculiarity, and physical relevance. The possibility of an ordered
excitonic state in this model was originally conjectured some 43 years 
ago\cite{Khomskii76},
and a brief review of  more recent literature can be found, {\it e.g.}, 
in Ref.
\onlinecite{prb12}. In particular, variegated analytical
and numerical methods were employed to investigate 
exciton condensation\cite{Batista04,condEFKM}, and more 
generally the BCS--BEC 
crossover\cite{BECEFKM}, in the EFKM.

The spinless Falicov--Kimball model proper\cite{FK} involves fermions $d_i$ and
$c_i$ in the localised and itinerant bands, interacting via a Coulomb
repulsion $U$ on-site:
\begin{equation}
{\cal H}=-\frac{t}{2}\sum_{\langle i j \rangle} \left(c^\dagger_i c_j +
c^\dagger_j c_i \right) + E_d \sum_i d^\dagger_i d_i +U \sum_i c^\dagger_i
d^\dagger_i d_i c_i\,,
\label{eq:FKM}
\end{equation}
where $E_d$ is the bare energy of the localised band. We are interested in
the case where $U$ is, broadly speaking, of the same order of magnitude as the
bare hopping amplitude $t$, and we choose the units
where $t$ and the period of the ($d$-dimensional
hypercubic) lattice are equal to unity. We also set $\hbar=k_B=1$.

In order to stabilise the state with a large on-site hybridisation at $T=0$, 
\begin{equation}
\Delta_i \equiv |\Delta_i| {\rm e}^{{\rm i}\varphi_i} = \langle c^\dagger_i d_i \rangle\,,
\label{eq:Deltagen}
\end{equation}
one must {\it extend} the Falicov--Kimball model by adding a 
perturbation of general form\cite{Czycholl99,Farkasovsky08,Schneider08,prb12}
\begin{eqnarray}
\delta {\cal H}=&&\!\!\!\!\!\!\left\{-\frac{t^\prime}{2}\sum_{\langle i j \rangle} 
d^\dagger_i d_j + V_0 \sum_i c^\dagger_i d_i -\frac{V_1}{2}\sum_{\langle i j \rangle}\left( 
c^\dagger _i d_j + c^\dagger_j d_i \right)-\right .\nonumber \\
&&\!\!\!\!\!\!\left. -\frac{V_2}{2} 
\sum_{\langle i j \rangle}\left( [\vec{R}_j-\vec{R_i}]\cdot \vec{\Xi}
\right)
\left(c^\dagger _i d_j - c^\dagger_j d_i \right) \right\}+ {\rm H. c.}\,,
\label{eq:pert}
\end{eqnarray}
where $t^\prime$ is the $d$-band hopping and $V_0$, bare  on-site
hybridisation.
$V_1$ ($V_2$) is the spatially-even (odd) nearest-neighbour hybridisation,
as appropriate for the case where the two original bands have the same 
(opposite) parity. $\vec{R}_i$ is the radius-vector of a site $i$, and 
$\vec{\Xi}=\sum_{\alpha=1}^{d}\hat{\mathbf x}_\alpha$, sum of 
Cartesian unit vectors. For $t^\prime=0$, the net Hamiltonian, 
Eqs. (\ref{eq:FKM}) and (\ref{eq:pert}), coincides with that of a 
spinless periodic Anderson model, while in the opposite case of 
$V_{0,1,2}=0$ the EFKM becomes identical with the asymmetric Hubbard model, 
where the hopping coefficients for spin-up ($c_i$) and spin-down ($d_i$) 
electrons differ, and $E_d$ in Eq. (\ref{eq:FKM}) is proportional to the Zeeman splitting.

In a broad range of values of parameters of Eq. (\ref{eq:FKM}), 
including {\it any} of the four 
terms in Eq. (\ref{eq:pert}) with an appropriate sign ({\it i.e.}, $t^\prime <0$, $V_0<0$, $V_1$ 
with $V_1 E_d<0$, or $V_2$ of any sign) 
 would result at $T=0$ in an ordered excitonic state
with a uniform $|\Delta_i|=\Delta$ and $\varphi_i=0$ 
(when only $t^\prime$ differs from $0$,  $\varphi_i$ can take any 
constant value; we choose the latter to be equal to zero). This is a 
mixed-valence state with uniform band 
occupancies,
\begin{equation}
n_{c,i}\equiv \langle c^\dagger_i c_i\rangle = n_c\,,\,\,\,n_{d,i}\equiv 
\langle d^\dagger_i d_i\rangle = n_d\,.
\end{equation}
The absolute value of the corresponding 
perturbation parameter must be larger than a certain critical value 
($|t^\prime_{cr}|$, etc.). Depending on 
the parameters
of the Hamiltonian, Eq. (\ref{eq:pert}), the value of $\Delta$ (at least at 
half-filling, $n= n_c+n_d=1$) may be large, $\Delta\stackrel{<}{\sim} 1/2$. With decreasing
perturbation strength ({\it e.g.,} 
the parameter
$|t^\prime|$ is decreased toward $|t^\prime_{cr}|$) the value of $\Delta$ 
does not tend to zero. Rather, at a critical point 
(such as $|t^\prime|= |t^\prime_{cr}|$) a new,
presumably charge-ordering order parameter arises via a second-order 
phase transition\cite{Farkasovsky08,prb12,Apinyan}, destroying the uniformly 
ordered excitonic state. The critical value $t^\prime_{cr}$ 
[as well as critical values of the hybridisations $V_0$, $V_1$, or $(V_2)^2$] 
turns out to be numerically small, some two orders of magnitude smaller than the
bare hopping $t$. Therefore, a useful insight can often be gained by either
treating $\delta {\cal H}$ perturbatively or even technically neglecting its 
effects by keeping only the leading-order term in the calculation.

The behaviour of the system at finite $T$ is crucially dependent on the two 
energy scales characterising the ordered excitonic state at $T=0$. The first one
is the hybridisation-induced energy gap, notably the {\it indirect} one,
which in cases where $U$ is not very large can
be roughly estimated as
\begin{equation}
  G \sim 2 U^2 \Delta^2/d
\label{eq:gapest}
\end{equation}
(see Sec. \ref{sec:HF}; note that the bandwidth of the unhybridised
itinerant band equals $2d$, twice dimensionality of the system), 
and can be an order of magnitude smaller than
the direct gap, 
\begin{equation}
u = 2 U \Delta\,.
\label{eq:dirgap}
\end{equation}
While the value of $G$ at $T=0$ determines the 
{\it crossover} temperature $T_*$, a much smaller scale of the low-lying 
collective excitations\cite{prb12,Jerome} controls critical temperature $T_{cr}$ of 
the {\it ordering transition} (corresponding to the Bose--Einstein condensation
of the excitonic gas). The value of $T_{cr}$ can be 
estimated\cite{prb12} as $(T_{cr})^2\sim |t^\prime|(|t^\prime|-|t^\prime_{cr})$ 
[or $(T_{cr})^2\sim V_2^2 (V_2^2-V_{2,cr}^2)$, etc., when 
hybridisation\cite{extfield} 
dominates $\delta {\cal H}$]. While the excitonic long-range order is lost 
already at $T=T_{cr}$ (where the phases $\varphi_i$ become disordered), the average
value $\Delta(T)$ of $|\Delta_i|$ remains finite, and the state of the
system can be termed {\it disordered electronic insulator}. It is also
variously called ``excitonic liquid'' or ``excitonic gas'' (as opposed to
``excitonic condensate'' at $T< T_{cr}$), as the relatively stable excitons 
persist in equilibrium without a condensate. Since this state
is not associated with a symmetry breaking, it fades away via a smooth 
crossover with increasing $T$ beyond $T_*$, when the thermal fluctuations of
$|\Delta_i|$ become comparable to $\Delta(T)$. Above $T_*$, 
excitons can no longer be considered stable, as they are being formed and 
destroyed rapidly in the course of fluctuations.

Historically, the investigations of EFKM at finite temperatures started 
with extending the pioneering Hartree--Fock decoupling 
approach of Ref. \onlinecite{Khomskii76} to finite $T$. However, this 
method misses the
lower energy scale altogether (also at $T=0$), yielding a second-order phase 
transition at a certain $T_*\sim G$, above which $\Delta(T)$ vanishes 
(see, {\it e.g.}, Ref. \onlinecite{Schneider08}). On the other hand,  
qualitative picture outlined in the previous paragraph is  substantiated 
by a more advanced self-consistent treatment of Ref. \onlinecite{Apinyan}.
Still, it appears that due to the restrictions of a specific mean-field approach  used in the latter reference (involving functional integrals technique with certain topological
complications stemming from the nature of the phase variable $\varphi_i$), its conclusions 
imply a distinct transition at $|T_*|$, as opposed to a smooth crossover expected on symmetry
grounds.

As already mentioned, it appears
highly desirable to try and construct a more intuitive treatment of a 
single-site type. In addition, one expects that the behaviour of the system in
the most interesting crossover regime around $T_*$ is strongly affected by 
the short-range fluctuations, which might not be dealt with accurately within 
the long-wavelength (continuum)  approach of Ref. \onlinecite{Apinyan}. 
Finally, 
one can anticipate that once an adequate single-site mean-field scheme is 
developed for the EFKM, it can be adapted for the entire family of related 
systems, as discussed in the beginning of this section.

In constructing our finite-temperature single-site mean field approach, we make
use of the known properties of the conventional Hartree--Fock 
solution\cite{Khomskii76,Farkasovsky08,prb12,Schneider08} for the EFKM. 
These are summarised in Sec. 
\ref{sec:HF}, where we also outline our general strategy, which requires 
taking into account thermal fluctuations of the local quantities $\Delta_i$ and 
$n_{d,i}$. While the values of hybridisation and band occupancies can be 
deduced from the  (fluctuating) on-site density matrix (OSDM),
our Hamiltonian is non-local, and in order to calculate the energy cost 
of a local fluctuation one needs a fuller knowledge of the quantum state of
the system. The form of the  wave functions corresponding to such local 
fluctuations is obtained, under broad assumptions, in Sec. \ref{sec:osdm}. 
The emergent correspondence between the OSDM and the states of the system is 
also used in order to find the phase-space volume corresponding to a local 
fluctuation. While finding the suitable integration measure in the space 
of quantum states appears complicated, an established notion\cite{Bengsston} 
of  measure
in the space of density matrices (Bures measure) can be readily adapted to the
case at hand. This is accomplished in Sec. \ref{sec:Bures}, completing the
description of our mean-field scheme. We note that the development in Secs. 
\ref{sec:osdm} and \ref{sec:Bures} appears rather general, and may prove 
useful beyond the Hartree--Fock approximation for the wave functions, utilised 
elsewhere throughout the paper.
 
The actual application of the technique introduced in Secs. \ref{sec:HF}--\ref{sec:Bures} begins in Sec. \ref{sec:low-T} with 
the analysis of the low-temperature behaviour, including the
ordering transition at $T_{cr}$. While in this case one does not expect any
single-site approach to yield an accurate description, we do find a second-order
phase transition with  the value of $T_{cr}$ controlled by the parameters of the
perturbation, Eq. (\ref{eq:pert}).

The behaviour of the EFKM in the high-temperature phase-disordered state, 
including
the crossover region at $T \sim T_*$, is considered in Sec. \ref{sec:high-T}. 
It appears that
the results obtained there are both reliable (except when the approach fails 
due to the
underlying Hartree--Fock approximation becoming invalid, Sec. 
\ref{subsec:gamma}) and new,
providing the first quantitative description of the crossover region in 
the EFKM. This description of the phase-disordered
state appears rather workable from the point of view of, {\it e.g.}, 
prospective calculation of the
transport properties.

In Sec. \ref{sec:exp}, the emerging picture is discussed in the context 
of the ongoing experimental
search for excitonic insulators.
While the experimental situation is still uncertain (see, {\it e.g.,}, a brief
review of recent literature on ${\rm Ta_2 Ni Se_5}$ in 
Sec. \ref{subsec:Ta2NiSe5}), this is likely to change in the near future, 
enabling a more meaningful comparison with the theoretical insights.
We also include a rather qualitative treatment of collective excitations 
(amplitude mode, Sec. \ref{subsec:Higgs}) in light of recent 
experiments\cite{Kogar2017}. 

One can expect that potential applications of the technique developed in 
this paper
extend beyond  those rather limited aspects considered in 
Secs. \ref{sec:high-T} and \ref{sec:exp}, both for the 
EFKM and in the context of other systems. This issue is, among others, 
discussed in Sec. \ref{sec:conclu}.

The reader interested mostly in our {\it results} for the behaviour of the EFKM
at finite temperature might want to skip the description of the formalism in
Secs. \ref{sec:osdm}--\ref{sec:Bures}. On the other hand, those interested 
specifically in the OSDM-based mean-field {\it technique} could, at a first 
reading, proceed directly from Sec. \ref{sec:Bures} to Sec. \ref{sec:conclu}.

Overall, the discussion in the  paper is rather self-contained, as the 
Appendices supply necessary technical details for Secs. \ref{sec:osdm}, 
\ref{sec:low-T}, and \ref{sec:high-T}. 
While some preliminary considerations were reported earlier in Ref. \onlinecite{physicab18}, 
the technique used there is largely inadequate. Hence, 
Ref. \onlinecite{physicab18} is completely superseded by the present paper.

\section{SINGLE-SITE MEAN-FIELD SCHEME AND THE HARTREE--FOCK SOLUTION}
\label{sec:HF}

An ordered excitonic insulator state at $T=0$ is characterised by the uniform
values of $n_{c,i}$, $n_{d,i}$, and (real positive) $\Delta_i$. At a finite temperature, 
these begin to fluctuate, and as long as $T$ is not too low, can be treated
as classical fluctuating quantities (see further discussion in 
Sec. \ref{sec:Bures} below). Given any distribution of local phases $\varphi_i$,
we can perform a gauge transformation,
\begin{equation}
d_i=\tilde{d}_i {\rm e}^{{\rm i}\varphi_i}\,,
\label{eq:gauge}
\end{equation}
which yields real $\langle c^\dagger_i \tilde{d}_i \rangle$ while leaving 
the unperturbed Falicov--Kimball Hamiltonian (\ref{eq:FKM}) 
invariant. The perturbation (\ref{eq:pert}) now reads as
\begin{eqnarray}
&&\delta {\cal H}=-\frac{t^\prime}{2}\sum_{\langle i j \rangle} 
\tilde{d}^\dagger_i \tilde{d}_j {\rm e}^{{\rm i}(\varphi_j-\varphi_i)}+ 
V_0 \sum_i c^\dagger_i \tilde{d}_i {\rm e}^{{\rm i}\varphi_i}-\nonumber \\
&&-\frac{V_1}{2}\sum_{\langle i j \rangle}\left( 
c^\dagger _i \tilde{d}_j{\rm e}^{{\rm i}\varphi_j} + 
c^\dagger_j \tilde{d}_i {\rm e}^{{\rm i}\varphi_i}\right)-
\label{eq:pert2}\\
&& -\frac{V_2}{2} 
\sum_{\langle i j \rangle}\left\{ (\vec{R}_j-\vec{R_i})\cdot \vec{\Xi}
\right\}
\left(c^\dagger _i \tilde{d}_j {\rm e}^{{\rm i}\varphi_j}- c^\dagger_j 
\tilde{d}_i {\rm e}^{{\rm i}\varphi_i}\right) + {\rm H. c.}\,.
\nonumber
\end{eqnarray}
We now proceed with the standard Hartree--Fock decoupling of the interaction
term in Eq. (\ref{eq:FKM}), replacing
\begin{eqnarray}
&&{\cal H}\rightarrow {\cal H}_{mf}=-\frac{t}{2}\sum_{\langle i j \rangle} \left(c^\dagger_i c_j +
c^\dagger_j c_i \right) + E_d \sum_i \tilde{d}^\dagger_i \tilde{d}_i + \nonumber \\
&&+U \sum_i \left\{n_{d,i}c^\dagger_i  c_i+
n_{c,i}\tilde{d}^\dagger_i  \tilde{d}_i-|\Delta_i|\left(c^\dagger_i  \tilde{d}_i+
\tilde{d}^\dagger_i  c_i \right) -\goth{n}_{\goth{d},i}  \right\} \nonumber\\
&&\label{eq:Hmf}
\end{eqnarray}
with the double occupancy on-site,
\begin{equation}
\goth{n}_{\goth{d},i} \equiv \langle c^\dagger_i
d^\dagger_i d_i c_i \rangle\,,
\label{eq:ngoth} 
\end{equation}
given by the mean-field expression, $\goth{n}_{\goth{d},i}=n_{d,i}n_{c,i}-|\Delta_i|^2$. This yields a quadratic Hamiltonian with fluctuating local parameters. 
While
these fluctuations will be taken into account later in a 
self-consistent way, presently we
make use of  {\it virtual-crystal} approximation, averaging both
Eqs. (\ref{eq:FKM}) and  (\ref{eq:pert2}) over the thermal fluctuations of
$n_{c,i}$, $n_{d,i}$, $|\Delta_i|$, and $\varphi_i$. In the spirit of a 
single-site mean-field theory, we assume that  fluctuations on different
sites are mutually uncorrelated. The latter implies that, for example, 
\begin{equation}
\langle  {\rm e}^{{\rm i}(\varphi_j-\varphi_i)} \rangle_T =\cos^2 \kappa\,,\,\,\,
\label{eq:expaver}
\end{equation}
where
\begin{equation}
\cos \kappa \equiv \langle \cos \varphi_i \rangle_T\,,
\label{eq:coskappa}
 \end{equation}
and the subscript $T$ in $\langle ... \rangle_T$ denotes averaging over 
the local thermal fluctuations. 

The resultant uniform virtual crystal will play the r\^{o}le of our mean-field background. The net
virtual crystal Hamiltonian [including the perturbation, Eq. (\ref{eq:pert2})] 
is readily diagonalised as
\begin{equation}
\!\!\!{\cal H}_{vc}\!\!\!=\!\!\!\sum_{\vec{k}}\left[(\epsilon^{(1)}_{\vec{k}}-\mu)f^\dagger_{1,\vec{k}}
f_{1,\vec{k}}+(\epsilon^{(2)}_{\vec{k}}-\mu)f^\dagger_{2,\vec{k}}
f_{2,\vec{k}}\right] - U\! N \langle\goth{n}_\goth{d}\rangle_T\,.
\label{eq:Hvc}
\end{equation}
Here, $\mu$ is the chemical potential, $N$ is the number of sites in the 
lattice, and $\langle \goth{n}_\goth{d}\rangle_T\equiv\langle\goth{n}_{\goth{d},i} \rangle_T$, average
double occupancy $\goth{n}_{\goth{d},i}$ on-site. 
The mean-field energies are given by
\begin{eqnarray}
\epsilon^{(1,2)}_{\vec{k}}&=&\frac{1}{2}\left(E_d+Un+\epsilon_{\vec{k}}+
t^\prime \epsilon_{\vec{k}}\cos^2
\kappa \right)  \mp\frac{W_{\vec{k}}}{2} \,,
\label{eq:mfspectrum} \\
W_{\vec{k}}&=&\sqrt{(\xi_{\vec{k}}+t^\prime \epsilon_{\vec{k}}\cos^2
\kappa )^2+4|U\Delta-V_{\vec{k}}|^2} \,
\label{eq:Wk}
\end{eqnarray}
with $\Delta=\langle|\Delta_i|\rangle_T$,
\begin{equation}
\epsilon_{\vec{k}}=
-\sum_{\alpha=1}^d \cos k_\alpha\,,\,\,\xi_{\vec{k}}=E_{rd} -\epsilon_{\vec{k}}\,,
\label{eq:epsilon}
\end{equation}
and  $E_{rd}=E_d + U (n_c- n_d)$ (here again, $n_{c,d}=\langle n_{c,d;i}\rangle_T$), renormalised
 relative energy of the localised band. The Fourier component of effective 
bare hybridisation is given by 
\begin{equation}
V_{\vec{k}}=\cos \kappa \times \left\{\begin{array}{ll}
V_0+V_1 \epsilon_{\vec{k}},\,\,\,&\mbox{even}\\ 
{\rm i} V_2 \lambda_{\vec{k}},\,&\mbox{odd}\end{array}\,,
\right.
\,\,\,\lambda_{\vec{k}}=-\sum_{\alpha=1}^d \sin k_\alpha.
\label{eq:lambda}
\end{equation}
(depending on the relative parity of the orbitals). The value of the indirect
gap $G$ in the virtual-crystal spectrum is obtained as a difference between
$\epsilon_2$ at $\vec{k}=0$ and $\epsilon_1$ at the corner of the Brillouin zone. Neglecting the perturbation $\delta {\cal H}$, we find
\begin{equation}
  G=\frac{1}{2}\left[\sqrt{(E_{rd}-d)^2+4U^2\Delta^2}+\sqrt{(E_{rd}+d)^2+4U^2\Delta^2}\right]-d\,,
\label{eq:gap}
\end{equation}
which in the limit of $|E_{rd}|,U\Delta \ll d$ yields Eq. (\ref{eq:gapest}).

The original fermionic 
operators,
\begin{equation}
c_i=\frac{1}{\sqrt{N}} \sum_{\vec{k}}{\rm e}^{{\rm i}\vec{k}\vec{R}_i }
c_{\vec{k}}\,,\,\,\, d_i=\frac{1}{\sqrt{N}} {\rm e}^{{\rm i}\varphi_i}
\sum_{\vec{k}}{\rm e}^{{\rm i}\vec{k}\vec{R}_i }
\tilde{d}_{\vec{k}}\,,
\label{eq:Fourier}
\end{equation}
are expressed in terms
of the mean-field quasiparticle operators $f_{1,\vec{k}}$ and  $f_{1,\vec{k}}$
with the help of
\begin{eqnarray}
c_{\vec{k}}&=&\sqrt{\eta_+(\vec{k})}\,f_{1,\vec{k}}+
\sqrt{\eta_-(\vec{k})}\,f_{2,\vec{k}}\,,  
\label{eq:diag1} \\
&~& \nonumber \\
\tilde{d}_{\vec{k}}\frac{U\Delta-V_{\vec{k}}}{|U\Delta-V_{\vec{k}}|}&=&
\sqrt{\eta_-(\vec{k})}\,f_{1,\vec{k}}- \sqrt{\eta_+(\vec{k})}\,f_{2,\vec{k}}\,,
\label{eq:diag2}
%
\end{eqnarray}
where
\[ \eta_\pm(\vec{k}) = \frac{1}{2} \left(1\pm \frac{\xi_{{\vec{k}}}+t^\prime \epsilon_{\vec{k}}\cos^2
\kappa }{W_{\vec{k}}}\right)\,.\]
We now readily find the average values over the canonical ensemble of mean-field fermions ({\it i.e.,} over the Fermi distribution of the mean-field
carriers), denoted $\langle ... \rangle_F$: 
\begin{widetext}
\begin{eqnarray}
\Delta_i^{(0)} &\equiv& \langle c^\dagger_i d_i \rangle_F={\rm e}^{{\rm i} \varphi_i}\Delta^{(0)} = {\rm e}^{{\rm i} \varphi_i}\frac{1}{N} \sum_{\vec{k}}
\Delta_{\vec{k}}\,,\,\,\,\Delta_{\vec{k}}= 
\frac{U \Delta-V^*_{\vec{k}}}{W_{\vec{k}}}\left(n^1_{\vec{k}}-n^2_{\vec{k}}\right)\,,
\label{eq:Delta} \\
n_c^{(0)} &\equiv& \langle c^\dagger_i c_i \rangle_F=\frac{1}{N} \sum_{\vec{k}}n^c_{\vec{k}}\,,\,\,\,n^c_{\vec{k}}= 
\eta_+(\vec{k})n^1_{\vec{k}}+
\eta_-(\vec{k})n^2_{\vec{k}} 
\,,
%
%
\label{eq:nc} \\
n_d^{(0)} &\equiv& \langle d^\dagger_i d_i \rangle_F=\frac{1}{N} 
\sum_{\vec{k}}n^d_{\vec{k}}\,,\,\,\, n^d_{\vec{k}}= 
\eta_-(\vec{k})n^1_{\vec{k}}+
\eta_+(\vec{k})n^2_{\vec{k}}\,, 
%
%
\label{eq:nd} 
\end{eqnarray}
\end{widetext}
where 
\[n^{1,2}_{\vec{k}}= \left({{\rm e}^{\frac{\epsilon_{\vec{k}}^{(1,2)}-\mu}{T}}+1}\right)^{-1} \]
are the Fermi distribution functions in two quasiparticle bands. The actual 
values of parameters $|\Delta_i|$, $n_{c,i}$, and $n_{d,i}$ on-site fluctuate: $|\Delta_i|= \Delta^{(0)}+ \delta |\Delta_i|$, etc. 
The mean-field self-consistency conditions for the average quantities 
$\Delta$, $n_{c}$, and $n_{d}$  [which enter the r.\ h.\ s. of Eqs. (\ref{eq:Delta}--\ref{eq:nd})] take the form
\begin{equation}
\Delta=\Delta^{(0)}+\langle \delta|\Delta| \rangle_T\,,\,\,
n_{c,d}=n_{c,d}^{(0)}+\langle \delta n_{c,d} \rangle_T\,.
\label{eq:mfe0}
\end{equation}
Together with Eq. (\ref{eq:coskappa}) this closes the mean-field scheme\cite{selfcons}. The procedure for evaluating the probability of an on-site 
fluctuation and calculating 
thermal average values will be outlined in the following Secs. 
\ref{sec:osdm}--\ref{sec:Bures}. If one is interested in reviewing the 
{\it results} of this approach, he or she should now proceed to Sec.
\ref{sec:low-T}.

It is  worthwhile to remind the reader that here we encountered three
distinct types of average values: in addition to $\langle ...\rangle$ 
(quantum mechanical average), we also used $\langle ...\rangle_F$ (canonical 
average over distribution of Hartree--Fock quasiparticles) and 
$\langle ...\rangle_T$ (average over the thermal fluctuations on-site). 
We will be using this notation throughout the rest of the paper.

\section{LOCAL FLUCTUATIONS AND THE ON-SITE DENSITY MATRIX}
\label{sec:osdm}

Let us consider a single site (located at origin) in the virtual-crystal
background. There are four quantum states $|s_n\rangle$ available on-site:
\begin{eqnarray}
|s_1\rangle&\equiv& |c\rangle=c^\dagger_0 |0\rangle\,,\,\,
|s_2\rangle\equiv |d\rangle=d^\dagger_0 |0\rangle\,,\,\,\\
|s_3\rangle &=& |0\rangle\,,\,\, 
|s_4\rangle\equiv |cd\rangle=c^\dagger_0 d^\dagger_0 |0\rangle\,,\,\, 
\nonumber
\end{eqnarray}
including two singly occupied states, vacuum state $|0\rangle$, and the 
doubly occupied state, $|cd\rangle$. In the absence of thermal fluctuations
of the on-site parameters, the thermal on-site density matrix (OSDM) is given by
\begin{equation}
\rho^{(0)}_{mn}=\langle\rho^{QM}_{mn}\rangle_F
\equiv\sum_{|\Psi\rangle}\rho^{QM}_{mn}(\Psi)P(\Psi)\,.
\label{eq:odsme}
\end{equation}
Here, the summation is over all basic many-body eigenfunctions $|\Psi \rangle$ 
of the  averaged Hamiltonian. 
Presently, we can choose these to be eigenfunctions of both the virtual-crystal 
({\it i.e.,} averaged Hartree--Fock) Hamiltonian
 (\ref{eq:Hvc}), and of the net particle number operator, $\hat{N}$. 
The matrix
\begin{equation}
\rho^{QM}_{mn}(\Psi) =\langle \Psi \left(|s_n\rangle \langle s_m 
|\right)\Psi\rangle
\label{eq:qmdm}
\end{equation}
is the regular quantum-mechanical OSDM calculated for the state
$|\Psi\rangle$, and
\begin{equation}
P(\Psi)=\frac{1}{Z} \langle \Psi|{\rm e}^{-\frac{{\cal H}_{vc}-\mu \hat{N}}{T}}
|\Psi \rangle\,
\label{eq:PPsi}
\end{equation}
is the canonical probability of this state. $Z$ is the partition function. 

Each eigenvector $|\Psi\rangle$ can be represented as a sum of four 
mutually orthogonal terms, 
\begin{eqnarray}
|\Psi\rangle=&&\!\!\!\!\!\!A_c(\Psi) |c\rangle |\Phi_c(\Psi)\rangle + A_d(\Psi)
|d\rangle |\Phi_d(\Psi)\rangle + \nonumber\\
&&\!\!\!\!\!\!+A_0(\Psi) |0\rangle |\Phi_0(\Psi)\rangle +A_{cd}(\Psi) 
|cd\rangle |\Phi_{cd}(\Psi)\rangle\,,
\label{eq:decomp0}
\end{eqnarray}
where $|\Phi_i\rangle$ are $|\Psi\rangle$-dependent normalised wavefunctions 
defined
on all the $N-1$ sites away from our central site $0$, and 
$|A_c|^2+|A_d|^2+|A_0|^2+|A_{cd}|^2=1$. Owing to the different
net electron numbers on these sites, we have
\begin{equation}
\langle \Phi_0|\Phi_{cd}\rangle=\langle \Phi_0|\Phi_{c,d}\rangle=
\langle \Phi_{cd}|\Phi_{c,d}\rangle=0\,.
\end{equation}
Therefore multiplying $A_0$, $A_{cd}$, or both $A_c$ and $A_d$ by a phase factor
does not affect $\hat{\rho}^{(0)}$ -- only the relative phase of the first and 
second  terms on 
the r.\ h.\ s. of Eq. (\ref{eq:decomp0}) appears in the OSDM. 
An obvious equality
\begin{equation}
d_0d_0^\dagger c_0^\dagger c_0+ c_0c_0^\dagger d_0^\dagger d_0+d_0 c_0 
c_0^\dagger d_0^\dagger+ c_0^\dagger d_0^\dagger d_0 c_0 =1
\label{eq:obvious}
\end{equation}
allows to perform the decomposition (\ref{eq:decomp0}) explicitly by writing
\begin{equation}
A_c |c\rangle |\Phi_c\rangle=d_0d_0^\dagger c_0^\dagger c_0|\Psi\rangle\,,
\label{eq:Acetc}
\end{equation}
etc. Indeed, each term in Eq. (\ref{eq:obvious}) projects upon a single local
state $|s_i\rangle$, and the r.\ h.\ s. of Eq. (\ref{eq:Acetc}) contains all 
those terms in $|\Psi\rangle$ which correspond to the site 0 being occupied by a
$c$-band electron in the absence of a $d$-band one. 
It follows that
\begin{eqnarray}
\!\!\!\!\!A_c |0\rangle|\Phi_c\rangle\!\!&=&\!\!c_0d_0d_0^\dagger 
|\Psi\rangle\,,\,\,\,A_d |0\rangle|\Phi_d\rangle=
d_0c_0c_0^\dagger|\Psi\rangle\,, \label{eq:Phi1}\\
\!\!\!\!\!A_0 |0\rangle|\Phi_0\rangle\!\!&=&\!\!d_0c_0 c_0^\dagger d_0^\dagger
|\Psi\rangle\,,\,\,\,
A_{cd} |0\rangle |\Phi_{cd}\rangle=d_0 c_0|\Psi\rangle\,.\label{eq:Phi2}
\end{eqnarray}
Substituting Eq. (\ref{eq:decomp0}) into Eq. (\ref{eq:odsme}) and using 
anticommutation relationships for the fermion operators on-site yields
\begin{eqnarray}
&&n_c^{(0)}-\goth{n}^{(0)}_\goth{d} = \rho^{(0)}_{11}=\langle |A_c|^2 
\rangle_F\,,\label{eq:odsme1a}\\
&&n_d^{(0)}-\goth{n}^{(0)}_\goth{d} = \rho^{(0)}_{22}=\langle |A_d|^2 
\rangle_F\,,
\label{eq:odsme1b}\\
&&\goth{n}^{(0)}_\goth{d}= \rho^{(0)}_{44}=\langle |A_{cd}|^2 \rangle_F\,,
\label{eq:odsme1bb}\\
&&\rho^{(0)}_{33}=\langle |A_{0}|^2\rangle_F=1+\goth{n}^{(0)}_\goth{d}-n_c^{(0)}-
n_d^{(0)}\,,
\label{eq:odsme1c}\\
&&
\Delta_0={\rm e}^{{\rm i}\varphi_0}\Delta^{(0)}=\rho^{(0)}_{21}=
\langle A_c^*A_d\Big(\langle \Phi_c  |\Phi_d \rangle\Big)\rangle_F\,.
\label{eq:odsme1d}
\end{eqnarray}
Here, the subscript ``F'' again implies canonical average over all 
virtual-crystal  eigenstates
$|\Psi\rangle$.
In the 
Hartree--Fock approximation, the states $|\Psi\rangle$ are merely products of 
operators $f^\dagger_{1,\vec{k}}$ and  $f^\dagger_{2,\vec{k}}$ acting on the 
overall vacuum $|{\rm vac} \rangle$ of the system, and
Eqs. (\ref{eq:odsme1a}--\ref{eq:odsme1d}) are 
readily verified with the help of Eqs. (\ref{eq:Fourier}--\ref{eq:diag2}),
(\ref{eq:Delta}--\ref{eq:nd}), and (\ref{eq:Phi1}--\ref{eq:Phi2}). It is 
equally easy to obtain the standard
Hartree--Fock result,
\begin{equation}
\goth{n}_\goth{d}^{(0)}=n_c^{(0)}n_d^{(0)}-[\Delta^{(0)}]^2\,.
\label{eq:doubleHF}
\end{equation} 

In writing Eq. (\ref{eq:odsme1d}),  we made allowance for a phase-disordered 
state with an arbitrary phase $\varphi_0$ of $\langle c^\dagger_0 d_0 \rangle$, 
which perhaps needs a clarification. The operators $\tilde{d}^\dagger_i$ are
obtained from $f^\dagger_{(1,2),\vec{k}}$ (used to construct the state  
$|\Psi\rangle$) with the help of Eq. (\ref{eq:diag2}), followed by a 
Fourier transform. The phases $\varphi_i$ of the operators $d_i$ can then
be assigned arbitrarily according to Eq. (\ref{eq:gauge}), or alternatively 
one can continue
working in terms of operators $\tilde{d}_i$, inserting the same values of  
$\varphi_i$ 
in Eq.(\ref{eq:pert2}). The state $|\Psi\rangle$ is an eigenstate of the full 
mean-field 
Hamiltonian ${\cal H}_{MF} + \delta {\cal H}$ [see Eqs. (\ref{eq:pert2}) and 
(\ref{eq:Hmf})] 
averaged over thermal fluctuations of these phases
and of other parameters [which is but a site representation of the 
virtual-crystal
Hamiltonian (\ref{eq:Hvc})].

Since the Hartree--Fock quasiparticles form an ideal Fermi gas, the
fluctuations of all the on-site quantities over the canonical distribution of 
the many-body eigenfunctions $|\Psi\rangle$ {\it vanish} in a large system ({\it
i.e.}, for $N \rightarrow \infty$; see Appendix \ref{app:average}). 
Hence, at least in the Hartree--Fock approximation, we can use Eqs. (\ref{eq:odsme1a}--\ref{eq:odsme1d}) to
substitute in Eq. (\ref{eq:decomp0})   
\begin{eqnarray}
\!\!\!\!\!A_c \rightarrow A_c^{(0)} \equiv & 
\sqrt{\langle |A_c|^2 \rangle_F}\,,\,\,\,
A_d \rightarrow A_d^{(0)} \equiv & {\rm e}^{{\rm i}\varphi_0}
\sqrt{\langle |A_d|^2 \rangle_F}\,, \nonumber \\
\!\!\!\!\!A_0 \rightarrow A_0^{(0)} \equiv & 
\sqrt{\langle |A_0|^2 \rangle_F}\,,\,\,\,
A_{cd} \rightarrow A_{cd}^{(0)} \equiv & {\rm e}^{{\rm i}\varphi_0}
\sqrt{\langle |A_{cd}|^2 \rangle_F}\,,
\nonumber\\
\label{eq:A0subst}
\end{eqnarray}
Here, our choice of relative phases, which corresponds
to a real $\langle \Phi_c|\Phi_d \rangle_F$, is a matter of convenience 
and reflects the choice of phases of the states $|\Phi_i\rangle$. Once the 
latter are fixed, this also fixes {\it all} the relative phases of $A_i$. 
This is 
because the Hamiltonian, ${\cal H}+ \delta{\cal H}$, is a non-local 
operator (unlike the OSDM). We will see that varying the phases of $A_i$ 
generally affects the average energy. 
 

From Eqs. (\ref{eq:odsme1a}--\ref{eq:odsme1d}) we 
observe that single-site {\it thermal} fluctuations (distinct from the 
Fermi-distribution fluctuations discussed in the previous paragraph), i.e., 
deviations of the OSDM from
$\hat{\rho}^{(0)}$ of Eq. (\ref{eq:odsme}), are obtained by varying both the 
complex coefficients $A_i$ in Eq. (\ref{eq:decomp0}), and the 
scalar product 
$\langle \Phi_c  |\Phi_d \rangle_F$. The latter, however, is  inconvenient 
as it implies changes to states $|\Phi_{c,d} \rangle$ and makes the procedure
convoluted. Therefore, it is expedient to use operators $a^\dagger$ and 
$b^\dagger$ which diagonalise $\hat{\rho}^{(0)}$:
\begin{eqnarray}
\!\!\!\!\!\!c^\dagger_0&\!\!=&\!\!\cos \frac{\beta^{(0)}}{2} a^\dagger-
\sin \frac{\beta^{(0)}}{2} 
b^\dagger\,,\label{eq:onsitediag1}\\ 
\!\!\!\!\!\!d^\dagger_0&\!\!\equiv&\!\!{\rm e}^{-{\rm i}\varphi_0} 
\tilde{d}^\dagger_0=
{\rm e}^{-{\rm i}\varphi_0}\sin \frac{\beta^{(0)}}{2} a^\dagger+
{\rm e}^{-{\rm i}\varphi_0}\cos \frac{\beta^{(0)}}{2} b^\dagger\,,
\label{eq:onsitediag2}\\
&\,&\tan \beta^{(0)}=\frac{2 \Delta^{(0)}}{n_c^{(0)}-n_d^{(0)}}\,.
\label{eq:onsitediag3}
\end{eqnarray}
While obviously $|ab\rangle\equiv a^\dagger b^\dagger |0\rangle = 
\exp{({\rm i}\varphi_0)}|cd\rangle$,
the singly-occupied part of the decomposition (\ref{eq:decomp0}) is 
re-written as
\begin{equation}
A_c |c\rangle |\Phi_c\rangle + A_d
|d\rangle |\Phi_d\rangle=A_a(\Psi) |a\rangle |\Phi_a(\Psi)\rangle + A_b(\Psi)
|b\rangle |\Phi_b(\Psi)\rangle\,
\end{equation}
with $|a\rangle=a^\dagger|0\rangle$ and $|b\rangle=b^\dagger|0\rangle$. 
This in turn yields
\begin{eqnarray}
&&(A_{a,b}^{(0)})^2\equiv\langle |A_{a,b}|^2 \rangle_F= \frac{1}{2}(n_c^{(0)}+
n_d^{(0)}-2\goth{n}_\goth{d}^{(0)}) \pm \nonumber\\
&&\,\,\,\,\,\,\pm\frac{1}{2}\sqrt{(n_c^{(0)}-n_d^{(0)})^2+4[\Delta^{(0)}]^2}\,,
\label{eq:Aab}\\
&&\langle A_a^*A_b\Big(\langle \Phi_a  |\Phi_b \rangle\Big)\rangle_F=0\,,
\label{eq:Aab1}
\end{eqnarray}
where the last equation implies that $|\Phi_a \rangle$ and $|\Phi_b \rangle$
are orthogonal ``on average'' (again with vanishing canonical fluctuations), 
which is precisely what is needed.
Finally, the first two terms on the r.\ h.\ s. of Eq. (\ref{eq:decomp0}) 
can be re-expressed with the help of
\begin{eqnarray}
\!\!\!\!\!\!\!\!\!\!\!\!\!\!\!A_{a}^{(0)}|0\rangle|\Phi_{a}\rangle&\!\!\!\!=&
\!\!\!\!\left(\!\!\cos \frac{\beta^{(0)}}{2}c_0 d_0 d_0^\dagger+
{\rm e}^{-{\rm i}\varphi_0}\sin \frac{\beta^{(0)}}{2}d_0 c_0 c_0^\dagger\!\!\right)
\!\!|\Psi\rangle, 
\label{eq:Phia}\\
\!\!\!\!\!\!\!\!\!\!\!\!\!\!\!A_{b}^{(0)}|0\rangle|\Phi_{b}\rangle&\!\!\!\!=&
\!\!\!\!\left(\!\!
{\rm e}^{-{\rm i}\varphi_0}\cos \frac{\beta^{(0)}}{2}d_0 c_0 c_0^\dagger-
\sin \frac{\beta^{(0)}}{2}c_0 d_0 d_0^\dagger\!\!\right)
\!\!|\Psi\rangle
\label{eq:Phib}
\end{eqnarray}
[cf. Eqs. (\ref{eq:Phi1}--\ref{eq:Phi2})], resulting in
\begin{eqnarray}
|\Psi\rangle=&&|c\rangle \left(\cos \frac{\beta^{(0)}}{2}A_{a}^{(0)}
|\Phi_{a}\rangle
-\sin \frac{\beta^{(0)}}{2}A_{b}^{(0)}|\Phi_{b}\rangle\right)+ \nonumber\\
&&+{\rm e}^{{\rm i}\varphi_0}|d\rangle\left(\sin \frac{\beta^{(0)}}{2}
A_{a}^{(0)}|\Phi_{a}\rangle+ \cos \frac{\beta^{(0)}}{2}A_{b}^{(0)}
|\Phi_{b}\rangle\right)  
+ \nonumber\\
&&+A_0^{(0)} |0\rangle |\Phi_0\rangle +A_{cd}^{(0)} |cd\rangle 
|\Phi_{cd}\rangle\,.
\label{eq:decomp1}
\end{eqnarray}
The vectors $|\Phi_{a,b}(\Psi)\rangle$ on the r.\ h.\ s. can be expressed 
directly via Eqs. (\ref{eq:Aab}) and (\ref{eq:Phia}--\ref{eq:Phib}), 
whereas $|\Phi_0\rangle$ and
$|\Phi_{cd}\rangle$ are similarly calculated using Eqs. (\ref{eq:Phi2}) and 
(\ref{eq:odsme1bb}--\ref{eq:odsme1c}). 

A single-site fluctuation (a fluctuation of OSDM at site 0) corresponds to a 
change of 
coefficients in Eq. (\ref{eq:decomp1}), as detailed in Appendix \ref{app:su4}. 
For every state 
$|\Psi\rangle$ this yields a perturbed state $|\tilde{\Psi}\rangle$, 
characterised by the parameters $\beta$, $\phi$, $\theta_i$, and $\gamma_i$:
\begin{widetext} 
\begin{eqnarray}
&&\!\!\!\!\!\!\!\!\!\!\!\!|\tilde{\Psi}(\beta,\phi,\theta_1,\theta_2,\theta_3,
\gamma_1,\gamma_2,\gamma_3)\rangle=
{\rm e}^{{\rm i} \gamma_1}\cos \theta_2 \cos \theta_3 \left[|c\rangle 
\left({\rm e}^{{\rm i} \gamma_2}\cos \frac{\beta}{2}\cos \frac{\theta_1}{2}
|\Phi_a(\Psi)
\rangle -{\rm e}^{-{\rm i} \gamma_2} \sin \frac{\beta}{2}\sin \frac{\theta_1}{2}
|\Phi_b(\Psi)\rangle\right)+
{\rm e}^{{\rm i}\phi}|d\rangle \right. \times \nonumber \\
&&\!\!\!\!\!\!\!\!\!\!\!\!\times\!\!\! \left. \left({\rm e}^{{\rm i} \gamma_2}
\sin 
\frac{\beta}{2}\cos \frac{\theta_1}{2}
|\Phi_a(\Psi)\rangle+{\rm e}^{-{\rm i} \gamma_2}\cos 
\frac{\beta}{2}\sin \frac{\theta_1}{2}   
|\Phi_b(\Psi)\rangle\right)\right]\!\!\!+
\sin \theta_2 \cos \theta_3|0\rangle |\Phi_0(\Psi)\rangle+
{\rm e}^{{\rm i}(2\gamma_1+\phi+\gamma_3)}\sin \theta_3  |cd\rangle 
|\Phi_{cd}(\Psi)\rangle\,, 
\label{eq:decompfluctgen}
\end{eqnarray}
with $0 \leq \beta \leq \pi$, $0 \leq \phi, \gamma_{1,3} \leq 2\pi$, and 
$0\leq \theta_{1,2,3}, |\gamma_2| \leq \pi/2$ (see Appendix \ref{app:su4}). 
Here, we will be  interested for the most part 
in the case of 
half-filling,  considering only fluctuations $|\tilde{\Psi}\rangle$ that 
preserve the site occupancy, $n=n_c+n_d=1$ (see Sec. \ref{sec:Bures} for 
further discussion). This implies $|A_0|=|A_{cd}|$ for both $|\Psi\rangle$ and 
$|\tilde{\Psi}\rangle$, i.e., $\sin \theta_2=\tan \theta_3$. In this case,
Eq. (\ref{eq:decompfluctgen}) takes a simpler form
\begin{eqnarray}
&&\!\!\!\!\!\!\!\!\!\!\!\!|\tilde{\Psi}(\beta,\phi,\theta_1,\theta_3,
\gamma_1,\gamma_2,\gamma_3)\rangle=
{\rm e}^{{\rm i} \gamma_1}\sqrt{\cos 2\theta_3}\left[ |c\rangle 
\left({\rm e}^{{\rm i} \gamma_2}\cos \frac{\beta}{2}\cos \frac{\theta_1}{2}
|\Phi_a(\Psi)\rangle - 
{\rm e}^{-{\rm i} \gamma_2}\sin \frac{\beta}{2}\sin \frac{\theta_1}{2}
|\Phi_b(\Psi)\rangle\right)+
{\rm e}^{{\rm i}\phi} |d\rangle  \times \right. \nonumber\\
&&\!\!\!\!\!\!\!\!\!\!\!\!\left. \left({\rm e}^{{\rm i} \gamma_2}\sin 
\frac{\beta}{2}\cos \frac{\theta_1}{2}|\Phi_a(\Psi)\rangle+
{\rm e}^{-{\rm i} \gamma_2}\cos \frac{\beta}{2}\sin \frac{\theta_1}{2}   
|\Phi_b(\Psi)\rangle\right) \right]+
\sin \theta_3\Big[ |0\rangle |\Phi_0(\Psi)\rangle+ 
{\rm e}^{{\rm i}(2\gamma_1+\phi+\gamma_3)} |cd\rangle 
|\Phi_{cd}(\Psi)\rangle\Big] \,, 
\label{eq:decompfluct}
\end{eqnarray}
\end{widetext}
where  $0 \leq \beta \leq \pi$, $0 \leq \phi,\gamma_{1,3} \leq 2\pi$,
$0\leq \theta_{1}, |\gamma_2| \leq \pi/2$, and $0\leq \theta_{3} \leq \pi/4$. 
As explained in Appendix \ref{app:su4}, Eq. (\ref{eq:decompfluctgen})  
is obtained using an SU(4) 
transformation in the four-dimensional space of vectors 
$|a\rangle |\Phi_a\rangle$,
$|b\rangle |\Phi_b\rangle$, $|0\rangle |\Phi_0\rangle$, and 
$|cd\rangle |\Phi_{cd}\rangle$, 
followed by an SU(2) 
transformation of the on-site states $|a\rangle$ and $|b\rangle$. Eq.
(\ref{eq:decompfluctgen}) [or similarly Eq. (\ref{eq:decompfluct})] can be 
re-written in the form 
\begin{equation}
\!\!\!\!\!\!\!\!\!\!\!\!|\tilde{\Psi}(\beta,\phi,\theta_1,\theta_2,\theta_3,
\gamma_1,\gamma_2,\gamma_3)\rangle\!\!=\!\!\hat{S}
(\beta,\phi,\theta_1,\theta_2,\theta_3,\gamma_1,\gamma_2,\gamma_3) |\Psi\rangle,
\label{eq:defS}
\end{equation}
with the expression for the operator $\hat{S}$ given in Appendix 
\ref{app:cost}. 
 The operator $\hat{S}$ is unitary 
``on average,'' $\langle \Psi | \hat{S}^\dagger\hat{S} | \Psi \rangle_F=1$, 
which can be verified directly.

The values of 
parameters $\beta$, $\phi$ and $\theta_i$ are the same for all unperturbed 
eigenstates  $|\Psi\rangle$. For the $n=1$ case of Eq. (\ref{eq:decompfluct}) 
the unperturbed states $|\Psi\rangle$
 are recovered, $|\tilde{\Psi}\rangle=|\Psi\rangle$, at $\gamma_{1,2,3}=0$, 
$\beta=\beta^{(0)}$, $\phi=\varphi_0$, $\theta_{1,3}=\theta_{1,3}^{(0)}$, where
\begin{equation}
\cos \theta_1^{(0)}=\frac{\sqrt{(1-2n_d^{(0)})^2+4(\Delta^{(0)})^2}}
{1-2\goth{n}_\goth{d}^{(0)}}\,,\,\,\,
\sin\theta_3^{(0)}=\sqrt{\goth{n}_{\goth{d}}^{(0)}}\,
\label{eq:theta130}
\end{equation} 
[for the $n \neq 1$ case, see Eqs. 
(\ref{eq:0anglesgen1}--\ref{eq:0anglesgen2})].
The fluctuation  of wave functions is translated into a 
fluctuation of OSDM, 
which is calculated as [cf. Eq. (\ref{eq:odsme})]
\begin{equation}
\rho_{mn}(\beta,\phi,\theta_i,\gamma_i)=\sum_{|\Psi\rangle}\rho^{QM}_{mn}
(\tilde{\Psi})P(\Psi)\,.
\label{eq:osdmfluct}
\end{equation}
We readily find that the form of OSDM, corresponding to Eq. 
(\ref{eq:decompfluctgen}) 
or (\ref{eq:decompfluct}), coincides with expressions given below in 
Sec. \ref{sec:Bures} [see Eqs. (\ref{eq:osdmgen}) and (\ref{eq:osdmn1}) 
respectively; the physical meaning of quantities $\gamma_i$, which 
do {\it not} affect the density matrix, will be discussed in Sec. 
\ref{subsec:gamma}]. As for
the {\it energy cost} of the  local fluctuation, it can be evaluated via
\begin{equation}
\!\!\!\!\!\!\!\!\delta E(\beta,\phi,\theta_i,\gamma_i)\!\!\!=\!\!\!
\langle\tilde{\Psi}| {\cal H}_{mf}+ \delta{\cal H}|
\tilde{\Psi} \rangle_{T',\varphi_0}-\langle{\Psi}| {\cal H}_{mf}+ \delta{\cal H}|
{\Psi} \rangle_{T',\varphi_0}
\label{eq:Efluct}
\end{equation}
Here, the average $\langle...\rangle_{T'}$ includes, in addition to the 
Fermi distribution averaging 
$\langle...\rangle_F$, also taking the average value over thermal 
fluctuations of the background, 
{\it i. e.} over thermal fluctuations on all sites other than our central 
site. In addition, it is 
convenient to add to the $\langle...\rangle_{T'}$ also an averaging over 
the phase $\varphi_0$, which is random
and obeys Eq. (\ref{eq:coskappa}). We recall that  $\varphi_0$ is the 
value of $\phi$ at site 0 before 
the fluctuation; it enters Eqs. 
(\ref{eq:decompfluctgen}--\ref{eq:decompfluct}) via $|\Phi_{a,b}\rangle$ 
(see Appendix 
\ref{app:cost} for details). 


Note that Eq. (\ref{eq:Efluct}) is written for the  mean field 
Hamiltonian, $ {\cal H}_{mf}+ \delta{\cal H}$. In the first term in 
Eq. (\ref{eq:Efluct}), the average 
values which enter the Hartree--Fock 
expression for the interaction energy in Eq. (\ref{eq:Hmf}) should be 
evaluated in the
perturbed state  $|\tilde{\Psi} \rangle$. 

In the important case when the fluctuation does not change the values of the 
three angles $\theta_i$, {\it i.e.}, when
$\theta_i=\theta_i^{(0)}$, the operator $\hat{S}$ is unitary not only 
``on average'' (see above), but also precisely\cite{diffsites}: 
$ \hat{S}^\dagger\hat{S}=1$.
In this situation, Eq. (\ref{eq:Efluct}) can be conveniently recast as
\begin{equation}
\delta E(\beta,\phi,\theta_i^{(0)},\gamma_i)=
\langle \Psi | \hat{S}^\dagger \left[{\cal H}+ \delta{\cal H}, \hat{S} 
\right]| \Psi \rangle_{T',\varphi_0} \,. 
\label{eq:Efluct2}
\end{equation}
Furthermore, in this case   $\hat{S}$ both commutes with the interaction term in ${\cal H}$, 
Eq.(\ref{eq:FKM}),
and does not change the average value of the mean-field interaction term in
Eq. (\ref{eq:Hmf}), which term therefore does not contribute to $\delta E$.

In the opposite case of $ \hat{S}^\dagger\hat{S}\neq 1$ (when the thermal 
fluctuations of $\theta_i$ are taken into account), taking the average over 
thermal fluctuations of the background in Eq. 
(\ref{eq:Efluct}) is problematic, because fluctuations
on different sites are no longer fully independent (a fluctuation of the OSDM
at site $i$ affects the value of the OSDM at site $j$). However, in 
Sec. \ref{sec:high-T} below we will provide a tentative argument to the 
effect that this
averaging almost does not affect the mean-field solution, so that one can use a 
simpler equation,
\begin{equation}
\!\!\!\delta E
\approx
\langle\tilde{\Psi}| {\cal H}_{mf}+ \delta{\cal H}|
\tilde{\Psi} \rangle_{F,\varphi}-\langle{\Psi}| {\cal H}_{mf}+ \delta{\cal H}|
{\Psi} \rangle_{F,\varphi}\,,
\label{eq:Efluct3}
\end{equation}
where the average is taken only over the thermal fluctuations of the phases 
$\varphi_i$ at all sites and over the Fermi distribution.
We recall that the phases $\varphi_i$ are detached from the fermionic degree 
of freedom of the Hartree -- Fock quasiparticles [see Sec. \ref{sec:HF}, 
beginning with Eq. (\ref{eq:gauge})], hence
we did not need to fully take the phase degree of freedom into account when 
constructing the representation (\ref{eq:decomp1}) of an eigenstate 
$|\Psi\rangle$ (where we, however, made allowance for an arbitrary 
$\varphi_0$). 
These phases {\it do} affect the energy via $\delta {\cal H}$, 
Eq. (\ref{eq:pert2}). 

Let us pause and briefly discuss the meaning of equations 
(\ref{eq:decompfluctgen}--\ref{eq:Efluct}). It will 
be expedient to consider first the case of a half-filled EFKM ($n=1$) at a 
relatively low temperature, $T \ll G$ [see Eq. (\ref{eq:gap})], when the lower 
quasiparticle band is filled and thermal excitations of quasiparticles 
across the gap  
freeze out. Then there remains only one term in the sum on the r.\ h.\ s. of
Eq. (\ref{eq:odsme}), corresponding to a fully occupied lower mean-field band,
\begin{equation}
| \Psi \rangle= | \Psi_0 \rangle \equiv \prod_{\vec{k}} f^\dagger_{1,\vec{k}} 
|{\rm vac} \rangle\,,
\end{equation} 
which can be decomposed  according to Eq. (\ref{eq:decomp1}). The states 
$|s_n\rangle |\Phi_{n'} \rangle$, which appear on the r.\ h.\ s., 
are eigenstates
of the particle number operator, and their structure is very similar to that 
of the original state $| \Psi_0 \rangle$. In fact, they are very close to being 
eigenfunctions of the Hamiltonian, solving the real-space Schr\"odinger 
equation everywhere except at the central site, $i=0$, and at neighbouring 
sites. In the case of the correct eigenfunction $| \Psi_0 \rangle$, 
the contributions of all such states should be ``stitched together'' at $i=0$,
which is achieved by the proper choice of coefficients in Eq. 
(\ref{eq:decomp1}). 
In general, these coefficients determine the OSDM, and vice versa. 
Hence, Eqs. (\ref{eq:decompfluctgen}--\ref{eq:decompfluct}) correspond to a 
situation whereby OSDM 
fluctuates while the average energy per site away from the central site (and 
neighbouring sites) stays constant, and the fundamental ``building blocks'' 
$|\Phi_n\rangle$ of the wave function $| \Psi_0 \rangle$ are kept intact. 
The state $|\tilde{\Psi}\rangle$ is not an eigenstate of the Hamiltonian, 
{\it i.e.}, quantum  mechanics dictates that the defect created at $i=0$ 
should eventually spread and dissipate, but we assume that this process 
(which involves redistribution of slow-moving fermions $d$) is 
slow in comparison to the thermal fluctuations of OSDM. The energy of 
this variational state can still be calculated on average, 
see Eq. (\ref{eq:Efluct}).
We note that 
calculating a quantum mechanical density matrix, Eq. (\ref{eq:qmdm}), 
for {\it any} state  (and not only for an eigenstate) is a legitimate 
operation. Overall, we conjecture that this kind of procedure is the closest 
analogue of a Weiss-type mean field for the case when itinerant carriers are 
present.  

Away from half-filling, or when temperature is sufficiently high to allow for 
quasiparticles populating the upper band, the system (in the absence of 
single-site fluctuations) can be found in one of the possible eigenstates 
$|\Psi\rangle$ with a probability $P (\Psi)$, as given by Eq. (\ref{eq:PPsi}).
Once an on-site fluctuation occurs (adiabatically), this state is deformed 
according 
to Eq. (\ref{eq:decompfluctgen}), and we wish to calculate the momentary value
of the  OSDM before the (deformed) state $|\tilde{\Psi}\rangle$ evolves
quantum mechanically, and certainly before the statistical probability of this
evolving state is adjusted via thermalisation.
Thus,  the contribution of the state $|\tilde{\Psi}\rangle$ to the 
(thermal) OSDM, Eq. (\ref{eq:osdmfluct}), clearly comes with the original 
weight $P (\Psi)$. Finally,
the fact that the values of parameters $\beta$, $\phi$, and $\theta_i$  in 
Eq. (\ref{eq:decompfluctgen}) are the same for all $|\Psi\rangle$, ensures that 
the 
thermal distribution away from the central site (relative contributions of 
different original $|\Psi\rangle$'s to the mutually orthogonal ``sectors'' 
$|\Phi_n\rangle$) remains undisturbed.

To summarise, our results in this section establish a one-to-one 
correspondence between the local fluctuations (i.e., thermal fluctuations of
the OSDM) and the deformations of the many-body wavefunctions. This allows
to calculate the energy cost $\delta E$ of a given fluctuation,
and hence the probability of such fluctuation, $w \propto \exp(-\delta E/T)$.
However, we still need to know the phase volume corresponding to each
fluctuation, or, in other words, the integration measure in the space of
parameters $\beta$, $\phi$, and $\theta_i$. This issue will be addressed in the
following section. 

\section{DENSITY MATRIX PARAMETRISATION AND THE BURES MEASURE}
\label{sec:Bures}

Our objective is to construct a single-site mean-field description for the 
EFKM at finite temperatures. To this end, in the previous section we analysed
the fluctuations of the on-site density matrix in the mean-field background. 
In order to proceed with the calculation of the average values, we need to 
determine the corresponding integration measure. In other words, we must 
learn to integrate over fluctuating variables, when these variables are 
elements of a density matrix, {\it i.e.,} form a peculiar mathematical object.

In Sec. \ref{sec:osdm} we also saw that the local fluctuations of the 
many-body wavefunctions, and hence of the OSDM, can be described in terms of
angular parameters $\beta$, $\phi$, and $\theta_{1,2,3}$ (additional 
wavefunction parameters $\gamma_{1,2,3}$ do not affect the OSDM). 
Here we will arrive
at exactly the same parametrisation of the OSDM, Eqs. (\ref{eq:osdmgen}) and 
(\ref{eq:osdmn1}), in a direct way, without analysing the wave functions of the
system.  

Taking into account that this is not a very familiar subject, we will first 
mention some general notions and results\cite{Bengsston}, and then show 
how these are adapted to the case at hand. An ${\cal N} \times {\cal N}$ 
positive-definite Hermitian matrix $\hat{\cal M}$ can be parametrised as
\begin{equation}
{\hat{\cal M}}={\hat{\cal U}} \hat{\Lambda} \, {\hat{\cal U}}^\dagger\,,
\label{eq:matgen}
\end{equation}
where $\hat{\Lambda}$ is a diagonal matrix of positive eigenvalues $\lambda_i$ 
($i=1,...,{\cal N}$), and ${\hat{\cal U}}$ is an SU(${\cal N}$) unitary 
matrix. While the 
question how to perform an integration over the 
elements of $\hat{\cal U}$ in principle 
has a ready answer, due to the existence of a well-defined 
Haar measure $d \Omega^{H}_{\cal N}$ in 
SU(${\cal N}$), integration over the eigenvalues $\lambda_i$ does present a 
difficulty. It is immediately clear that the corresponding integration measure 
must show a 
non-trivial dependence on the eigenvalues $\lambda_i$, vanishing whenever any 
two eigenvalues coincide, $\lambda_m=\lambda_n$. This is due to the 
fact that the matrix $\hat{\Lambda}$ 
(and hence ${\hat{\cal M}}$) will then be invariant under the action of the 
corresponding SU(2) subgroup of the SU(${\cal N}$) (acting on these two 
eigenvalues only; this corresponds to an invariance of a 2 $\times$ 2 unity matrix 
under unitary transformations). The presence of these ``inefficient'' (in 
terms of varying ${\hat{\cal M}}$) transformations should then be 
compensated by the measure of the  $\lambda_i$ integration vanishing at the 
 point  $\lambda_m=\lambda_n$. 

The appropriate  {\it Bures measure} $d\Omega_B$ for 
integration in the 
space of matrices $\hat {\cal M}$ is constructed based on an assumption that 
an infinitesimal distance $ds_B$ between two matrices ${\hat{\cal M}}$ and
${\hat{\cal M}}+ \delta{\hat{\cal M}}$ is given by the 
{\it Bures metric}\cite{Bures1969}, which can be cast in the form\cite{Hall98}
\begin{equation}
(ds_B)^2=\sum_{j=1}^{\cal N} \frac{(d\lambda_j)^2}{\lambda_j} + 4 \sum_{j < k} 
\frac{(\lambda_i-\lambda_k)^2}{\lambda_i+\lambda_k}\left[ (dx_{jk})^2 + 
(dy_{jk})^2 \right]\,.
\end{equation}
Here, the quantities $dx_{jk}$ and $dy_{jk}$ are real and imaginary parts of the
matrix element ${\cal U}_{jk}$ in Eq. (\ref{eq:matgen}) for the case of an 
infinitesimal unitary transformation, and the basis $j,k$ is chosen in such a 
way that  $\hat {\cal M}$ is diagonal. If we also add a requirement that the 
trace of the matrix ${\hat{\cal M}}$ should be equal to unity, 
$\sum_i \lambda_i=1$ (which merely 
introduces the delta function in the following 
equation\cite{Hall98,Slater99,Sommers03}), the expression for
the Bures measure reads as\cite{Hall98,Byrd01}:
\begin{eqnarray}
d\Omega_B= \delta\left(\sum_{i=1}^{\cal N} \lambda_i-1 \right)&& 
\left[\prod_{j<k}4\frac{(\lambda_i-\lambda_k)^2}{\lambda_i+\lambda_k} \right]
\times \nonumber \\
&&\times\left[\prod_{i=1}^{\cal N} \frac{d \lambda_i}{\sqrt{\lambda_i}}\right]d 
\Omega^{H}_{\cal N}\,.
\label{eq:Buresgen}
\end{eqnarray} 

In our case, the density matrix $\hat{\rho}$ is a 4 $\times$ 4 one, built on the local 
states $|c \rangle$,  $|d \rangle$, $|0 \rangle$, and $|cd \rangle$ (in this 
order). Furthermore, our Hamiltonian preserves the total number of electrons, 
{\it and} we are using the basic wavefunctions of the whole system, which 
 diagonalise the 
particle number operator (unlike, {\it e.g.,} the BCS wave functions). In this 
case, those off-diagonal elements of $\hat{\rho}$ 
which involve at least one of the states $|0 \rangle$ and $|cd \rangle$, being 
also off-diagonal in the electron number on-site, must vanish. Hence, the
only off-diagonal elements of $\hat{\rho}$ which may be present are $\rho_{12}$ 
and  $\rho_{21}=\rho_{12}^*$. Therefore the matrix ${\hat{\cal U}}$ in Eq. 
(\ref{eq:matgen}) must take the form 
\begin{equation}
{\hat{\cal U}}=\left(\begin{array}{cccc}
{\rm e}^{-{\rm i} \phi/2} \cos \frac{\beta}{2} & -{\rm e}^{-{\rm i} \phi/2 }\sin 
\frac{\beta}{2} & 0 & 0 \\
~ & ~ & ~ & ~ \\
{\rm e}^{{\rm i} \phi/2} \sin \frac{\beta}{2} & {\rm e}^{{\rm i} \phi/2} 
\cos \frac{\beta}{2} & 0 & 0 \\
0 & 0 & 1 & 0 \\
0 & 0 & 0 & 1

\end{array}\right),
\end{equation}
with an SU(2) matrix (omitting the additional phase parameter which cancels 
out in the final expression for $\hat{\rho}$) in 
the 
upper left quadrant. The Bures distance then reads as
\begin{equation}
(ds_B)^2=\sum_{j=1}^{4} \frac{(d\lambda_j)^2}{\lambda_j}+
4  
\frac{(\lambda_1-\lambda_2)^2}{\lambda_1+\lambda_2}\left[ (dx_{12})^2 + 
(dy_{12})^2 \right]\,,
\end{equation}
and the first product on the r.\ h.\ s. of Eq. (\ref{eq:Buresgen}) is 
replaced with a single factor,
\[4\frac {(\lambda_1-\lambda_2)^2}{\lambda_1+\lambda_2}\,.\] We then 
parametrise
the four eigenvalues according to
\begin{eqnarray}
\lambda_1&=&r\cos^2 \theta_3 \cos^2 \theta_2 \cos^2 \frac{\theta_1}{2}\,,
\,\,\,\,\nonumber\\
\lambda_2&=&r\cos^2 \theta_3 \cos^2 \theta_2 \sin^2 \frac{\theta_1}{2}\,, 
\nonumber\\
\lambda_3&=&r\cos^2 \theta_3 \sin^2 \theta_2\,, \,\,\,\,
\lambda_4=r\sin^2 \theta_3\,.
\end{eqnarray}
Substituting these into Eq. (\ref{eq:Buresgen}), working out the Jacobian and 
performing the integration over real positive $r$, we arrive at
\begin{equation}
d\Omega_B=\frac{32}{\pi^3}\cos^4 \theta_3 \cos^3 \theta_2 \cos^2 \theta_1 
\sin \beta d \theta_1 d \theta_2 d\theta_3 d\beta d \phi\,.
\label{eq:Bures4}
\end{equation}
The five angles in Eq. (\ref{eq:Bures4}) vary within the ranges
\begin{equation}
0 \leq \theta_1, \theta_2, \theta_3 \leq \frac{\pi}{2}\,,\,\,\,
0 \leq \beta \leq \pi\,,\,\,\,
0 \leq \phi \leq 2\pi \,,
\end{equation}
and in writing Eq. (\ref{eq:Bures4}) we renormalised the overall pre-factor in 
such a way that
$\int d\Omega_B=4$, the net number of states on-site. Eq. (\ref{eq:matgen}) 
yields the OSDM in the form
\begin{widetext}
\begin{equation}
\hat{\rho}=\left(\begin{array}{cccc}
\frac{1}{2} \cos^2 \theta_3 \cos^2 \theta_2 (1+\cos \theta_1 \cos \beta) &
\frac{1}{2}{\rm e^{-{\rm i} \phi}}\ \cos^2 \theta_3 \cos^2 \theta_2 
\cos \theta_1 \sin \beta & 0 & 0 \\
~ & ~ & ~ & ~ \\
\frac{1}{2}{\rm e^{{\rm i} \phi}}\ \cos^2 \theta_3 \cos^2 \theta_2 \cos \theta_1 
\sin \beta & \frac{1}{2} \cos^2 \theta_3 \cos^2 \theta_2 (1-\cos \theta_1 
\cos \beta)& 0 & 0\\
0 & 0 & \cos^2 \theta_3 \sin^2 \theta_2 & 0 \\
0 & 0 & 0& \sin^2 \theta_3 
\end{array}\right).
\label{eq:osdmgen}
\end{equation}
\end{widetext}

The angles $\theta_{1,2,3}$, $\beta$, and $\phi$ will be treated as 
fluctuating classical variables, akin to  Euler angles in 
the familiar spin-coherent states technique\cite{Assa} for an insulating 
magnet. This is expected to be 
qualitatively correct as long as thermal fluctuations are sufficiently strong. 
We note that at  very low temperatures (well below the ordering temperature
$T_{cr}$) {\it any} 
single-site treatment would be inadequate. 

For given values of the angles, the 
quantum  average value of an on-site operator $\hat{\cal O}$ can be
 read off Eq. (\ref{eq:osdmgen}) according to
\begin{equation}
{\cal O} (\beta,\phi,\theta_1,\theta_2,\theta_3)= \sum_{i,j=1}^{4}\rho_{ij}
{\cal O}_{ji}\,.
\end{equation}
For example,
\begin{eqnarray}
&&\tilde{n}_c(\beta,\theta_1,\theta_2,\theta_3)=\rho_{11}+ \rho_{44}=\nonumber \\
&&=\frac{1}{2} \cos^2 \theta_3 \cos^2 \theta_2 (1+\cos \theta_1 \cos \beta)+  
\sin^2 \theta_3\,, \label{eq:tildencgen}\\
&&\tilde{n}_d(\beta,\theta_1,\theta_2,\theta_3)=\rho_{22}+ \rho_{44}=\nonumber \\
&&=\frac{1}{2} \cos^2 \theta_3 \cos^2 \theta_2 (1-\cos \theta_1 \cos \beta)+  
\sin^2 \theta_3\,,  \label{eq:tildendgen}
\end{eqnarray}
etc. Thermal fluctuations of the OSDM  lead to fluctuations of the band 
occupancies on-site, and the tilde accents on the l.\ h.\ s. of Eqs. 
(\ref{eq:tildencgen}--\ref{eq:tildendgen}) serve to 
distinguish these fluctuating quantities from their average values 
[see Eq. (\ref{eq:mfe0})].

It can be assumed that local fluctuations of the net carrier occupancy on-site, 
$\tilde{n}=\tilde{n}_c+\tilde{n}_d$, are suppressed by a strong 
electrostatic interaction 
(not explicitly included in our model), hence we only need to consider those 
fluctuations 
which do not disturb the value of $n$, with the integration measure
\begin{equation}
d\Omega_B(n)=\delta(\cos^2 \theta_3 \cos^2 \theta_2+ 2 \sin^2 \theta_3-n) 
d\Omega_B\,.
\label{eq:Buresn}
\end{equation}
We find that the total number of states on-site available for a given $n$ is
\begin{equation}
I(n) \equiv \int d\Omega_B(n)=\frac{8}{\pi} \log \left| 
\frac{\sqrt{2-n}+\sqrt{n}}{\sqrt{2-n}-\sqrt{n}} \right| - 
\frac{8}{\pi}\sqrt{n}\sqrt{2-n}\,.
\label{eq:In}
\end{equation}
Throughout the rest of this paper we shall restrict ourselves to the 
half-filled case, $n=1$.  
Then, the value of $\theta_2$ in the integrand should be substituted according to
\begin{equation}
\sin \theta_2=\tan \theta_3\,.
\label{eq:theta2}
\end{equation}
whereas the integration measure, $d\Omega \equiv d\Omega_B(1)$, can be 
obtained by performing 
the integral over $\theta_2$ in Eq. (\ref{eq:Buresn}):
\begin{eqnarray}
d\Omega  &=&\frac{1}{A(\tau)} \cot 2 \theta_3 \cos^2 \theta_3 \cos^2 
\theta_1 \sin \beta d \theta_1 d \theta_3 d \beta d \phi\,,
\nonumber \\
A(\tau)&=&\frac{1}{4}\pi^2 (2 |\log \tau| - \log 2 -1)\,.
\label{eq:Bures}
\end{eqnarray}
The integration should be performed over the range
\begin{equation}
0 \leq \theta_1 \leq \frac{\pi}{2}\,,\,\, \tau \leq \theta_3 \leq 
\frac{\pi}{4}\,,\,\,
0 \leq \beta \leq \pi\,,\,\,\,0 \leq \phi \leq 2\pi
\label{eq:Buresrange}
\end{equation}
for a small but finite value of $\tau>0$, which then should be taken to zero in 
the final expressions for thermal average values. This procedure is required
due to a logarithmic divergence arising from the singularity of the measure 
$d\Omega$ at $\theta_3 \rightarrow 0$; the latter in turn reflects the 
logarithmic singularity of the quantity $I(n)$ [Eq. (\ref{eq:In})] at $n=1$. 
The measure in 
Eq. (\ref{eq:Bures}) has been re-normalised according to $\int d \Omega=1$. 

Using Eq. (\ref{eq:theta2}), we find the final expression for the OSDM,
\begin{widetext}
\begin{equation}
\hat{\rho}=\left(\begin{array}{cccc}
\frac{1}{2} \cos 2 \theta_3  (1+\cos \theta_1 \cos \beta) &
\frac{1}{2}{\rm e^{-{\rm i} \phi}}\ \cos 2 \theta_3  \cos \theta_1 \sin \beta & 
0 & 0 \\
~ & ~ & ~ & ~ \\
\frac{1}{2}{\rm e^{{\rm i} \phi}}\ \cos 2 \theta_3 \cos \theta_1 \sin \beta & 
\frac{1}{2} \cos  2\theta_3   (1-\cos \theta_1 \cos \beta)& 0 & 0\\
0 & 0 & \sin^2 \theta_3 & 0 \\
0 & 0 & 0& \sin^2 \theta_3 
\end{array}\right),
\label{eq:osdmn1}
\end{equation}
\end{widetext}
and hence for the (fluctuating)
local physical quantities at $n=1$:
\begin{eqnarray}
&& \!\!\!\!\!\!\!\!\!\!\!\!\!\!\! \langle c^\dagger_0 c_0\rangle \equiv  
\tilde{n}_c(\beta,\theta_1,\theta_3)=\frac{1}{2}+\frac{1}{2}\cos 2 \theta_3 
\cos \theta_1 \cos \beta\,,\\
&& \!\!\!\!\!\!\!\!\!\!\!\!\!\!\! \langle d^\dagger_0 d_0\rangle \equiv 
\tilde{n}_d(\beta,\theta_1,\theta_3)= \frac{1}{2} - \frac{1}{2} \cos 2 
\theta_3 \cos \theta_1 \cos \beta\,, \\
&& \!\!\!\!\!\!\!\!\!\!\!\!\!\!\! \langle c^\dagger_0 d_0\rangle \equiv 
{\rm e}^{{\rm i} \varphi} \tilde{\Delta}(\beta,\theta_1,\theta_3) = \rho_{21}= 
\nonumber \\
&&\,\,\,=\frac{1}{2}{\rm e}^{{\rm i}\phi}\cos 2 \theta_3 \cos \theta_1 \sin 
\beta\,,
\label{eq:tdelta}
\\
&& \!\!\!\!\!\!\!\!\!\!\!\!\!\!\!\langle c^\dagger_0 d^\dagger_0 d_0 c_0\rangle 
\equiv \tilde{\goth{n}}_{\goth{d}}(\beta,\theta_1,\theta_3)= \rho_{44}=\sin^2 
\theta_3\,.
\label{eq:tdouble}
\end{eqnarray}

Should one desire to consider only those fluctuations which respect the 
Hartree--Fock condition, $\tilde{\goth{n}}_\goth{d} = \tilde{n}_c 
\tilde{n}_d - \tilde{\Delta}^2$, 
an additional restriction is introduced, fixing the
value of $\theta_3$:
\begin{equation}
\sin^2 \theta_3= \frac{\sin \theta_1- \sin^2 \theta_1}{2 \cos^2 \theta_1}\,.
\label{eq:HFtheta}
\end{equation}
The integration measure, Eq. (\ref{eq:Bures}), is then multiplied 
by an appropriate delta function. A more mathematically rigorous procedure 
might yield also an additional $\theta_1$-dependent prefactor, but since, in 
the regime of interest, the 
average
value of $\theta_1$ is typically mid-range (away from potential singularities),
this is unlikely to affect the results.

We are finally in a position to complete our mean-field scheme, as outlined in 
Sec. \ref{sec:HF}. At a fixed density $n=1$, there are only 
three\cite{selfcons} independent
mean-field parameters, $\Delta$, $\cos \kappa$, and $n_d$ (with $n_c=1-n_d$).
We first note that the three phases $\gamma_i$ in Eq.
(\ref{eq:decompfluct}) do not
affect the density matrix and should be integrated over, with the measure and
ranges
\begin{equation}
d \Gamma=\frac{d\gamma_1 d \gamma_2 d \gamma_3}{4\pi^3}\,,\,\,\,
0\leq \gamma_{1,3} \leq 2 \pi,\,\,\,-\frac{\pi}{2} \leq \gamma_2\leq 
\frac{\pi}{2}
\label{eq:Gammameasure}
\end{equation}
(see Appendix \ref{app:su4}).
Writing the probability of a local fluctuation of the OSDM as
\begin{eqnarray}
w(\beta,\phi,\theta_1,\theta_3)&=&\frac{1}{Q}\int {\rm e}^{-\frac{1}{T} 
\delta E(\beta,\phi,\theta_1,\theta_3, \gamma_1,\gamma_2, \gamma_3)}d \Gamma,
\\
Q &\equiv& \int {\rm e}^{-\frac{1}{T} 
\delta E} d\Gamma d \Omega 
\end{eqnarray}
[see Eqs. (\ref{eq:Efluct}) and (\ref{eq:Bures}--\ref{eq:Buresrange})], 
we can substitute in Eqs. (\ref{eq:coskappa}) and (\ref{eq:mfe0})
\begin{eqnarray}
&&\!\!\!\!\!\!\!\!\!\!\!\!\!\!\!\cos \kappa = \int \cos \phi \, 
w(\beta,\phi,\theta_1,\theta_3) d\Omega\,,\\
&&\!\!\!\!\!\!\!\!\!\!\!\!\!\!\!\langle \delta|\Delta|\rangle_T= 
\int \left[\tilde{\Delta}(\beta,\theta_1,\theta_3)-
\Delta^{(0)}\right] w (\beta,\phi,\theta_1,\theta_3)d\Omega 
\,,
\label{eq:dDeltaTgen}\\
&&\!\!\!\!\!\!\!\!\!\!\!\!\!\!\!\langle \delta n_d\rangle_T= 
\int \left[\tilde{n}_d(\beta,\theta_1,\theta_3)-
n_d^{(0)}\right] w (\beta,\phi,\theta_1,\theta_3) d\Omega 
\,. 
\label{eq:dndTgen}
\end{eqnarray}
Self-consistency is ensured due to the dependence of  
$\Delta^{(0)}$ and $n_d^{(0)}$ [as well as  $\delta E$, which is affected
due to the structure of wave functions in the
averaging procedure in Eq. (\ref{eq:Efluct})] on 
$\Delta$, $\cos \kappa$, and $n_d$ [see Eqs.(\ref{eq:Delta}), (\ref{eq:nd}), 
and (\ref{eq:mfe0})].

We shall now turn to implementing this approach 
and studying the properties of resultant  mean-field solution in two different 
temperature 
regimes. In the simplified treatment which follows, we will be interested only
in the thermal fluctuations of the phase $\phi$ and of the angle $\beta$ 
(the latter affecting in turn the quantity $n_c-n_d$ and the
absolute value $\Delta$ 
of the hybridisation), while assuming that all other variables are frozen at
their respective virtual-crystal values, $\theta_i=\theta_i^{(0)}$ and 
(except in Sec. \ref{subsec:gamma}) $\gamma_i=0$. Formally, this corresponds 
to multiplying the integration
measure by the appropriate delta functions. Since the values 
$\theta_i^{(0)}$ lie somewhere in the middle of the integration range ({\it 
i.e.,} away from any singularities)
and the measure of integration over $\gamma_i$, Eq. (\ref{eq:Gammameasure}), 
is featureless, we can expect
that no qualitatively important effects are left out of our results for
$\Delta(T)$ and $\cos \kappa$. Nevertheless, these neglected fluctuations
can result in an additional $T$-dependent term in the specific heat.

\section{LOW-TEMPERATURE ORDERING TRANSITION}
\label{sec:low-T}

We begin with the low-temperature regime of $T\stackrel{<}{\sim} T_{cr}$. 
In this region, only fluctuations of the phase $\varphi_i$ are expected 
to be appreciable (and therefore there are altogether three self-consistency 
equations to solve,
for $n_d$, $\Delta$ and $\varphi$). Nevertheless, in order to provide 
connexion with the
discussion of the high-temperature regime we will also allow for small 
fluctuations of  
$\beta$ [which in turn lead to fluctuations of the absolute value $\Delta$ 
of hybridisation 
$\langle c^\dagger_0 d_0 \rangle$, see Eq. (\ref{eq:tdelta})]. Neglecting the
small fluctuations of both $\theta_i$ and $\gamma_i$, we use 
Eqs. (\ref{eq:Efluct2}) and 
(\ref{eq:commwithH}--\ref{eq:commwithdH}) to find the energy cost $\delta E$ of
a single-site fluctuation of $\phi$ and $\beta$:
\begin{eqnarray}
&&\!\!\!\!\!\!\!\!\delta E = 2 (\cos \kappa -\cos \phi) 
\left[t^\prime l_d^{(0)} \cos \kappa- V_0 \Delta^{(0)}+V_1 
l_\Delta^{(0)}+\right. \nonumber \\
&&\!\!\!\!\!\!\!\!\left.\vphantom{+V_1 l_\Delta^{(0)}}+V_2 m \right]+\delta 
\beta (\cos \kappa -\cos \phi)\left[ V_1 l_c^{(0)}+t^\prime l_\Delta^{(0)} 
\cos \kappa-\right. V_0\times \nonumber \\
&&\!\!\!\!\!\!\!\!\times \left. \left(n^{(0)}_{c}-n^{(0)}_{d}\right)
\right]
+\frac{1}{4}(\delta \beta )^2 \left[l_c^{(0)}+E_d\left(n_c^{(0)}-n_d^{(0)}\right)
 \right]\,,
\label{eq:delowT}
\end{eqnarray}
where $\delta \beta\equiv \beta - \beta^{(0)}$ is assumed small. 
In writing the (small) second and third terms in Eq. (\ref{eq:delowT}), we
omitted contributions of higher order in $t^\prime$ and $V_i$. 
The four real quantities $l_c^{(0)}$, $l_d^{(0)}$, $l_\Delta^{(0)}$, and $m$ are 
defined as
\begin{eqnarray}
l_{c,d}^{(0)}=-\frac{1}{N}\sum_{\vec{k}}n_{\vec{k}}^{c,d}\epsilon_{\vec{k}}\,,&
\,\,\,&
l_{\Delta}^{(0)}=-\frac{1}{N}\sum_{\vec{k}}\Delta_{\vec{k}}\epsilon_{\vec{k}}\,,
\nonumber\\
m&=&-\frac{{\rm i}}{N}\sum_{\vec{k}}\Delta_{\vec{k}}\lambda_{\vec{k}}
\label{eq:deflm}
\end{eqnarray}
[see Eqs. (\ref{eq:epsilon}), (\ref{eq:lambda}), and 
  (\ref{eq:Delta}--\ref{eq:nd})]. As for the phase-space integration
measure [Eq. (\ref{eq:Bures})], to leading order it reduces to just
$d\phi d \beta$.

We begin with discussing the first term in Eq. (\ref{eq:delowT}), {\it i.e.}, 
the
$\delta \beta=0$ case. The expression in square brackets is of the first order
in parameters $t^\prime$, $V_0$, and $V_1$, and of second order in $V_2$. The 
latter is so because in the expression for $m$ the quantity $\Delta_{\vec{k}}$ is
even in momentum at $V_2=0$  [see Eq. (\ref{eq:Delta})] whereas 
$\lambda_{\vec{k}}$ is always odd, hence to leading order $m$ is proportional 
to $V_2$:
\begin{equation}
m=V_2 \tilde{m}\cos \kappa\,,\,\,\,\,\tilde{m}=\frac{1}{N} \sum_{\vec{k}} 
\frac{(\lambda_{\vec{k}})^2}{W_{\vec{k}}}\left( n^1_{\vec{k}}-n^2_{\vec{k}} \right)
\end{equation}
[see Eq. (\ref{eq:Wk})].
Accordingly, the first three terms in the brackets can be readily obtained 
within the first-order perturbation theory in $\delta {\cal H}$, although
strictly speaking, our expression in Eq.  (\ref{eq:delowT}) includes 
self-consistent corrections  (note that the perturbation $\delta {\cal H}$ 
also gives rise to small changes in $l^{(0)}_{c,d}$ and $\Delta^{(0)}$). 
The $V_2$ term, however, can {\it not} be obtained as a second-order 
perturbative correction, as the latter includes the effects of wave-function
readjustment away from our central site and therefore cannot be used to 
construct a viable single-site mean-field scheme.

We also note that the first term in Eq. (\ref{eq:delowT}) is similar to that
obtained in a Weiss description of an XY magnet. Specifically, $\cos \kappa$ 
plays the r\^{o}le of magnetisation, the sum of exchange terms is loosely 
paralleled by
\begin{equation}
J=2t^\prime l_d^{(0)} +2V_2^2\tilde{m},
\end{equation} 
and 
\begin{equation}
H=-2V_0 \Delta^{(0)} +2V_1 l_\Delta^{(0)}
\label{eq:Hlow}
\end{equation}
is the
``external field''. This similarity is an expected one, given the U(1) 
nature of the order parameter $\varphi$, yet we note that the direct 
correspondence between the EFKM and an XY magnet, as outlined above, occurs
in the single-site treatment to leading order in $\delta {\cal H}$, but not
necessarily beyond that.

\begin{figure}
\includegraphics{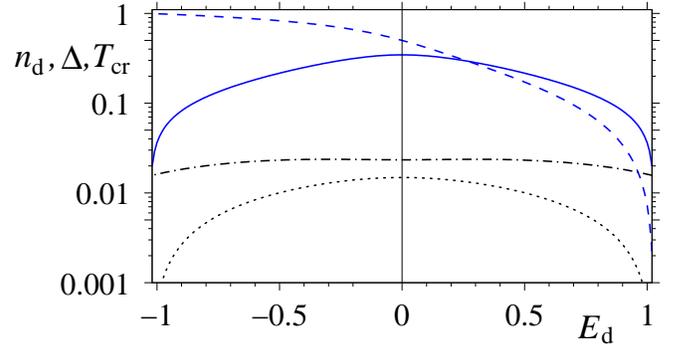}
\caption{\label{fig:Tcr} (colour online) The mean-field critical 
temperature $T_{cr}$ for a 2D EFKM at half-filling (n=1) with $U=1$ and 
the perturbation 
$t^\prime=-0.045$ (dotted line) or $V_2=0.15$ (dashed-dotted line). 
Solid and
dashed lines show the unperturbed values of $\Delta$ and $n_d$ at $T=0$. The BEC scenario is realised as long as $T_{cr}$ is much smaller than crossover
temperature $T_*$, which implies also $T_{cr} \ll U\Delta$.
}
\end{figure}

Analysing Eq. (\ref{eq:delowT}) to leading order in $\delta {\cal H}$, it is 
easy to see that the effect of the second term is negligible (in particular,
$\langle \delta \beta \rangle_T \rightarrow 0$), hence, this term can be 
omitted.  
Furthermore, at low $T$ one can neglect the difference between $\Delta^{(0)}$ 
and $\Delta$, etc., and also write [using Eqs. (\ref{eq:Delta}--\ref{eq:nd})]
\begin{equation}
l_\Delta^{(0)} \approx - E_d \Delta\,.
\label{eq:ldeltalowT}
\end{equation}
The self-consistently conditions of the mean-field theory are given by Eqs. 
(\ref{eq:mfe0}) {\it at zero T} (when the fluctuation terms on the r.\ h.\ s. 
vanish), supplemented with Eq. (\ref{eq:coskappa}) for $\cos \kappa$.
All the statistical properties (such as average values and standard deviations
of $\phi$ and $\delta \beta$) are readily expressed in terms of 
imaginary-argument Bessel functions $I_n$. For example, the partition function
takes the form
\begin{eqnarray}
Z_0 &=& {\rm const} \cdot I_0\left(\frac{J \cos \kappa+H}{T} \right) \sqrt
{\frac{T}{l_c+ E_d(n_c-n_d)}} \times \nonumber \\
&&\times \exp\left\{-(J \cos \kappa+H)\frac{\cos \kappa}{T}
\right\} 
\label{eq:Z0} 
\end{eqnarray}
[where the pre-factor includes also the constants originating from the 
integration measure (\ref{eq:Bures})], etc. We find
\begin{equation}
\langle (\delta \beta)^2 \rangle_T=\frac{2 T}{l_c^{(0)}+ E_d(n_c-n_d)}\,,
\label{eq:dbetalow}
\end{equation}
an expected linear (in $T$) behaviour. At $H=0$, an ordering transition 
takes place
at $T_{cr}=J/2$ (see Fig. \ref{fig:Tcr}), with $\cos \kappa$ vanishing above 
$T_{cr}$ and
\[\cos \kappa = \left(2 \frac{T_{cr}-T}{T_{cr}} \right)^{1/2}\,,\,\,\,
0<\frac{T_{cr}-T}{T_{cr}} \ll 1 \,.\]

\begin{figure}
\includegraphics{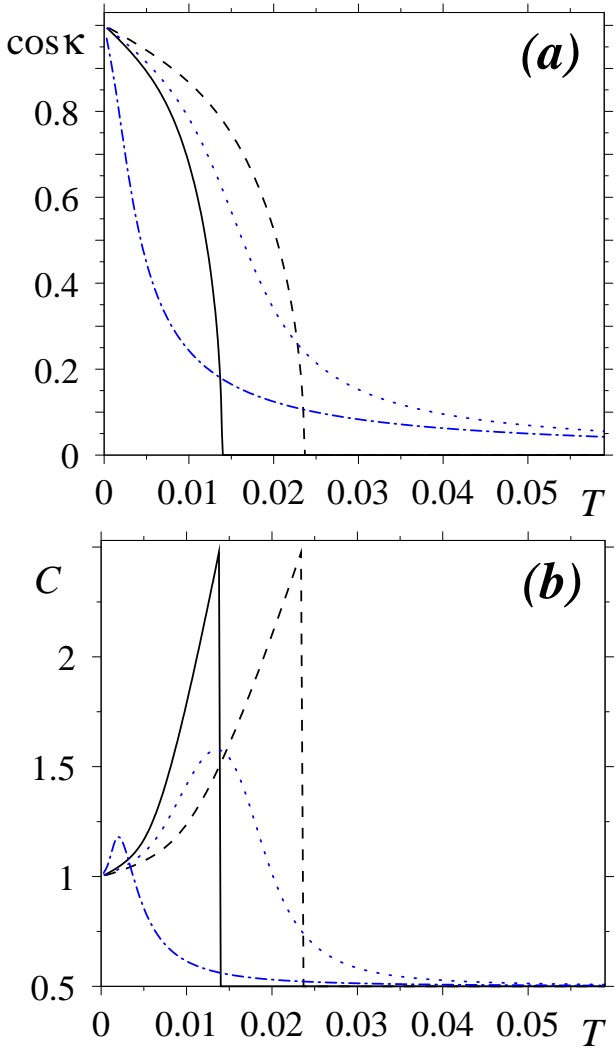}
\caption{\label{fig:lowT} (colour online) Mean-field temperature dependence
of $\cos \kappa$ {\it (a)} and specific heat {\it (b)} for a 2D EFKM with $U=1$ and
$E_d=0.2$ at $n=1$ 
in the low-temperature regime. Solid and dashed lines correspond, respectively,
to $t^\prime=-0.045$ ($T_{cr} \approx 0.014$) and $V_2 = 0.15$ 
($T_{cr} \approx 0.024$). Dashed-dotted line corresponds to $V_0=-0.008$, and the 
dotted one to combined $t^\prime=-0.045$ and $V_1=-0.04$; 
from Eq. (\ref{eq:Hlow})
one finds that for both of the latter two cases $H\approx 0.005$. 
}
\end{figure}

At $H>0$, the phase transition is replaced by a smooth crossover, with 
$\cos \kappa$ asymptotically vanishing at high temperatures:
\begin{equation}
\cos \kappa \approx \frac{H}{2T-J}\,,\,\,\,\,\,T \gg J,H.
\label{eq:highTkappa}
\end{equation}
(note the similarity to the Curie--Weiss law).
More generally (but still to leading order in $\delta {\cal H}$), $\cos \kappa$ throughout the $T\ll T_*$ range
solves the equation
\begin{equation}
I_0\left(\frac{J\cos \kappa +H}{T}\right) \cos \kappa = I_1\left(\frac{J\cos \kappa +H}{T}\right)\,,
\label{eq:numcos}
\end{equation}
and should be found numerically (see Fig. \ref{fig:lowT} {\it a}). 

At low temperatures, $T\stackrel{<}{\sim}T_{cr}$, and for $n=1$ (when the excitonic
gap is present at the chemical potential), the contribution of fermionic 
degrees of freedom to entropy is exponentially small and can be neglected. 
Thus the entropy can be evaluated as
$S= \log Z_0$ [see Eq. (\ref{eq:Z0})], and the {\it specific 
heat} as $C=T \partial S/\partial T$. Using also Eq. (\ref{eq:numcos}), we find 
\begin{equation}
C=-(J \cos \kappa + H) \frac{\partial \cos \kappa}{\partial T} +\frac{1}{2}\,.
\label{eq:ClowT}
\end{equation}
At $H=0$, it suffers a negative jump of $\Delta C = -2$ at $T_{cr}$, whereas at
$H>0$ and at temperatures $T\gg J,H$, 
\begin{equation}
C \approx \frac{4TH^2}{(2T-J)^3}+\frac{1}{2}\,.
\end{equation}

Numerical results for $C$ are shown in Fig. \ref{fig:lowT} b.
The finite 
value of $C=1$ obtained at 
$T \rightarrow 0$ is an expected artefact of treating the $\phi$ and $\beta$ 
degrees of freedom classically. This value includes a $T$-independent (at $T \ll T_*$)
contribution of $1/2$, originating from the small fluctuations of $\beta$, see
Eq. (\ref{eq:dbetalow}). Taking into account small fluctuations of 
other classical degrees
of freedom, which were assumed frozen [such as $\theta_i$ and $\gamma_i$ in 
Eq.(\ref{eq:decompfluctgen})] will yield  additional constant terms in the 
specific heat. On the other hand, treating all the degrees of freedom as 
quantum would not affect the value of $C$ at higher $T$, whereas at 
$T \rightarrow 0$ one would obtain the correct result, $C \rightarrow 0$. 
Obviously, a proper description in the latter regime should be based on the 
analysis of the low-energy, long-wavelength excitations 
(cf. Ref. \onlinecite{prb12}), rather than on a single-site approach as 
considered presently.

The numerical results shown in Fig. \ref{fig:lowT}  were obtained as outlined above.
First, the  $T=0$ mean-field equations in the absence of 
$\delta {\cal H}$ were solved, producing the values of $\Delta$ and $n_{c,d}$
(see  Fig. \ref{fig:Tcr}).
These are substituted into the leading-order 
Eq. (\ref{eq:numcos}), yielding $\cos \kappa$ as a function of temperature 
(Fig. \ref{fig:lowT} {\it a}),
and Eq. (\ref{eq:ClowT}) then gives the specific heat 
(Fig. \ref{fig:lowT} {\it b}).

A more exact solution to the mean-field equations would require taking into
account the subleading terms in powers of $\delta {\cal H}$, which in turn 
depend on $\cos \kappa$ both directly and self-consistently. However, such 
treatment is
unwarranted here, in view of the obvious limitations of our approach at low $T$.
In reality, the results obtained in this section are in any case only as good as a 
single-site description of an XY model in the low-temperature and critical 
regions would be (note also that the 
competition\cite{Farkasovsky08,prb12,Batista02} between different phases at
$T=0$ implies that the system is frustrated). In other words, they have  a 
rough qualitative 
validity, missing a number of important features and strongly overestimating 
the stability of the ordered phase (and the value
of $T_{cr}$).

Indeed, for the values of $U$ and $E_d$ used in Fig. \ref{fig:lowT}, the 
analysis\cite{prb12} of low-energy spectra at $T=0$ gives the minimal absolute
value of $t^\prime$ required to stabilised a uniform ordered phase as 
$|t^\prime_{cr}|\approx 0.04$. Hence we estimate that for $t^\prime = 0.045$, 
which barely exceeds this, the actual value of $T_{cr}$ should be at least 
an order of magnitude smaller than $T_{cr} \approx 0.014$ shown in  
Fig. \ref{fig:lowT}.
The critical values of hybridisations\cite{prb12}, $V_{0,cr}\approx-0.096$ 
and $V_{2,cr}\approx 0.21$ are greater than those used in Fig. \ref{fig:lowT},
implying that in reality the ordering transition (which perhaps also takes place at a much lower
temperature) is a transition into a competing charge-ordered state, and not into the uniform 
phase. 

Physically, the reason for these inaccuracies is that in this regime an 
important r\^{o}le is played
by the low-energy, long-wavelength collective excitations\cite{prb12} 
(phase mode, as opposed
to the amplitude mode discussed in Sec. \ref{subsec:Higgs} below), 
which cannot be treated adequately within a single-site approach. Furthermore,
the actual behaviour may depend on the dimensionality of the system 
(as it does for the XY model, with 2D being a special 
case due to the possibility of vertex formation\cite{Kosterlitz}), 
which is also overlooked
in a single-site treatment.  Noting that these shortcomings are shared
by the available descriptions of the EFKM ordering transition, including Refs.
\onlinecite{Apinyan,condEFKM}, we omit further discussion of the literature.

However, we expect that these complications are restricted to
the low-temperature range of $T \stackrel{<}{\sim} T_{cr}$, whereas 
at higher $T$
(where short-range fluctuations become more prominent) 
one can hope to obtain a more faithful picture.   

\section{PHASE-DISORDERED EXCITONIC INSULATOR AND THE HIGH-TEMPERATURE 
CROSSOVER}
\label{sec:high-T}

Presently, we will consider the high-temperature regime of a fully 
phase-disordered excitonic insulator at $T \gg T_{cr}$. 
In this case, $\cos \kappa \rightarrow 0$ [see Eq. (\ref{eq:highTkappa})] 
and therefore the perturbation 
$\delta {\cal H}$ vanishes on average, 
$\langle \delta {\cal H} \rangle_T=0$ (the latter equality holds to leading
order in $T_{cr}/T$ and becomes exact in the case where $V_0=V_1=0$). 
Hence, formally $\delta {\cal H}$ 
does not affect the virtual-crystal Hamiltonian
(\ref{eq:Hvc}), nor indeed any quantity arising in our single-site 
mean-field description. While this writer 
believes that 
physically the perturbation is nevertheless essential for the validity of 
the qualitative scenario presented here, this is 
not the place for an in-depth discussion of this potentially controversial 
issue. Very briefly, we expect that the actual physical situation is 
reminiscent of that at $T=0$, when a finite, but 
small, value of 
$t^\prime$ or $V_{0,1,2}$ in Eq. (\ref{eq:pert}) is 
required\cite{prb12,Farkasovsky08} to stabilise 
the state with a uniform value of $\Delta>0$ and 
a uniform $\phi$, yet once such a state is stable, the higher-energy
properties of the mean-field solution (such as the magnitude of $\Delta$)
to leading order do not depend on the perturbation.
The difference here is that in the case of disordered $\phi$ the actual phase
transition  at the critical value of a 
perturbation 
parameter\cite{prb12,Farkasovsky08} (found to take place at $T=0$) 
should be replaced by a smooth crossover (where the value of $\Delta>0$ 
saturates once the perturbation strength exceeds a certain characteristic 
scale), 
since the system does not 
undergo a symmetry change.

We further note that even if the 
perturbation 
is not sufficiently strong to stabilise a uniform ordered excitonic 
insulator at $T=0$ and additional charge ordering appears 
at low temperatures ({\it i.e.}, if the value of the corresponding perturbation
parameter is less than the critical one), the analysis in this 
section is still likely 
to be relevant for the behaviour of the system at 
higher $T$, when both phase and charge orders melt. While this issue merits 
further study, it also falls beyond the scope of this work.

As we already mentioned in the Introduction, the phase-disordered excitonic 
insulator state does not break any symmetry, and therefore increasing 
temperature further should result in a decrease of $\Delta=\frac{1}{2} \langle 
\cos 2 \theta_3 \cos \theta_1 \sin \beta\rangle_T$ [see Eq. (\ref{eq:tdelta})] 
via a smooth 
crossover. We are not specifically interested in the situation where 
the average value of
$\theta_1$ approaches $0$ [corresponding to a band insulator without mixed 
valence, see Eqs. (\ref{eq:tdouble}--\ref{eq:HFtheta})] or $\pi/2$ (this corresponds to the high-temperature 
limit of the two equally populated bands and $\Delta\rightarrow0$, see below).
Elsewhere, 
weak or moderate 
fluctuations of $\theta_1$ around its average value
do not affect the average values of $\Delta$ or $n_d$ and add  little to the 
qualitative picture. Treating these
fluctuations would also necessitate a straightforward but cumbersome 
calculation, as one cannot use a simpler formula, Eq. (\ref{eq:Efluct2}).
Therefore, we will treat the angles $\theta_1$ and $\theta_3$ 
as frozen at their 
virtual-crystal values $\theta_i^{(0)}$.
Fluctuations of the angle $\beta$, 
on the contrary, can affect the average
value of $\Delta$, decreasing it when the average $\beta$ is close to 
$\pi/2$ and increasing
$\Delta$ whenever the end points $0$ or $\pi$ are approached.

We also assume that the values of phases $\gamma_i$ in 
Eq. (\ref{eq:decompfluct}) are still frozen at $\gamma_i=0$; the effect 
of fluctuations of $\gamma_i$ will be discussed in Sec. \ref{subsec:gamma}. 
Then, the energy cost of a single-site fluctuation
of the angle $\beta$ can be deduced from Eq. (\ref{eq:commwithH}) as
\begin{eqnarray}
\delta E(\beta)&=& 4 l_c \sin^2 \frac{\beta-\beta_0}{4} + 2 l_\Delta 
\sin \frac{\beta-\beta_0}{2}+ 
\label{eq:dET} \\
&&+\frac{E_d}{2} \sqrt{n^2-4\goth{n}^{(0)}_{\goth{d}}} \left(
\cos \beta^{(0)}-\cos \beta \right)\,.
\nonumber
\end{eqnarray} 
By construction, the value of $\delta E$ vanishes
at $\beta=\beta^{(0)}$ (unperturbed virtual crystal). The quantities 
$l_c$ and $l_\Delta$ 
obey the self-consistency conditions,
\begin{eqnarray}
\!\!\!l_c&\!\!\!=&\!\!\!\int_0^\pi \!\!\!\left(-l^{(0)}_\Delta \sin
\frac{\beta-\beta^{(0)}}{2}+l_c^{(0)}\cos\frac{\beta-\beta^{(0)}}{2}\right) 
w(\beta) \sin \beta d \beta\,, 
\nonumber \\&&\label{eq:dlcT}\\
\!\!\!l_\Delta&\!\!\!=&\!\!\!\int_0^\pi \!\!\!\left(-l^{(0)}_d 
\sin\frac{\beta-\beta^{(0)}}{2}+l_\Delta^{(0)}\cos
\frac{\beta-\beta^{(0)}}{2}\right) 
w(\beta) \sin \beta d \beta\,, 
\nonumber\\&&\label{eq:dldeltaT}
\end{eqnarray}
[see Appendix \ref{app:cost}, Eqs. (\ref{eq:dlcTg}--\ref{eq:dldeltaTg})], with
the values of $l^{(0)}_{c,d,\Delta}$ given by Eqs. (\ref{eq:deflm}), and
\begin{equation}
w(\beta)=\frac{1}{Q}\,{\rm e}^{-\delta E(\beta)/T}\,,\,\,\,Q=\int_0^\pi 
{\rm e}^{-\delta E(\beta)/T} \sin \beta d\beta\,.
\label{eq:probT}
\end{equation}
Here the integration measure, $\sin \beta d \beta$, again comes from Eq. 
(\ref{eq:Bures}).

As the temperature is 
lowered toward $T_{cr}$, the fluctuations of $\beta$ become small, and one
can expand Eq. (\ref{eq:dET}) in powers of $\delta \beta =\beta-\beta^{(0)}$. 
At the same time, the quantities $l_c$ and $l_\Delta$ approach their 
respective virtual-crystal values, $l_c^{(0)}$ and $l_\Delta^{(0)}$. 
Using also Eq. (\ref{eq:ldeltalowT}), we find for this low-$T$ region of the
phase-disordered state
\begin{equation}
\delta E (\delta \beta) \approx \frac{1}{4} \left[ l_c^{(0)}+ 
E_d (n^{(0)}_c-n^{(0)}_d)
\right](\delta \beta)^2\,,
\end{equation}
matching the last term in Eq. (\ref{eq:delowT}). 

For all temperatures above $T_{cr}$, we can now use Eqs. (\ref{eq:dET}) and 
(\ref{eq:probT}) to explicitly
write Eqs. (\ref{eq:dDeltaTgen}--\ref{eq:dndTgen})
as
\begin{eqnarray}
&&\!\!\!\!\!\!\!\!\langle \delta \Delta\rangle_T=
\frac{1}{2}\sqrt{n^2-4\goth{n}^{(0)}_{\goth{d}}}
\int_0^\pi(sin \beta-\sin \beta^{(0)}) w(\beta)\sin \beta d\beta\,,
\nonumber\\
&&\label{eq:dDeltaT} \\
&&\!\!\!\!\!\!\!\!\langle \delta n_d\rangle_T=\frac{1}{2}
\sqrt{n^2-4\goth{n}^{(0)}_{\goth{d}}}
\int_0^\pi(\cos \beta^{(0)}-\cos \beta) w(\beta)\sin \beta d\beta\,.
\nonumber\\
&&\label{eq:dndT}  
\end{eqnarray}
Standard deviations of $\Delta$ and $n_d$ from their average values,
\begin{eqnarray}
\sigma_\Delta&=&\left[\langle  \left(\delta \Delta-\langle \delta 
\Delta\rangle_T\right)^2
\rangle_T\right]^{1/2}\,, \label{eq:Deltasd}\\
\sigma_d&=&\left[\langle  \left(\delta n_d-\langle \delta n_d\rangle_T\right)^2
\rangle_T\right]^{1/2}\,,
\end{eqnarray}
can be evaluated in a similar way.
The mean-field scheme is closed by substituting 
(\ref{eq:dDeltaT}--\ref{eq:dndT}) into Eqs. (\ref{eq:mfe0}). The four resultant
self-consistency equations [including also Eqs. 
(\ref{eq:dlcT}--\ref{eq:dldeltaT})] are 
readily solved numerically.

\begin{figure*}
\includegraphics{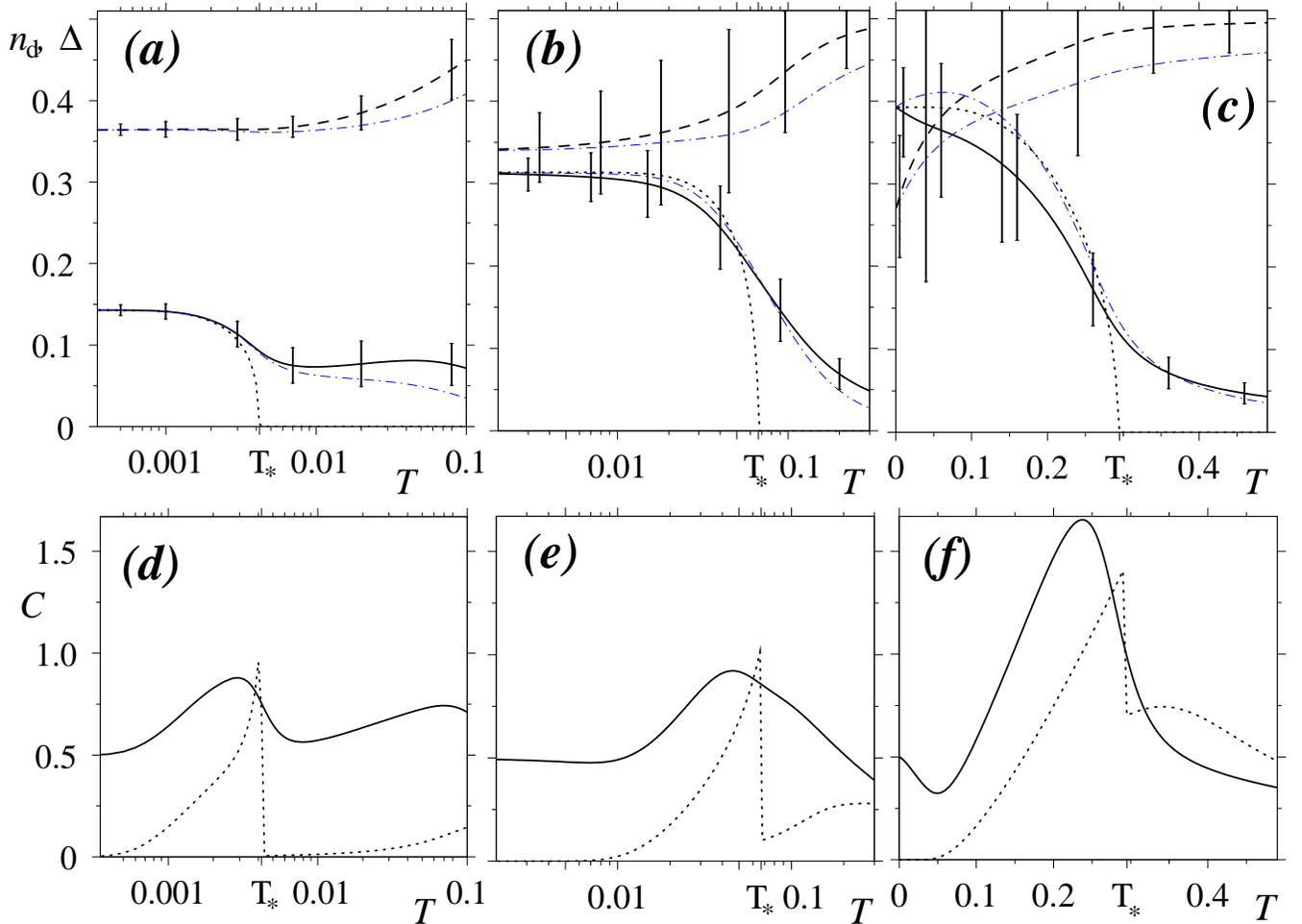}
\caption{\label{fig:highT} (colour online) Single-site mean-field solution 
for a 2D EFKM 
at $n=1$ in the phase-disordered region, $T>T_{cr}$,  for $E_d=0.2$ and $U=0.5$ 
(panels {\it a,d}), $U=1$
({\it b,e}), and $U=2$ ({\it c,f}). In panels {\it a,b,c}, the solid and 
dashed lines
show the values of $\Delta$ and $n_d$, with the error bars corresponding 
to the standard
deviations. The dashed-dotted lines represent the virtual-crystal 
contributions 
$\Delta^{(0)}$ and $n_d^{(0)}$, whereas the dotted lines show the 
Hartree--Fock solution 
for $\Delta$ 
which neglects the effect of thermal fluctuations on-site. 
In {\it d,e,f}, solid lines
show the specific heat $C$ obtained for the full mean-field solution, 
whereas the dotted lines correspond to neglecting the thermal fluctuations.}  
\end{figure*}

Typical results, obtained for a two-dimensional system at $n=1$ are shown in 
Fig. \ref{fig:highT}. Since the ordering transition temperature is determined
by the parameters of the perturbation $\delta {\cal H}$ 
(see Sec. \ref{sec:low-T}) and, at least within the present approach, can be
arbitrarily small, we may carry out our computation for the phase-disordered 
state at any finite $T$ while assuming $T \gg T_{cr}$. The three values of 
$U$ used in Fig. \ref{fig:highT} correspond to the cases of weak, 
intermediate, and strong 
coupling. The latter terminology refers {\it not} to the ratio between the
crossover temperature $T_*$ and $T_{cr}$, but rather to the properties of
the uniform mean-field solution of the pure $FKM$ at $T \rightarrow 0$, 
specifically to the value of the double occupancy on-site 
$\goth{n}_\goth{d}^{(0)}$ [Eq. (\ref{eq:doubleHF})]. For $U=0.5$, $U=1$, and 
$U=2$ we obtain, respectively, $\goth{n}_\goth{d}^{(0)}\approx 0.21$, $0.13$, and
$0.04$ in the limit of low $T$.

Neglecting  thermal fluctuations of the OSDM\cite{HFcalc}, one obtains a purely 
Hartree--Fock result for $\Delta(T)$, which in Figs. \ref{fig:highT} {\it a,b,c}
is represented by the dotted line. It incorrectly predicts a 
second-order phase transition at a certain temperature, which we will instead
identify as the crossover temperature $T_*$. We find $T_*\approx 0.0042$ for 
$U=0.5$, $T_* \approx 0.067$ for $U=1$, and $T_* \approx 0.29$ for $U=2$.
As expected, the value of the indirect gap $G$ at $T=0$
[$G \approx .00525$, $0.102$, and $0.647$, respectively,
  see Eq. (\ref{eq:gap})] yields a correct order-of-magnitude estimate of
$T_*$, although it is worth noting that the fit worsens with increasing $U$.
The latter is due to the fact that for larger $U$, the dominant contribution to
the temperature dependence of energy comes not from the smearing of the Fermi
distribution and the resultant particle-hole excitations across the gap, but
rather from the changes of the average interaction energy per site,
$U\goth{n}_\goth{d}=U\goth{n}_\goth{d}^{(0)}=U(n_c^{(0)} n_d^{(0)}-
|\Delta^{(0)}|^2)$. Indeed, for $U=1$ and $U=2$ the quantity
\begin{equation}
\langle \Delta E_{int} \rangle_F \sim U \left( \goth{n}_\goth{d}^{(0)}(T_*)-
\goth{n}_\goth{d}^{(0)}(0)\right)\,,
\label{eq:Tstar}
\end{equation}
(the fluctuation-induced increase of the interaction energy from $T=0$ to
$T=T_*$) gives a perfect estimate for $T_*$. On the other hand, it is actually
{\it negative} for the weakly interacting case of $U=0.5$, where the net energy
change is dominated by the effects of Fermi distribution smearing, and hence
the single-particle gap $G$ yields a rather accurate estimate for $T_*$.

The values
of $\Delta(T)$ and $n_d(T)$, obtained within our single-site mean-field 
approach, are illustrated by the solid and dashed lines, respectively, with the 
value of $\Delta(T)$ showing a smooth downturn in a broad region around $T_*$.
In the $U=0.5$ case (Fig. \ref{fig:highT} {\it a}), this is followed by an
upturn, due to the increasing thermal fluctuations.
Since the contribution of the small-$\Delta$ region is suppressed by the 
factor $\sin \beta$ in the measure
[see Eq. (\ref{eq:dDeltaT})], these lead to the overall increase in $\Delta$. 
Indeed, in this region we see the increase of both the standard deviation, 
Eq. (\ref{eq:Deltasd})  (shown by the error bars), and of the difference 
$\langle \delta \Delta \rangle_T$ between the net $\Delta$ and
its virtual-crystal part, $\Delta^{(0)}$, represented by the dashed-dotted 
line. Then, the value of $\Delta(T)$ passes through a broad maximum and 
begins its decrease to a higher-temperature mean-field solution, where both 
orbitals are equally populated [note that the monotonously increasing $n_d(T)$ 
is now approaching $1/2$] while $\Delta$ vanishes (whereby $\goth{n}_\goth{d}$ 
will reach its maximal value of $1/4$).
This regime is formally possible only at $T{>}E_d$. Indeed, in the absence
of $\Delta$, there are two unhybridised  Hartree  bands, dispersive and
localised, and if these are equally populated the energy difference between
their respective centres equals $E_d$; on the other hand, the two band
occupancies can approach each other only when the temperature is large in 
comparison with this energy difference. Due to  suppression 
of
the fluctuations of $\Delta$, this configuration
minimises the free energy at sufficiently high $T$.  

Yet, it is clear that this ``high-temperature limit'' with 
$\Delta \rightarrow 0$  and $\sigma_\Delta \rightarrow 0$ is an 
artefact of our assumption that the fluctuations of $\theta_i$ can be 
omitted. Indeed, we observe that the case of 
$n_c=n_d$ and $\Delta=0$ corresponds to $\theta^{(0)}_1=\pi/2$ [see Eqs. 
(\ref{eq:theta130}) and (\ref{eq:osdmn1})].
Since this is an endpoint of the variation range for $\theta_1$, the thermal
fluctuations of this parameter will be asymmetric and will shift its average
to lower values,
increasing the difference $n_c-n_d$ and the fluctuations (and hence the average
value) of $\Delta$. Moreover, an additional factor of $\cos^2 \theta_1$ in the
phase-space measure, Eq. (\ref{eq:Bures}), which reduces the relative
contribution of the region near $\theta_1=\pi/2$ to the partition function,
guarantees that the fluctuations of $\theta_1$ will be large, once $T$ becomes
comparable to the energy scale associated with such fluctuations (which should
be the largest of $E_d$ and $U$, possibly with a prefactor).
Thus, we expect that the value of $\Delta(T)$ passes through a minimum 
at $T\stackrel{<}{\sim}{\rm max}(E_d,U)$ and begins to increase due to 
an overall increase of the thermal fluctuations at higher $T$. However, 
as explained in
Sec.\ref{subsec:gamma} below, this region is in any case out of reach for us, 
at least
within the present version of our mean-field approach.

Interestingly, the increase of $n_d(T)$ and decrease of $\Delta(T)$  at
$T<T_*$ lead to the value of $\goth{n}_\goth{d}^{(0)}(T_*)$ being about the 
same, $\goth{n}_\goth{d}^{(0)} = 0.21 \pm 0.01$, in all three cases of $U=0.5$, 
$U=1$, and $U=2$.
At $T>T_*$, the intermediate region of increasing $\Delta(T)$ is absent at
higher $U$ (Fig. \ref{fig:highT} {\it b,c}), for the following two reasons:
(i) the corresponding values of $T$ are much larger, due to higher $T_*$,
and therefore closer to the high-temperature regime, where the present 
calculation predicts a strong decrease of $\Delta$  (see the discussion above); 
(ii) the
value of $\Delta$ is larger, therefore thermal fluctuations are nearly 
symmetric (the $\beta=0$ point is far away), and their contribution 
$\langle \delta \Delta \rangle_T$ to the net value of $\Delta$ is 
much smaller than in the low-energy case (and actually changes sign 
in the region above $T_*$). We note, however, that our results for 
higher $U$ (especially for $U=2$) become quantitatively unreliable in 
this region due to strong fluctuations of $\gamma_i$ (see Sec. 
\ref{subsec:gamma} below).

%

The {\it specific heat}  is calculated as $C=\partial \langle E \rangle_T/
\partial T$. Here, the average energy per site,
\begin{eqnarray}
\langle E \rangle_T&=& \langle {\cal H} \rangle_F+\langle \delta E \rangle_T=
-l_c^{(0)}+E_d n_d^{(0)}+Un_c^{(0)}n_d^{(0)}-
\nonumber \\
&&-U(\Delta^{(0)})^2+ \int_0^\pi\delta E(\beta) w(\beta)\sin \beta d\beta
\end{eqnarray}
[see Eqs. (\ref{eq:deflm})], is the sum of the virtual-crystal contribution 
and the average fluctuation 
energy $\langle \delta E \rangle_T$. The calculated values of $C$ (solid lines
in Fig. \ref{fig:highT} {\it d,e,f} ) approach
$0.5$ in the low-temperature limit (corresponding to the presence of one 
classical degree of freedom, $\beta$, and, in the case of Fig. \ref{fig:highT} {\it e}, to the higher-$T$ 
part of Fig. \ref{fig:lowT} {\it b}), show a broad maximum in the crossover
region $T\sim T_*$, and decrease at high temperatures [mirroring the decrease of
$\Delta(T)$]. In the weakly interacting case of Fig. \ref{fig:highT} {\it d},
the initial increase following the peak at $T\sim T_*$ corresponds to 
increasing $\Delta(T)$ in this region (see above). The dotted lines represent 
the Hartree--Fock results\cite{HFcalc} (as described above for Fig.  \ref{fig:highT} 
{\it a,b,c}), including only the contributions from fermionic degrees of 
freedom and
from the temperature dependence of the Hartree--Fock values of $\Delta$ and 
$n_d$. As expected, there is a negative jump at $T=T_*$, an artefact 
of neglecting the thermal fluctuations of the OSDM. 
It appears possible that the low-temperature limiting value 
$C \rightarrow 0.5$ is again an artefact
of the classical treatment of the fluctuations of $\beta$ and a proper quantum 
treatment would yield $C\rightarrow 0$ at $T\rightarrow 0$   
{\it both} for $T \ll T_{cr}$ (see Sec. \ref{sec:low-T}) and for 
the phase-disordered case of
$T \gg T_{cr} \rightarrow 0$   (but certainly not for $T \sim T_{cr}$). 
At all events, the classical description of $\beta$ becomes more adequate
with increasing $T$, and should be appropriate for most of the temperature 
range in Fig. \ref{fig:highT} (where the relevant scale is that of $T_*$).

\begin{figure}
\includegraphics{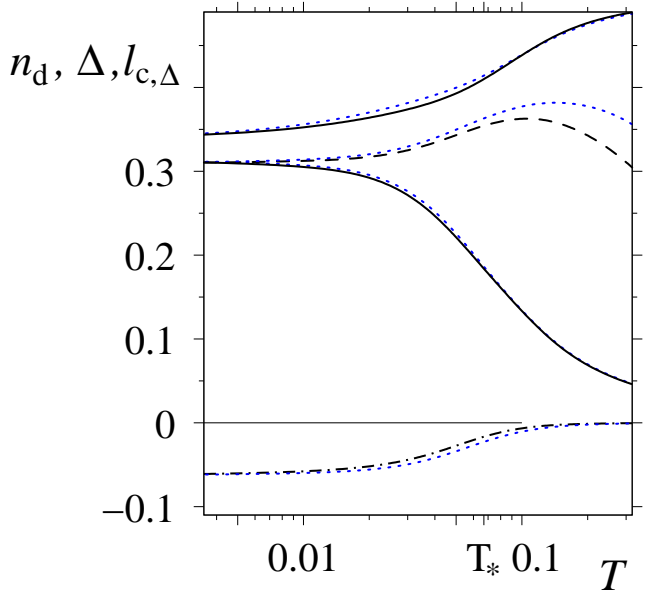}
\caption{\label{fig:lcdelta} (colour online) The effect of self-consistent 
renormalisation of $l_{c,\Delta}$ on the mean-field solution. Dashed and 
dashed-dotted lines show, respectively, the self-consistent values of $l_c$ 
and $l_\Delta$ for a half-filled 2D EFKM with $U=1$ and $E_d=0.2$. The 
adjacent dotted lines show the 
unrenormalised virtual-crystal terms $l_{c,\Delta}^{(0)}$ 
[Eqs. (\ref{eq:deflm})]
calculated for the same self-consistent solution. The upper and lower solid 
lines show the self-consistent values of $n_d$ and $\Delta$ (same data as in 
Fig. \ref{fig:highT} {\it b}), whereas the adjacent dotted lines correspond 
to the case where only two mean-field equations (\ref{eq:mfe0}) are solved and 
the resultant 
unrenormalised  $l_{c,\Delta}^{(0)}$ are substituted for  $l_{c,\Delta}$. }  
\end{figure}

Finally, we are now in a position to clarify the importance of the 
self-consistency conditions (\ref{eq:dlcT}--\ref{eq:dldeltaT}). 
As exemplified in Fig. \ref{fig:lcdelta}, the self-consistent renormalisation 
$\delta l_{c,\Delta}=l_{c,\Delta}-l_{c,\Delta}^{(0)}$ of the quantities 
$l_{c,\Delta}$ is 
rather small, its relative size increases moderately in the high-temperature 
region well above $T_*$. Importantly, if the self-consistency conditions
(\ref{eq:dlcT}--\ref{eq:dldeltaT}) are omitted altogether, and one solves
only {\it two} mean-field equations  (\ref{eq:mfe0}) for $n_d$ and $\Delta$ 
(substituting for $l_{c,\Delta}$ the values of $l_{c,\Delta}^{(0)}$, calculated for
the same $n_d$ and $\Delta$), this leads to a small shift in the resultant 
mean-field solution, $n_d(T)$ and $\Delta(T)$. This small change, which peaks 
in the region of $T\sim T_*$, appears negligible for all practical purposes.   
 
It seems reasonable to expect that this unimportance of Eqs. 
(\ref{eq:dlcT}--\ref{eq:dldeltaT}) is a general property. While presently we 
had no difficulty carrying out the full self-consistent calculation, this would 
have been problematic had we included also the fluctuations of $\theta_i$ 
(see Sec. \ref{sec:osdm} for discussion). However, if the error
introduced by substituting $l_{c,\Delta}^{(0)}$ in place of $l_{c,\Delta}$ is 
indeed insignificant, this would justify the use of a simpler
 Eq. (\ref{eq:Efluct3}) in place of 
a difficult Eq.(\ref{eq:Efluct}).

To summarise, our single-site mean-field approach yields a physically 
transparent description of the phase-disordered 
state of the EFKM above the low-temperature ordering transition. This 
includes, at least for the case of weak to moderate interaction
strength,  the crossover region of $T \sim T_*$ (see Sec. \ref{subsec:gamma}
regarding larger values of $U$). It appears that previously 
such a description has been lacking, at least in the context of the EFKM. 
Further discussion of  results obtained in this section will follow in 
Secs. \ref{sec:exp} and \ref{sec:conclu}.

\subsection{Validity of the Hartree-Fock approximation for the wave functions}
\label{subsec:gamma}

The quantities $\gamma_i$ are additional phase variables of the SU(4) rotation
(see Appendix \ref{app:su4}), which affect the wave function $|\tilde{\Psi}\rangle$ 
[Eqs. (\ref{eq:decompfluctgen}--\ref{eq:decompfluct})], but {\it not} the 
corresponding OSDM [Eqs. (\ref{eq:osdmgen}) and (\ref{eq:osdmn1})]. When either
$\gamma_1$ or $\gamma_3$ differs from zero, an electron hopping to or from the
central site acquires an additional phase which depends on the specific quantum
state concerned (at the central site). Similarly, $2\gamma_2$ is the phase
difference between the two singly occupied states which diagonalise the OSDM 
at the central 
site, and it affects both the phase carried away by a hopping electron and 
the hopping probability. 
Strong fluctuations of $\gamma_i$ would suggest a possibility of non-trivial
phase-related phenomena, such as strongly fluctuating flux through a plaquette. 

Importantly, the fluctuations of $\gamma_i$ cannot be incorporated into our 
self-consistent scheme, which relies on the underlying virtual crystal 
[Eq. (\ref{eq:Hvc})] and hence, on the 
corresponding Hartree--Fock wavefunctions $|\Psi\rangle$. The latter 
correspond to
all $\gamma_i$ being equal to zero everywhere, and there is apparently 
no way to 
include the fluctuations of $\gamma_i$ by merely renormalising the 
parameters of the
virtual crystal (and hence of $|\Psi\rangle$), as we did with the 
fluctuations of 
$\beta$ above, and with fluctuations of $\varphi$ in Sec. \ref{sec:low-T} 
[or as {\it can} be done with the fluctuations of $\theta_{1,3}$ under 
restriction 
(\ref{eq:HFtheta})]. 

On the other hand, there is no difficulty in calculating the energy cost of a 
local
fluctuation of both $\beta$ and $\gamma_i$ at site 0, provided that all 
$\gamma_i$ vanish
elsewhere. In addition to $\delta E(\beta)$ [Eq. (\ref{eq:dET})], the 
energy of such 
a fluctuation
acquires another term, $\delta E_\gamma(\beta,\gamma_1,\gamma_2,\gamma_3)$, 
given by Eq.
(\ref{eq:dETgamma}). One can proceed one step further and calculate the 
average value of
$\cos \gamma_i$ under such fluctuations as

\begin{eqnarray}
\!\!\!\!\!\!&&\!\!\!\!\!\!\langle \cos \gamma_i \rangle_T=\int_0^\pi 
\!\!\sin \beta d \beta \int_0^{2\pi}\!\!d\gamma_1\int_{-\pi/2}^{\pi/2}\!\!
d\gamma_2 \int_0^{2\pi} \!\!d \gamma_3 \tilde{w}\cos \gamma_i \,,
\nonumber\\
&&\label{eq:cosgamma}\\
\!\!\!\!\!\!&&\!\!\!\!\!\!\tilde{w}(\beta,\gamma_1,\gamma_2,\gamma_3)=
\frac{1}{\tilde{Q}}\,{\rm e}^{-[\delta E(\beta)+\delta 
E_\gamma(\beta,\gamma_1,\gamma_2,\gamma_3)]/T}\,,
\nonumber \\
\!\!\!\!\!\!&&\!\!\!\!\!\!\tilde{Q}=\int_0^\pi \!\!\sin \beta d \beta 
\int_0^{2\pi}\!\!d\gamma_1\int_{-\pi/2}^{\pi/2}\!\!d\gamma_2 \int_0^{2\pi} \!\!
d \gamma_3 \tilde{w}(\beta, \gamma_1,\gamma_2,\gamma_3)\,.
\nonumber
\end{eqnarray}

\begin{figure*}
\includegraphics{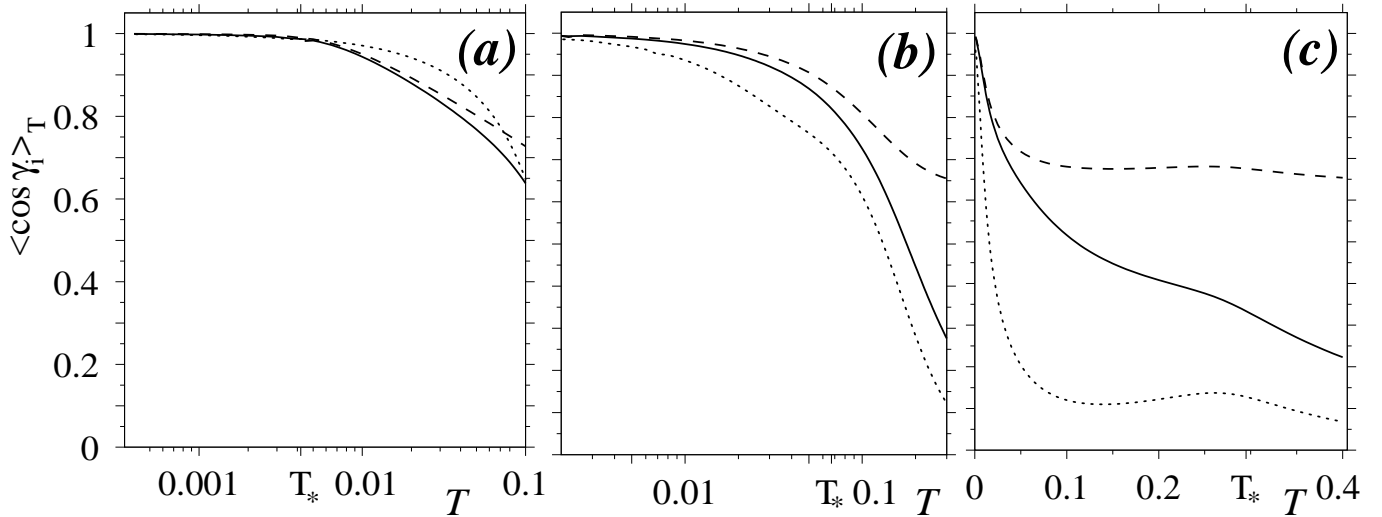}
\caption{\label{fig:gammas} Thermal fluctuations of phases $\gamma_i$ for a 
2D EFKM 
at $n=1$,  for $E_d=0.2$ and $U=0.5$ 
({\it a}), $U=1$
({\it b}), and $U=2$ ({\it c}). 
Solid, dashed, and dotted lines show $\langle \cos \gamma_i \rangle_T$ for 
$i=1,2$ and $3$, respectively. Standard deviation is about 
$1-\langle \cos \gamma_i \rangle_T$ in all cases. }  
\end{figure*}

In Fig. \ref{fig:gammas}, we plot the three quantities $\langle \cos \gamma_i 
\rangle_T$ for the three cases considered in Fig.
\ref{fig:highT} and corresponding to weak, moderate, and strong interaction 
$U$. While in the limit of $T\rightarrow 0$ all three values of $\gamma_i$ 
approach zero in all the three cases (attesting to the consistency of the 
underlying Hartree--Fock approximation at
low $T$), the behaviour at increased temperatures shows a marked dependence 
on the interaction strength. In the weak-coupling case of 
Fig. \ref{fig:gammas} {\it a}, the values of $\langle \cos \gamma_i \rangle_T$ 
remain above $0.65$ throughout the entire
range of the plot, including also the crossover region (at $T=T_*$ we 
find $\langle \cos \gamma_i \rangle_T\approx0.986\pm0.003$
for all $i$). With the fluctuations of $\gamma_i$ being either small or 
very moderate, we conclude that they indeed can be neglected, and the 
Hartree--Fock virtual-crystal treatment remains valid up to $T_*$ and beyond. 
At $U=1$ (see Fig. \ref{fig:gammas} {\it b}), the values of 
$\langle \cos \gamma_i \rangle_T$ at $T_*$ are $ 0.83$,       
$0.88$, and      $ 0.73$   for $i=1,2$ and $3$, respectively, 
hence, we expect that the Hartree--Fock picture remains valid 
and qualitatively reliable up to $T_*$, but not much beyond that. 
Thus, one concludes that in these two cases the consistency of our 
mean-field approach at $T\stackrel{<}{\sim}T_*$ is not in danger.

The situation is different for the strong-coupling case of $U=2$ 
(shown in Fig. \ref{fig:gammas} {\it c}), where the values
of $\langle \cos \gamma_i \rangle_T$ at $T_*$ are approximately 
$0.34$,  $ 0.68$, and   $ 0.13$, which suggests that the fluctuations
of $\gamma_i$ are no longer negligible in any sense. The Hartree--Fock 
description still remains applicable in the phase-disordered state, 
but at much lower temperatures: for example, at $T=0.03$  (and for $U=2$) 
we find  $\langle \cos \gamma_i \rangle_T \approx 0.72$,       $0.76$, 
and  $0.31$ for $i=1,2$, and $3$. The strong fluctuations of 
$\langle \cos \gamma_3 \rangle_T$ do not constitute an immediate cause 
for concern, as the phase $\gamma_3$ affects only the doubly-occupied 
component of the perturbed wave function (\ref{eq:decompfluct}), 
and the relative weight of this component at $T=0.03$ is still fairly 
small, with  $\goth{n}_\goth{d}^{(0)}\approx 0.05$. We speculate that 
the overall behaviour suggested by Fig. \ref{fig:highT} {\it c}, 
which is in line with the reliable results obtained for smaller $U$, 
is still {\it probably} correct, but this conjecture certainly lacks 
a solid justification.

Finally, we note that although the phases $\gamma_i$ do not directly affect
the OSDM (including the fluctuating values of $n_d$ and $\Delta$), taking into
account the fluctuations of $\gamma_i$ does modify the probability distribution
for the angle $\beta$, and hence does affect the {\it average} values
$\Delta(T)$ and $n_d(T)$ (as well as those of $l_{c,\Delta}$). In the 
region where our theory is applicable, this
effect is not very strong, reaching up to 10\% for $\Delta(T)$, and  about
1 \% for $n_d(T)$. The difference is most pronounced in the region 
where the fluctuations of $\Delta(T)$ are largest, thus falling well 
within the ``error bars'' on Fig. \ref{fig:highT} {\it a,b,c}. Likewise, 
the relative change of the net $l_{c,\Delta}$ seldom exceeds 15\% (whereas 
the values of $\delta l_{c,\Delta}$, see above, can
become several
times larger or smaller). Accordingly, when one substitutes $\tilde{w}$
given by Eq. (\ref{eq:cosgamma}) in place of $w$ in the Eqs. 
(\ref{eq:dDeltaT}--\ref{eq:dndT})
of the self-consistent calculation and includes additional integration over
$\gamma_i$, the resultant change of $n_d(T)$ and $\Delta(T)$ is within 10\%,
with no new features
(we checked this for $U=1$ and  $T{<}0.1$).

\section{Excitonic insulators -- experiment and theory}
\label{sec:exp}

\subsection{Experimental situation: The case of ${\rm Ta_2 Ni Se_5}$}
\label{subsec:Ta2NiSe5} 

Experimentally, specific heat $C$ was measured\cite{Fu2017} in an excitonic 
insulator candidate ${\rm Ta_2 Ni Se_5}$. In this compound, a phase transition 
is observed at $T=326$ K, accompanied by a symmetry change and by a peak in 
the $C(T)$ dependence; below the 
transition, the (direct) interband gap has a flattened momentum 
dependence\cite{Wakisaka09,Seki14}, suggestive of an excitonic insulator. There 
is an ongoing discussion as to whether the 
transition is primarily of electronic\cite{Mazza19} or lattice\cite{Watson19} 
origin, and it is clear that
electronic and lattice degrees of freedom are interdependent. 
Since the (long-range) 
lattice strain fields are presumably coupled to the phase degree 
of freedom in the electronic insulator (our $\varphi_i$), one expects 
that this increases the energies of collective excitations in the 
low-$T$ ordered phase. This might in turn push the value of the ordering 
transition temperature $T_{cr}$ upwards, shrinking or  
obliterating the 
phase-disordered intermediate region $T_{cr}<T<T_*$, discussed 
in Sec. \ref{sec:high-T} above. 

In is generally considered as plausible that the feature\cite{Fu2017} 
seen in $C(T)$ at $326$ K 
corresponds to this increased value of $T_{cr}$ (excitonic condensation); 
the transition is still second order,
although modified (in comparison to the one discussed in 
Sec. \ref{sec:low-T}) by a strong 
involvement of the lattice.  
The apparent absence of hybridisation
above the transition\cite{Watson19,Lee19} suggests that there is no further 
high-temperature crossover (our $T_*$) located in that region. 
{\it Mutatis mutandis}, this is 
a  BCS-type picture\cite{Lee19}, which is also consistent 
with the expectation\cite{Seki14,Sugimoto18} that the effective masses 
in the two bands are not 
very different. Yet, we note that a {\it phase-disordered} excitonic insulator 
with $T_*$ above $326$ K, while characterised by a small (fluctuating) 
hybridisation gap, on average would {\it not} violate the higher lattice 
symmetry; this opportunity, and the corresponding BEC behaviour, are 
discussed in  Refs. 
\onlinecite{Seki14,Sugimoto18}, which also identify  the transition 
at $326$ K  as the excitonic ordering temperature 
(our $T_{cr}$).

Another scenario would 
have the (lattice) transition at $326$ K accompany (and perhaps sharpen) 
the excitonic {\it crossover} 
[hence, $T_* \approx 326$ K, cf. the peak of $C(T)$ at $T\sim T_*$ in 
Fig.  \ref{fig:highT} {\it d,e,f} ], 
and the {\it excitonic ordering} take place 
at a lower $T_{cr}$; with the lattice symmetry breaking playing the role of
``external field'' $H$ in a corresponding XY model [cf. Eq. (\ref{eq:Hlow})], 
the transition at $T\sim T_{cr}$ and the associated  jump 
in $C$ would both be smeared and perhaps difficult 
to pinpoint\cite{200K} (cf. the dotted  line in 
Fig. \ref{fig:lowT} {\it b}, and the discussion in Sec. \ref{sec:low-T}).
This would constitute a pronounced excitonic BEC behaviour 
with a weak or moderate coupling to 
the lattice and $T_*/T_{cr} \gg 1$ or $T_*/T_{cr} \stackrel{>}{\sim} 1$. 
Note that (i) in principle, the existence of the condensate extends all the 
way up to $326$ K [cf. Eq. (\ref{eq:highTkappa}); in our theoretical analysis 
in Sec. \ref{sec:high-T}, carried out for $T\gg T_{cr}$, this effect 
was neglected]; (ii) the latter is not sufficient to identify $326$ K  as the
excitonic BEC transition temperature $T_{cr}$; this is similar to 
magnetisation being 
induced by an external field in a ferromagnet {\it above} the Curie point.

Finally, if hybridisation is 
dominated by the lattice effects at all temperatures, with excitonic pairing
providing a perturbative correction at low $T$, the system should not be 
termed an excitonic insulator. The experimental 
results\cite{Larkin17,Werderhausen18}, 
however, point to
 strong excitonic effects, which in turn affect the properties of phonons.

We  note that
the value of  excitonic order parameter directly influences the 
electrostatic 
properties of the system. 
In the case of ${\rm Ta_2 Ni Se_5}$, the gap is direct, and 
uniform ($\vec{k}=0$) measurements
of the appropriate electrostatic moment (possibly quadrupole rather than 
dipole one, in contrast to a simpler case considered in 
Refs.\onlinecite{Portengen,Batista04,pssb13}) and of the corresponding 
response should be 
performed in order to identify the correct scenario. 
Dynamical measurements might allow to distinguish between the 
lattice (slow) and excitonic (fast) contributions, as suggested also 
in Ref. \onlinecite{Mazza19}. In addition, further assistance in assigning
the value of $T_{cr}$ and (if distinct) of $T_*$ can be drawn from studying 
the spectra
of phase and amplitude
{\it collective modes}.

\subsection{Amplitude mode and amplitude susceptibility}
\label{subsec:Higgs}

The presence of a non-zero absolute value $\Delta$ of the on-site 
hybridisation implies the existence of a collective mode, corresponding to its
oscillations. Presently, this subject receives much attention both in the 
framework of general interest in such ``Higgs mode'' in solid state 
physics\cite{Higgs}, 
and in a more narrow context of prospective excitonic insulators. Indeed, an 
important
recent paper\cite{Kogar2017} is devoted to experimental identification of the 
amplitude mode in the case of dichalcogenide ${\rm 1}T-{\rm TiSe}_2$; 
presence of such a mode was also reported\cite{Werderhausen18} for  
${\rm Ta_2 Ni Se_5}$, where its fingerprint 
is seen in the phonon dynamics.  Therefore,
it appears important to discuss the insight which can be gained from our present
work in this regard.

If one neglects thermal fluctuations of the OSDM (including those of the phases
$\varphi_i$; this is the ``pure Hartree--Fock'' 
case discussed above, corresponding to the dotted lines in 
Fig. \ref{fig:highT}), 
the spectrum $\omega_{\vec{q}}$ of higher-energy plasmon excitation can be 
calculated along the lines of Ref.\onlinecite{prb12}. To zeroth order in 
$\delta {\cal H}$ and for the case of $\vec{q}=0$, the secular equation 
takes the form\cite{secular}:
\begin{equation}
\omega^2 \left\{\left[J_1(\omega)\right]^2+\left(u^2 -\omega^2\right)
\left[J_0(\omega)\right]^2\right\}=0\,
\label{eq:secularHF}
\end{equation}
[see Eq. (\ref{eq:dirgap})], where for $l=0,1$
\begin{equation}
J_l(\omega)=\frac{1}{N} \sum_{\vec{k}} \frac{\Delta_{\vec{k}}}{\Delta}
\frac{\xi_{\vec{k}}^l}{u^2 -\omega^2+ \xi_{\vec{k}}^2}
\label{eq:Jspec}
\end{equation}
[see Eqs. (\ref{eq:epsilon}) and (\ref{eq:Delta}); upon converting 
the r.\ h.\ s. of Eq. (\ref{eq:Jspec}) 
to an integral, principal value of the latter should be evaluated].

\begin{figure}
\includegraphics{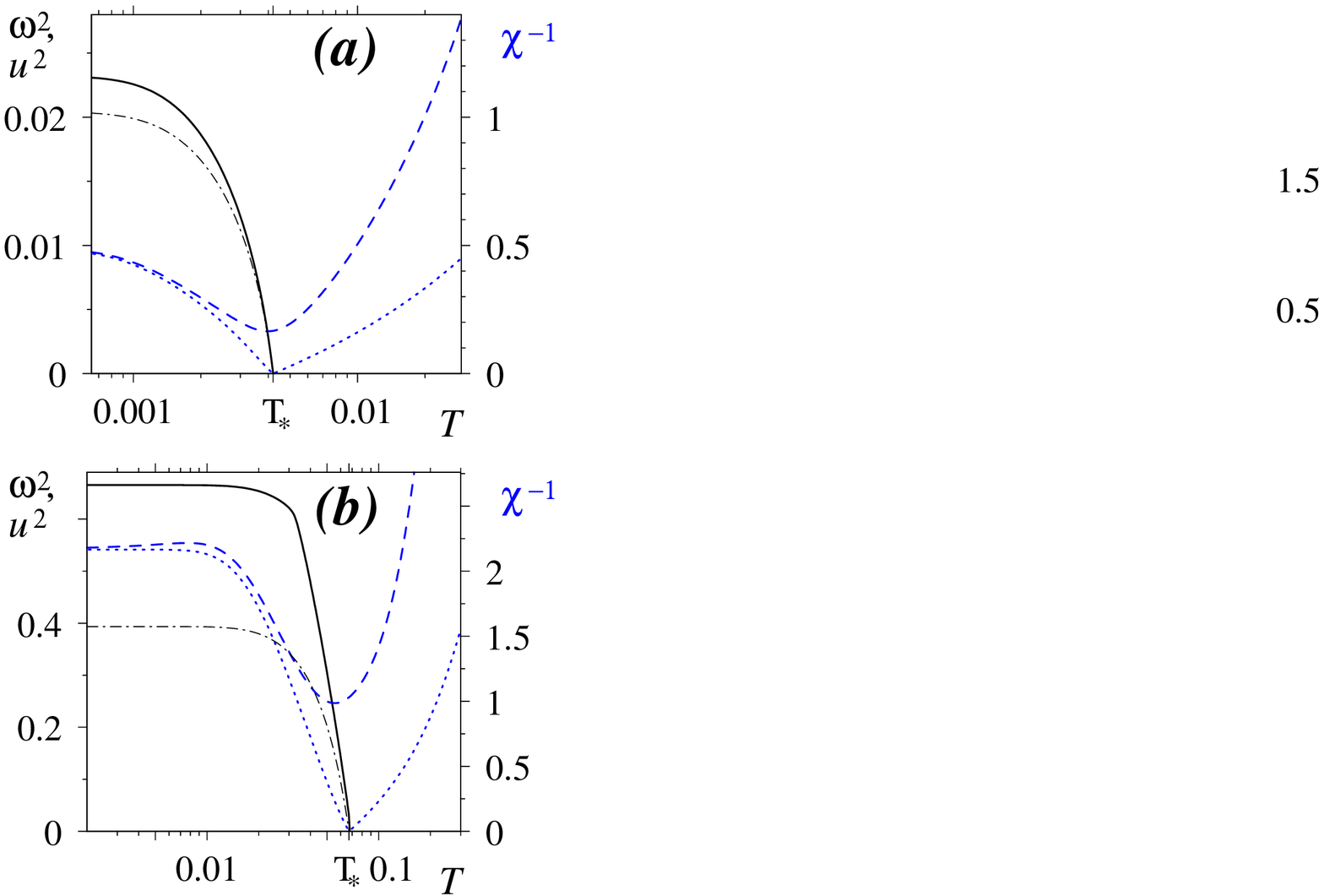}
\caption{\label{fig:Higgs} (colour online) Solid line (left scale) shows the 
amplitude mode energy squared for a half-filled 2D EFKM with $E_d=0.2$ and 
$U=0.5$ ({\it a}) or  $U=1$ ({\it b}), calculated at $\vec{q}=0$ and in 
the absence of the thermal fluctuations of the density matrix [Hartree--Fock, 
see Eq. (\ref{eq:secularHF})]. The dashed-dotted line (left scale) represents
$u^2$, square of the direct energy gap.
Dashed and dotted lines (right scale) correspond to an inverse susceptibility,
$1/\chi$. Dashed line includes the effect of thermal fluctuations of the OSDM.} 
\end{figure}

Eq. (\ref{eq:secularHF}) is valid below the Hartree--Fock critical point $T_*$
and has two solutions. Of these, $\omega=0$ corresponds to the phase mode,
vanishing in the unperturbed case of $\delta {\cal H}=0$ (for $T=0$, 
the perturbed case is investigated in detail in Ref. \onlinecite{prb12}). The
other solution, which must correspond to the amplitude mode, lies above the
direct gap $u=2U\Delta$, which means that it is likely to be strongly 
damped by the particle-hole excitations. Typical behaviour
of $\omega^2(T)$ is plotted in Figs. \ref{fig:Higgs} {\it a,b} with solid lines 
(left scale). As expected, it vanishes at $T\rightarrow T_*$.

For a relatively small $U=0.5$, the value of $\omega$ closely follows that of 
$u$ (see Fig. \ref{fig:Higgs} {\it a}). We note that smaller $U$ results also
in smaller values of $\Delta$ (cf. Fig. \ref{fig:highT}), leading to a strong 
overall decrease in $u=2U\Delta$. In this case, it appears that also away 
from $T_*$ the non-zero 
solution of Eq. (\ref{eq:secularHF}) is strongly affected by a somewhat complex
anomaly of $J_0$ [Eq. (\ref{eq:Jspec})], located at $u,\omega \rightarrow 0$. 
This is no longer the case for $U=1$ (see  Fig. \ref{fig:Higgs} {\it b}), where
$\omega$ is found to exceed $u$ significantly. 

Above $T_*$, the phase mode must disappear (as there is no corresponding 
symmetry breaking), whereas the amplitude mode is expected to recover to higher
energies (see below). However, it can no longer be represented as a linear 
combination of particle-hole excitations and hence cannot be calculated within 
the approach of Ref. \onlinecite{prb12}.

It is unclear whether this approach can be extended to include the thermal 
fluctuations of the OSDM, and whether the time-independent treatment of these,
as constructed in this paper, would be sufficient. At all events, the
energy scale of the amplitude fluctuations can be deduced from the value of
susceptibility $\chi=\partial \Delta/\partial {\goth F}$ with respect to a 
fictitious external scalar field ${\goth F}$, coupled to the {\it absolute 
value} of 
the hybridisation. If the amplitude mode is present, its energy squared, 
$\omega^2$, can be expected
to be roughly proportional to $1/\chi$, with the unknown $T$-dependent
coefficient affected by the quantisation of the fluctuations 
of $\Delta$ and by the precise form of  the excitation  wavefunction.
In order to calculate $\chi$, one must add the term
\begin{eqnarray}
\delta_{\goth F} {\cal H}_{mf} & = &-\frac{1}{2}{\goth F}\sum_i \left(c^\dagger_i
\tilde{d}_i+ \tilde{d}_i^\dagger c_i \right)= \nonumber \\
&=&-\frac{1}{2}{\goth F}\sum_{\vec{k}} \left(c^\dagger_{\vec{k}}
{d}_{\vec{k}}+ {d}_{\vec{k}}^\dagger c_{\vec{k}} \right)
\end{eqnarray}
[cf. Eq. (\ref{eq:gauge})] to the mean-field (virtual-crystal) 
Hamiltonian\cite{substit}, Eq. (\ref{eq:Hvc})
[see also Eq. (\ref{eq:Hmf})], and 
\begin{eqnarray}
\delta_{\goth F}E(\beta)&=&
-{\goth F}\left[\tilde{\Delta}(\beta)-\Delta^{(0)}\right]= \nonumber \\
&=&-\frac{1}{2}{\goth F}
\sqrt{n^2-{\goth n}_{\goth d}} \left(\sin \beta - \sin \beta^{(0)}\right)
\end{eqnarray}
[see Eq. (\ref{eq:tdelta})] to the single-site fluctuation
energy in the phase-disordered state, Eq. (\ref{eq:dET}). 
As before, our single-site approach
dictates that the contribution of $\delta {\cal H}$ [Eq. (\ref{eq:pert})], 
vanishes in the
phase-disordered regime above $T_{cr}$, hence, $\delta {\cal H}$ can be dropped 
altogether.

We first consider the purely Hartree--Fock case when the thermal fluctuations
are neglected\cite{HFcalc}, which corresponds to the dotted lines in Figs. \ref{fig:Higgs} 
{\it a,b} 
(right scale). The second-order phase transition is then located at $T=T_*$ 
(with $\Delta$ being the order parameter),
and the behaviour of $\chi(T)$ conforms to a simple Landau theory. 
At $T<T_*$, the free energy $F$ as a function of $\Delta$ has a minimum 
at $\Delta=\Delta^{(0)}>0$, resulting in a finite $\chi$. The value of 
$\Delta^{(0)}$ decreases with temperature, and both $\Delta^{(0)}$ and 
$\partial^2 F/\partial \Delta^2$ vanish at $T_*$, hence, $1/\chi$ vanishes
[as does the actual value of $\omega^2$, available in this case from Eq. 
(\ref{eq:secularHF})]. 
Thereafter, $\Delta^{(0)}$ remains equal to zero, whereas the second derivative
becomes finite, leading to a recovery of $1/\chi$. 

When the thermal fluctuations of the OSDM are included, the average $\Delta$ 
remains finite at all temperatures. Accordingly, the zero of  $1/\chi$ 
at $T=T_*$ is replaced by a broad minimum 
(see the dashed lines in Fig. \ref{fig:Higgs}). Note also a pronounced hardening
of amplitude fluctuations at higher $T$, due to the increase of the 
corresponding derivative of $F$ (the increase of thermal fluctuations pushes
the average $\Delta$ towards larger values, where the dependence of the energy
on hybridisation amplitude is more sharp).

Note that both curves merge in the limit of low $T$. This illustrates the fact
that, to leading order in $\delta {\cal H}$, the value of $\chi$ is 
unaffected by the phase ordering arising below $T_{cr}$. The effect of
the exciton condensation on $\chi$ is therefore confined to a 
small correction (subleading term), which vanishes above $T_{cr}$.  

In terms of collective excitation energies in the presence of thermal 
fluctuations, these results add up to a rather
coherent qualitative picture. While the {\it phase mode} softens at the 
low-temperature ordering transition, $T=T_{cr}$, and is absent anywhere 
above $T_{cr}$, the {\it amplitude mode} is only weakly affected by the 
excitonic condensation taking place at $T_{cr}$. The phase-mode spectrum
at $T<T_{cr}$ crucially depends\cite{prb12} on the perturbation, 
Eq. (\ref{eq:pert}), yet
the latter to leading order does not affect the amplitude mode, 
yielding a small correction\cite{smallcorr} only. At higher $T$, 
the amplitude mode 
energy shows a broad 
minimum in the crossover region, $T\approx T_*$, above 
which it increases
to the values which are higher than those of the lower-temperature
region, $T<T_*$. As to whether the amplitude excitation corresponds to an 
actual propagating mode or to a broadened resonance-type feature, this 
depends on the strength of damping and cannot be discussed here.

We are now in a position to compare these expectations to the experimental 
results for ${\rm 1}T-{\rm TiSe}_2$, reported in Ref. \onlinecite{Kogar2017}, 
and to their suggested
interpretation. The plasmon mode described there is identified as the amplitude
mode of an excitonic insulator, whereas the phase mode is either absent or not
detectable. When the temperature increases towards  $T_C \approx 190K$, the 
amplitude mode energy gradually decreases toward that of the low-energy phonon 
and possibly vanishes\cite{finiteq} at $T=T_C$ 
(the error bars are relatively large). At higher temperatures it rebounds and
becomes larger than in the low-$T$ region below $T_C$. The suggested 
interpretation\cite{Kogar2017} is that $T_C$ corresponds to a phase transition,
specifically -- to exciton condensation. This  implies the
BCS (as opposed to BEC) scenario, and appears plausible indeed, especially 
assuming that the effective masses of the two bands are not too 
different\cite{Monney2012} (the opposite situation would likely 
lead to the BEC physics).

However, positive identification of the excitonic condensate based on its 
excitation spectrum requires detecting a phase mode, which in this case would
exist only below $T_C$, softening and vanishing at the transition point. As 
this mode was not observed, it appears possible that the excitonic transition,
while lying low in energy, is pre-empted by a Peierls one.

We further note that the observed amplitude mode spectrum may also suggest 
another
possibility, viz., that of the BEC scenario as discussed in this work. 
Then, the broad minimum of the
mode energy around $T_C$ would correspond to the higher-temperature 
crossover (our $T_*$), and  {\it not} to the excitonic condensation 
(which might or might not take place at a lower temperature $T_{cr}$, 
below which the phase-mode energy would increase sharply). 
The superlattice reflections observed
below $T_C$ (see discussion in Ref. \onlinecite{Kogar2017}) would then be due 
to a 
structural change, perhaps with additional 
contribution from the stable exciton gas (as opposed to condensate), 
which arises at $T<T_*$ (see discussion in Sec. \ref{sec:intro}).
If the excitonic condensation temperature $T_{cr}$ is indeed below $T_C$, 
the former would again correspond to a smeared transition (see  
Sec. \ref{subsec:Ta2NiSe5} above; in this situation, one does not expect
the phase mode to completely disappear above $T_{cr}$, nor would its energy
exactly vanish at $T_{cr}$). 

Although the present discussion of the amplitude mode is clearly of a 
preliminary 
character, we expect that our conclusions are solid at the qualitative level. 
This applies
both to the overall temperature dependence of the mode energy 
and to the need for further measurements in order to clarify the experimental 
situation reported in Ref. \onlinecite{Kogar2017}.

\section{DISCUSSION AND OUTLOOK}
\label{sec:conclu}

Our mean-field treatment of the EFKM yields a physically transparent 
description of the excitonic insulator in a broad range of
temperatures and fully supports  the general expectations discussed 
in Sec. \ref{sec:intro}. In particular, we were able to characterise 
the phase-disordered state, including the crossover region, $T \sim T_*$, 
and such a quantitative description appears to represent a new development 
in the general context of correlated electron systems with 
interaction-induced pairing in the BEC regime. While at the low-temperature 
region around the ordering transition (exciton condensation) the theory 
has at best a rough qualitative accuracy due to shortcomings expected 
of any single-site treatment at low $T$, there is ample reason to 
expect higher reliability at larger $T$, except in the high-temperature 
region for the case of strong interactions. We therefore 
suggest that these results provide a sound base for a more detailed 
description of the phase-disordered excitonic insulator, including 
susceptibilities and transport properties. Presently, our results 
lead to two conclusions, and we suggests that these should be 
checked for those
compounds which are suggested as possible narrow-band (i.e., BEC 
rather than BCS) excitonic insulators.

First, in the phase-disordered state above the phase transition at 
$T_{cr}$ there exists a crossover temperature $T_*$, corresponding 
to a smooth decrease of the induced hybridisation $\Delta(T)$ and 
to an equally smooth peak in the specific heat
$C(T)$. The order-of-magnitude estimate of $T_*$ is given by the 
value of indirect energy gap $G$ at $T=0$ [Eqs. (\ref{eq:gap}) or 
(\ref{eq:gapest})] (but see further discussion in Sec. \ref{sec:high-T}). 
Second, while the value  of $\Delta(T)$ above $T_*$ is smaller than below, 
it is still of the same order of magnitude (i.e., {\it  not} small in 
absolute terms), and this situation persists until much higher temperatures. 
This is due to the thermal fluctuations of $\Delta$, which increase with 
$T$ and are naturally asymmetric (with $\Delta$ being defined as a positive 
quantity, its fluctuating value cannot dip below zero). In addition, 
the phase space where these fluctuations occur is built in such a way 
that the contribution of the small-$\Delta$ region is suppressed. 
Technically, this is represented by the factor $\sin \beta$ in the 
integration measure
in Eq. (\ref{eq:dDeltaT}) [or in  a more general Eq. (\ref{eq:Bures4})] 
and may be interpreted as a reminder about the fact that while within 
this temperature  region the phase $\phi$ of the induced hybridisation 
is completely disordered, it is still present as a physical variable. 
Indeed, had this not been
the case, the measure of the  fluctuations of $\Delta$ would correspond 
to O(1) rather than to SU(2), leading to a 
replacement\cite{physicab18} $\sin \beta d \beta \rightarrow d\beta$. 
Above $T_*$, the value of  $\Delta(T)$ initially continues to decrease with 
increasing $T$; it  is expected to 
pass through a minimum (possibly at rather high temperatures, when $T$ becomes 
comparable to the bandstructure energy scales, {\it viz.,} 
$T\stackrel{>}{\sim} E_d$, see Sec. \ref{sec:high-T}), which is 
followed by an increase driven by the increasing
thermal fluctuations of the OSDM.

In recent years, both theoretical\cite{condEFKM} and experimental
studies of purported excitonic insulators have been enjoying a pronounced
{\it renaissance}. In most  compounds  where
the excitonic behaviour was suggested (with a likely exception of 
${\rm Ta_2 Ni Se_5}$, see 
discussion in Sec. \ref{subsec:Ta2NiSe5} and Refs. 
\onlinecite{Fu2017,Wakisaka09,Seki14,Mazza19,Watson19,Lee19,Sugimoto18,Larkin17,Werderhausen18}), it involves a narrow (or massive) band.
These
candidate BEC excitonic insulators include, in addition to the familiar
samarium and thulium compounds\cite{SmTm,samariumth}, a dichalcogenide
semimetal\cite{1TTiSe2,Kogar2017} $1T$-TiSe$_2$, and also graphene multilayers 
at high
magnetic fields\cite{graphene}. Recent theoretical and experimental
contributions are typically focused on the ordered phase, or on the
ordering transition (``exciton condensation''), although it appears possible
that the features seen in some of the measurements actually correspond to
the higher-temperature crossover ($T_*$ rather than $T_{cr}$, see Secs. \ref{subsec:Ta2NiSe5} and \ref{subsec:Higgs}), and are
incorrectly attributed to $T_{cr}$. While the first experiments reporting
observations of a disordered excitonic insulator above $T_{cr}$ are already
available\cite{Monney2012,aboveTcr}, neither positive identification of this 
state nor a
direct comparison to our theory are possible at this stage. Experimentally,
the specific-heat measurements are still lacking for all compounds apart from 
${\rm Ta_2 Ni Se_5}$, while the theory should be
extended to clarify the role of other degrees of freedom (spin, lattice, and
charge ordering), which are clearly important\cite{condEFKM} at low $T$ and
perhaps may significantly modify the system properties also at higher
temperatures. Given the rapid development of the field, we expect a
significant progress in the near future.

Importantly, we anticipate that the formalism, developed in 
Secs. \ref{sec:osdm} and \ref{sec:Bures}, can be generalised to other 
systems with interaction-induced pairing. This includes both bulk 
systems (such as Kondo lattice and related models for the heavy-fermion 
systems) and lattice impurity models. While the latter are typically 
amenable to  much more refined theoretical treatments, constructing an 
OSDM-based mean-field approach still can be expected to yield new insights 
into the properties of the method and  possibly into those of the physical 
system as well. 

Therefore, comments are in order concerning some aspects of the newly developed 
OSDM-based  mean field formalism. Presently, we implemented it in 
Secs. \ref{sec:low-T} and \ref{sec:high-T} in a rather truncated form, 
only for the half-filled ($n=1$) two-dimensional case  and omitting 
fluctuations of the eigenvalues of
the OSDM (parameters $\theta_i$) and (aside from a brief qualitative 
discussion in Sec. \ref{subsec:gamma}) those of the phases $\gamma_i$ 
[the three  SU(4) phases which affect the wave function
(\ref{eq:decompfluct}), corresponding to a single-site fluctuation]. 
The dimensionality of the system is hardly important at the mean-field 
level, and especially in the phase-disordered state, where the short- to 
medium-range correlations are expected to dominate. Regarding the value 
of $n$, it appears that studying a system with any carrier density 
(as long as it supports pairing) should not present a difficulty, 
at least in principle. The same apparently applies to treating the 
fluctuations of $\theta_i$ and $\gamma_i$ in the case of correlated 
impurity (Kondo, Anderson, etc.) problems. On the other hand, fully 
including fluctuations of $\gamma_i$ (or {\it unrestricted} fluctuations 
of $\theta_3$) in a bulk system would require going beyond  the underlying 
Hartree--Fock approximation, and it is presently unclear whether and how 
this could be performed.

The question of including fluctuations of $\theta_i$ which preserve the 
Hartree--Fock condition, $\goth{n}_\goth{d}=n_c n_d-|\Delta|^2$ in the 
case of EFKM [see Eq. (\ref{eq:HFtheta}), valid at $n=1$], for a bulk 
system is more subtle. Strictly speaking, in this case the fluctuations 
of the OSDM on different sites are not mutually independent 
(see Sec. \ref{sec:osdm}), which precludes full self-consistency of a 
single-site mean-field approach. On the other hand, there is a good 
reason to expect that the correction introduced by this inter-dependency 
is small and can be neglected (see Sec. \ref{sec:high-T}). In this case, 
such fluctuations can be treated in a cumbersome but straightforward way, 
using a reduced equation (\ref{eq:Efluct3}). Yet, we suspect that such a 
calculation might not prove worthwhile, as the weak to moderate symmetric 
fluctuations of $\theta_i$ about their respective virtual-crystal average 
values are unlikely to significantly affect the results in the region of 
interest, $T\stackrel {<}{\sim}T_*$.

One might also view this issue in a rather more pedantic way: We set out 
to build an analogue of the Weiss-type mean-field theory for a system 
with itinerant carriers.  Considering all possible single-site fluctuations, 
we found that there exists a subclass of these, which allows in principle 
for a full self-consistency of this approach. This subclass includes the 
fluctuations of the OSDM parameters $\beta$ and $\phi$ (and  also $\gamma_i$), 
corresponding to $\hat{S}\hat{S}^\dagger=1$ (see Sec. \ref{sec:osdm}). 
Hence this technique may prove useful in analysing the physical systems 
where fluctuations of the transverse component of the
density matrix are expected to play a r\^{o}le, including excitonic 
insulators, heavy-fermion systems, superconductors and perhaps also 
spin-polarised systems where the (on-site) transverse spin dynamics 
is important. As for the opposite case when the OSDM is fully diagonal, 
the need for a Weiss-type treatment there is doubtful, as such  
``longitudinal'' problems are best addressed by more conventional means, 
including the analysis of plasmon spectra, etc.

\acknowledgements
The author takes pleasure in
 thanking  A. G. Abanov, R. Berkovits, A. V. Kazarnovski-Krol, 
B. D. Laikhtman, and M. D. Watson 
for discussions.
This work was supported by the Israeli Absorption Ministry.


\appendix

%

\section{On-site fluctuations in a lattice Fermi gas}
\label{app:average}

What follows is a rather obvious derivation, included here for completeness.
Consider for simplicity a single-band ideal Fermi gas on a lattice, with an 
arbitrary 
dispersion law. Let $\hat{F}$ be an on-site operator of the form
\begin{equation}
\hat{F}=\frac{1}{N} \sum_{\vec{k}}F(\vec{k})\hat{n}_{\vec{k}}\,,
\label{eq:Fgen}
\end{equation}
where $\hat{n}_{\vec{k}}=g^\dagger_{\vec{k}} g_{\vec{k}}$ is the occupancy,
$g_{\vec{k}}$ are the fermion annihilation operators, and the summation is
over the Brillouin zone. Our local operators  $c^\dagger_0 c_0$, 
 $d^\dagger_0 d_0$,  or  $c^\dagger_0 d_0$, whose average values yield band 
occupancies or hybridisation (see Secs. \ref{sec:HF}--\ref{sec:osdm}), 
all have  the general form (\ref{eq:Fgen}). In these cases, the function
$F({\vec{k}})$ contains also the coefficients of transformation from the 
Hartree--Fock quasiparticle operators $f_{a,{\vec{k}}}$ to the original 
fermions $c_{\vec{k}}$, $d_{\vec{k}}$, see Eqs. (\ref{eq:diag1}--\ref{eq:diag2}).
We find
\[\bar{F} \equiv \langle \hat{F} \rangle_F =\frac{1}{N}\sum_{\vec{k}}F(\vec{k})
{n}_{\vec{k}}\,,\]
where $n_{\vec{k}}$ is the Fermi distribution function, and likewise
\begin{eqnarray} \langle{\delta F}^2 \rangle_F&=&\langle(\hat{F})^2-
\left(\bar{F}\right)^2 \rangle_F =\nonumber \\
&=&\frac{1}{N^2}\sum_{\vec{k},\vec{p}} F(\vec{k}) F(\vec{p})
\langle\hat{n}_{\vec{k}} \hat{n}_{\vec{p}}-n_{\vec{k}}n_{\vec{p}}\rangle_F. 
\nonumber
\end{eqnarray}
Since the occupancies for different momenta are statistically independent, the
non-zero contribution comes only from the $\vec{k}=\vec{p}$ terms [where one
obtains the well-known formula\cite{Volume5} for the occupancy fluctuation,
$\langle (\delta n_{\vec{k}})^2 \rangle_F = n_{\vec{k}}(1-n_{\vec{k}})$]. Thus,
\begin{equation}
\sqrt{\langle{\delta F}^2 \rangle_F}=\left\{\frac{1}{N^2} \sum_{\vec{k}}
[F(\vec{k})]^2n_{\vec{k}}(1-n_{\vec{k}})\right\}^{1/2} \propto 
\frac{1}{\sqrt{N}}\,, 
\label{eq:neglect}
\end{equation}
and vanishes in the $N\rightarrow \infty$ limit. Generalisation for a 
two-band case and for higher-order local operators $\hat{F}$ is 
straightforward.

We emphasise that Eq. (\ref{eq:neglect}) applies to finite-temperature 
fluctuations in a 
canonical ensemble, and {\it not} to quantum-mechanical fluctuations of an 
observable in a given state $|\Psi\rangle$. 

\section{On-site fluctuations of the many-body wave function}
\label{app:su4}

We re-write Eq.(\ref{eq:decomp0}) as
\begin{eqnarray}
|\Psi\rangle=&&\!\!\!\!\!\!A_a^{(0)} |a\rangle |\Phi_a(\Psi)\rangle + A_b^{(0)}
|b\rangle |\Phi_b(\Psi)\rangle + \nonumber\\
&&\!\!\!\!\!\!+A_0^{(0)} |0\rangle |\Phi_0(\Psi)\rangle +|A_{cd}^{(0)}|\cdot 
|ab\rangle |\Phi_{cd}(\Psi)\rangle\,,
\label{eq:decomp1a}
\end{eqnarray}
where the states  $|\Phi_i(\Psi)\rangle$ are orthonormal. In the last term, we
introduced 
\begin{equation}
|ab\rangle\equiv a^\dagger b^\dagger|0\rangle = {\rm e}^{{\rm i} \varphi_0}|cd 
\rangle
\label{eq:stateab}
\end{equation}
 [see Eqs. (\ref{eq:A0subst}--\ref{eq:onsitediag2})]. The state of the system 
is affected by  
the choice of the on-site states $|c\rangle$ and $|d\rangle$ relative to 
$|a\rangle$ and $|b\rangle$ [cf. Eqs (\ref{eq:onsitediag1}--
\ref{eq:onsitediag2}); this affects the values of 
$\rho_{11}$, $\rho_{22}$, and $\rho_{12}=\rho_{21}^*$ but {\it not} the 
eigenvalues of  the OSDM, $\hat{\rho}$], and by varying the coefficients 
$A_i$ (which 
in turn changes the eigenvalues). The states $|\Phi_i\rangle$ relate only to
the rest of the system and are unaffected by the on-site fluctuations.

The changes of $A_i$ are described by SU(4) transformations in the 
four-dimensional space of orthonormal vectors $|a\rangle |\Phi_a\rangle$, 
$|b\rangle |\Phi_b\rangle$, $|0\rangle |\Phi_0\rangle$, and
$|ab\rangle |\Phi_{cd}\rangle$ (matrix equations that follow assume this order
of basic vectors). While one could act with an  SU(4) 
transformation ${\cal D}$ on the original
state (\ref{eq:decomp1a}), it is more convenient to choose a fixed initial
state, $|\psi_0\rangle=|ab\rangle |\Phi_{cd}(\Psi)\rangle $. As explained in 
Ref. \onlinecite{physicab18} (see also Ref. \onlinecite{Tilma2002}), one can 
then use a reduced 
form $\tilde{{\cal D}}$ of the 
 SU(4) transformation matrix, depending only on six Euler angles (instead of 
15): 
\begin{widetext}
\begin{equation}
\!\!\!\!\!\tilde{\cal D}\!\!=\!\!\!\left(\!\!\!\begin{array}{cccc} 
{\rm e}^{{\rm i}(\alpha_1+\alpha_2+\alpha_3)}\cos \frac{\theta_1}{2} \cos \theta_2
\sin \theta_3 & {\rm e}^{{\rm i}(\alpha_1-\alpha_2-\alpha_3)}\sin 
\frac{\theta_1}{2} & {\rm e}^{{\rm i}(\alpha_1+\alpha_2)}\cos \frac{\theta_1}{2} 
\sin \theta_2 & 
{\rm e}^{{\rm i}(\alpha_1+\alpha_2+\alpha_3)}\cos \frac{\theta_1}{2} 
\cos \theta_2 \cos \theta_3 \\ ~ & ~ & ~ & ~ \\
 -{\rm e}^{{\rm i}(-\alpha_1+\alpha_2+\alpha_3)}\sin\frac{\theta_1}{2}
\cos \theta_2 \sin \theta_3 & {\rm e}^{-{\rm i}(\alpha_1+\alpha_2+\alpha_3)}
\cos \frac{\theta_1}{2} & -{\rm e}^{{\rm i}(-\alpha_1+\alpha_2)}\sin 
\frac{\theta_1}{2} 
\sin \theta_2 & -{\rm e}^{{\rm i}(-\alpha_1+\alpha_2+\alpha_3)}\sin 
\frac{\theta_1}{2} \cos \theta_2 \cos \theta_3\\ ~ & ~ & ~ & ~ \\
-{\rm e}^{{\rm i}\alpha_3}\sin \theta_2 \sin \theta_3 & 0 &  \cos \theta_2 & 
-{\rm e}^{{\rm i}\alpha_3}\sin \theta_2 \cos \theta_3\\ ~ & ~ & ~ & ~ \\
-\cos \theta_3 & 0 & 0 &  \sin \theta_3 \end{array}\!\!\!\right) 
\label{eq:SU4gen}
\end{equation}
(with  $ 0 \leq \alpha_1, \theta_1 \leq \pi$, $0 \leq  \theta_{2,3} 
\leq {\pi}/{2}$, and $0 \leq \alpha_{2,3}\leq 2 \pi$).  
We find [up to an inconsequential overall phase factor of 
$-\exp({\rm i}\alpha_3)$]
\begin{eqnarray}
|\psi(\theta_1,\theta_2,\theta_3,\gamma_1,\gamma_2,\gamma_3) \rangle= 
\tilde{\cal D}|\psi_0\rangle&=&
{\rm e}^{{\rm i}(\gamma_1+\gamma_2)}\cos \frac{\theta_1}{2} \cos \theta_2 \cos 
\theta_3 |a\rangle |\Phi_a\rangle+{\rm e}^{{\rm i}(\gamma_1-\gamma_2)}
 \sin \frac{\theta_1}{2} \cos \theta_2 \cos \theta_3 |b\rangle |\Phi_b\rangle+
\nonumber\\
&&+
\sin \theta_2 \cos \theta_3|0\rangle |\Phi_0\rangle+
{\rm e}^{{\rm i}(2\gamma_1+\gamma_3)}\sin \theta_3  |ab\rangle 
|\Phi_{cd}\rangle\,, 
\label{eq:decomp2}
\end{eqnarray}
\end{widetext}
where $\gamma_1=\alpha_2-\pi/2$, $\gamma_2=\alpha_1-\pi/2$, and 
$\gamma_3=-\alpha_3-2\alpha_2$.
Eq. (\ref{eq:decomp2}) reduces to Eq. (\ref{eq:decomp1a}) when the angles 
$\theta_i$ are given by 
\begin{eqnarray}
&&\sin \theta_3= \sqrt{\goth{n}_\goth{d}^{(0)}}\,,\,\,\, \tan \theta_2=
\sqrt{ \frac{1+\goth{n}_\goth{d}^{(0)}-n_c^{(0)}-n_d^{(0)}}{n_c^{(0)}+n_d^{(0)}-
2\goth{n}_\goth{d}^{(0)}}}\,, \nonumber \\
&& \label{eq:0anglesgen1} \\
&&\cos \theta_1=\frac{\sqrt{(n_c^{(0)}-n_d^{(0)})^2+4 (\Delta^{(0)})^2}}
{n_c^{(0)}+n_d^{(0)}-2\goth{n}_\goth{d}^{(0)}}\,
\label{eq:0anglesgen2}
\end{eqnarray}
and $\gamma_1=\gamma_2=\gamma_3=0$.
Next, we perform a (modified) SU(2) rotation according to
\begin{eqnarray}
|a\rangle &=& {\rm e}^{{\rm i} \zeta}\left(\cos \frac{\beta}{2} |c\rangle + 
{\rm e}^{{\rm i}\phi}\sin \frac{\beta}{2}|d\rangle\right) \,, \label{eq:su2.1}\\
|b\rangle &=& {\rm e}^{-{\rm i} \zeta}\left( -\sin \frac{\beta}{2} |c\rangle + 
{\rm e}^{{\rm i}\phi}\cos \frac{\beta}{2}|d\rangle\right)\,,
 \label{eq:su2.2}
\end{eqnarray}
implying also $|ab\rangle ={\rm e}^{{\rm i}\phi} |cd \rangle$ [note that states
$|c\rangle$, $|d\rangle$, and $|cd\rangle$  here are different from the 
original ones in Eqs. (\ref{eq:decomp0}) or 
(\ref{eq:stateab})]. The parameter $\zeta$ in Eqs. 
(\ref{eq:su2.1}--\ref{eq:su2.1}) is additive with $\gamma_2$ in 
Eq. (\ref{eq:decomp2}) and can be set to zero, whereas $\phi$ and $\beta$ vary 
in the ranges $0 < \phi < 2\pi$, $0< \beta < \pi$. Substituting Eqs. 
(\ref{eq:su2.1}--\ref{eq:su2.2}) into (\ref{eq:decomp2}), we finally obtain 
Eq. (\ref{eq:decompfluctgen}); in order to avoid double counting, 
one must restrict the values of angle $\theta_1$ to the
interval $0 \leq \theta_1 \leq \pi/2$.

As noted in the Introduction, the issue of the integration over the space of
quantum states is cumbersome. However, in the specific case of phase factors
in Eq. (\ref{eq:decomp2})  symmetry considerations dictate that the
three quantities $\gamma_1+\gamma_2$, $\gamma_1-\gamma_2$, and 
$2\gamma_1+\gamma_3$ should all vary between $0$ and $2\pi$ (the values
differing by $2\pi$ are equivalent) with the uniform 
integration measure [agreeing with the results for the Haar measure of the
SU(4) transformation, Eq.(\ref{eq:SU4gen}) (see Ref. \onlinecite{physicab18})].
In this way,  we arrive at Eq. (\ref{eq:Gammameasure}).

\section{Energy cost of a single-site fluctuation}
\label{app:cost}

Here, we will use Eq. (\ref{eq:Efluct}) to derive the general expression 
for the energy cost of
a fluctuation, without introducing any 
restrictions on $n$. The case of 
$n=1$, considered in Secs. \ref{sec:low-T}  and \ref{sec:high-T},  
can be obtained with the help of Eq. (\ref{eq:theta2}).

Using explicit expressions for states $|\Phi_i\rangle$ and coefficients 
$A_i^{(0)}$ (see the main text), the operator $\hat{S}$ in Eq. (\ref{eq:defS}) 
can be written in the form
\begin{widetext}
\begin{equation}
\hat{S}=X_0 \hat{1} + X_1 {\rm e}^{-\rm{i} \varphi_0} c^\dagger_0 d_0 + 
X_2 {\rm e}^{\rm{i} \phi} d^\dagger_0 c_0 + (X_3-X_0)c^\dagger_0 c_0 + 
(X_4{\rm e}^{\rm{i} 
(\phi-\varphi_0)}-X_0) d^\dagger_0 d_0 + \left([X_5-X_4]{\rm e}^{\rm{i} 
(\phi-\varphi_0)}-X_3+X_0\right) c^\dagger_0 d^\dagger_0 d_0 c_0\,,
\label{eq:Sgen}
\end{equation}
where 
\begin{eqnarray}
X_0&=& \frac{\cos \theta_2 \cos \theta_3}{\sqrt{1+\goth{n}_{\goth{d}}^{(0)}-n_c^{(0)}-n_d^{(0)}}}\,,\\
X_1&=& {\rm e}^{{\rm i} \gamma_1}\cos \theta_2 \cos \theta_3 
\left(\frac{{\rm e}^{{\rm i} \gamma_2}}{A_a^{(0)}}\sin \frac{\beta^{(0)}}{2}
\cos \frac{\beta}{2} \cos \frac{\theta_1}{2}-
\frac{{\rm e}^{-{\rm i} \gamma_2}}{A_b^{(0)}}\cos \frac{\beta^{(0)}}{2}
\sin \frac{\beta}{2} \sin \frac{\theta_1}{2}\right)\,,
\\
X_2&=& {\rm e}^{{\rm i} \gamma_1}\cos \theta_2 \cos \theta_3 \left( 
\frac{{\rm e}^{{\rm i} \gamma_2}}{A_a^{(0)}}\cos \frac{\beta^{(0)}}{2}
\sin \frac{\beta}{2} \cos \frac{\theta_1}{2}-
\frac{{\rm e}^{-{\rm i} \gamma_2}}{A_b^{(0)}}\sin \frac{\beta^{(0)}}{2}
\cos \frac{\beta}{2} \sin \frac{\theta_1}{2}\right)\,,
\\
X_3&=& {\rm e}^{{\rm i} \gamma_1}\cos \theta_2 \cos \theta_3 \left( 
\frac{{\rm e}^{{\rm i} \gamma_2}}{A_a^{(0)}}\cos \frac{\beta^{(0)}}{2}
\cos \frac{\beta}{2} \cos \frac{\theta_1}{2}+
\frac{{\rm e}^{-{\rm i} \gamma_2}}{A_b^{(0)}}\sin \frac{\beta^{(0)}}{2}
\sin \frac{\beta}{2} \sin \frac{\theta_1}{2}\right)\,, \\
X_4&=& {\rm e}^{{\rm i} \gamma_1}\cos \theta_2 \cos \theta_3 \left( 
\frac{{\rm e}^{{\rm i} \gamma_2}}{A_a^{(0)}}\sin \frac{\beta^{(0)}}{2}
\sin \frac{\beta}{2} \cos \frac{\theta_1}{2}+
\frac{{\rm e}^{-{\rm i} \gamma_2}}{A_b^{(0)}}\cos \frac{\beta^{(0)}}{2}
\cos \frac{\beta}{2} \sin \frac{\theta_1}{2}\right)\,,
\end{eqnarray}
\end{widetext}
and 
\begin{equation}
X_5={\rm e}^{{\rm i}(\gamma_3+2 \gamma_1)}\frac{\sin \theta_3}
{\sqrt{\goth{n}_{\goth d}^{(0)}}}. 
\label{eq:X5}
\end{equation}
This expression for $\hat{S}$ should be substituted in Eq. (\ref{eq:Efluct}), 
leading to a somewhat tedious calculation. For our purposes in this 
paper, it will suffice to consider the case of $\theta_i=\theta_i^{(0)}$, when
one may use a simpler expression (\ref{eq:Efluct2}).
Since the latter involves averaging over the Hartree--Fock wave 
functions 
$|\Psi\rangle$, average values of the products of  Fermi operators in 
each term of the resultant expression decouple into pairwise averages, some of
which contain operators at different sites, viz., at our central site $0$
and at one of the neighbouring sites $a$. When the operator at site $a$ is
either $c_a$ or $c^\dagger_a$, the corresponding average does not involve the
phase $\varphi_a$ of the operator $d_a$, and we readily find 
\begin{eqnarray}
\frac{1}{2}\sum_a \langle c^\dagger_0 c_a \rangle_F=l_c^{(0)}\,,&\,\,\,&
\frac{1}{2}\sum_a \langle d^\dagger_0 c_a \rangle_F={\rm e}^{{-\rm i} \varphi_0}
l_\Delta^{(0)}\,,
\label{eq:lcdelta0}\\
\frac{1}{2}\sum_a \vec{a} \cdot \vec{\Xi}\langle c^\dagger_a d_0 
\rangle_F &=&-{\rm e}^{{\rm i} \varphi_0}m
\end{eqnarray}
[see Eqs. (\ref{eq:deflm}); the vector $\vec{\Xi}$ is defined following 
Eq. (\ref{eq:pert}), and the vector $\vec{a}$ connects the central site and 
the site $a$.] 

We begin with the {\it low-temperature case} of $T\stackrel{<}{\sim}T_{cr}$, 
when fluctuations of the 
quantities $\gamma_i$ and $\beta$ are very small. Averaging over the thermal 
fluctuations of the 
background  [included in $\langle ... \rangle_{T'}$ in Eqs. (\ref{eq:Efluct}) 
and 
(\ref{eq:Efluct2})] is then 
equivalent to averaging over the phases $\varphi_a$ on the neighbouring sites, 
which in turn obey 
Eq.(\ref{eq:coskappa}). 
In defining the coefficients $X_i$ above, we explicitly factored out the 
dependence on $\phi$ and $\varphi_0$, in 
order to facilitate averaging over $\varphi_i$. It can be carried out in 
two stages:

\noindent (i) The operator $\hat{S}$ [Eq. (\ref{eq:Sgen})] contains only 
the fermion 
operators at our central site $0$. Taking into account the form of the
Hamiltonian, ${\cal H}+ \delta {\cal H}$, this implies that each term in 
the operator $\hat{S}^\dagger \left[{\cal H}+ \delta{\cal H}, \hat{S} 
\right]$ [see  Eq. (\ref{eq:Efluct})] contains at most one of the operators 
$d_a$
or $d^\dagger_a$ at one of the neighbouring sites $a$. Therefore we can 
readily take the average value over the fluctuations of $\varphi_a$, 
replacing $\exp(\pm {\rm i} \varphi_a)$ with $\cos \kappa$ 
[see Eq. (\ref{eq:coskappa})]. Since $\langle ... \rangle_{T'}$ is but a 
combination
of  canonical averaging $\langle ... \rangle_F$ and averaging over 
$\varphi_a$, we find 
\begin{eqnarray}
\frac{1}{2}\sum_a \langle d^\dagger_a d_0 \rangle_{T'}=
{\rm e}^{{\rm i} \varphi_0}l_d^{(0)}\cos\kappa\,,&\,\,\,&
\frac{1}{2}\sum_a \langle d^\dagger_a c_0 \rangle_{T'}=l_\Delta^{(0)}\cos\kappa\,,
\nonumber\\
\frac{1}{2}\sum_a \vec{a} \cdot \vec{\Xi}\langle d^\dagger_a c_0 
\rangle_{T'}&=&m\cos\kappa\,,
\end{eqnarray}
with $l_d^{0)}$ defined in Eqs. (\ref{eq:deflm}). In addition, we note that 
\[ \sum_a \langle c^\dagger_0 c_a \rangle_F \vec{a} \cdot \vec{\Xi} =
\sum_a\langle d^\dagger_0 d_a \rangle_{T'} \vec{a} \cdot \vec{\Xi} =0\,.\]

(ii) Now, inspecting every term in $\langle \Psi | \hat{S}^\dagger 
[{\cal H}+ \delta{\cal H}, 
\hat{S}] |\Psi \rangle_{T'}$ we find that it is either independent of 
$\varphi_0$ or 
linear in $\exp(\pm {\rm i}\varphi_0)$. In the latter case, averaging over 
fluctuations of 
$\varphi_0$ amounts to replacing $\exp(\pm {\rm i}\varphi_0)$ 
with $\cos \kappa$. This completes the averaging over the phases $\varphi_i$ 
in Eq. (\ref{eq:Efluct2}).

In this way, we find the two terms in the expression (\ref{eq:Efluct2}) for 
$\delta E$ (averaged over fluctuations of {\it all} $\varphi_i$):
\begin{widetext}
\begin{eqnarray}
\!\!\!\!\!\!\!\!\!\langle \Psi |\hat{S}^\dagger \left[{\cal H}, 
\hat{S}\right] |\Psi \rangle_{F,\varphi}&=&\left[|X_1|^2-|X_2|^2-|X_3-X_0|^2+
|X_4-X_5|^2\right](l_c^{(0)} n_d^{(0)} -l_\Delta^{(0)} \Delta^{(0)})+
\left[ X_2 X_4^*-X_1X_3^* \right](l_\Delta^{(0)}+E_d \Delta^{(0)})+\nonumber \\
&& +2 {\rm {Re}}\left\{X_1^*(X_0-X_3)+X_2^*(X_5-X_4)
\right\}(l_\Delta^{(0)} n_c^{(0)} -l_c \Delta^{(0)})
+ 
2 {\rm Re}\left\{X_1(X_3^*-X_0) \right\}l_\Delta^{(0)}+\nonumber\\
&&+\left[|X_2|^2+|X_3-X_0|^2 \right]l_c^{(0)} + 
|X_1|^2E_d(\goth{n}_{\goth{d}}^{(0)}-n_d^{(0)})-|X_2|^2E_d(
\goth{n}_{\goth{d}}^{(0)}-n_c^{(0)})\,,
\label{eq:commwithH}
\end{eqnarray}
which does not depend on either $\phi$ or $\cos \kappa$,
and
\begin{eqnarray}
&&\!\!\!\!\!\!\!\!\langle \Psi |\hat{S}^\dagger \left[\delta{\cal H}, 
\hat{S}\right] |\Psi \rangle_{F,\varphi}=\left\{\left[(-X_0^2-|X_1|^2+|X_2|^2+
|X_3|^2-|X_4|^2+|X_5|^2)\cos \kappa+
2{\rm Re}((X_0X_4-X_5X_3^*){\rm e}^{{\rm i}\phi})\right]
\times \right. \nonumber \\
&&\!\!\!\!\!\!\!\times (l_d^{(0)} n_c^{(0)}-l_\Delta^{(0)} \Delta^{(0)})+ 
\left[ (X_1 X_3^*+X_4 X_2^*)
\cos \kappa - 2 {\rm Re}\left(X_0 X_2 {\rm e}^{{\rm i}\phi}\right)\right]
l_\Delta^{(0)}
+\left[(X_0^2+|X_1|^2+|X_4|^2) \cos \kappa-2 {\rm Re}\left(X_0 X_4 
{\rm e}^{{\rm i}\phi}\right)\right]l_d^{(0)}+\nonumber \\
&&\left. \!\!\!\!\!\!\!+2{\rm Re}\left[(X_0X_2+X_5X_1^*){\rm e}^{{\rm i}\phi}-
(X_1 X_3^*+X_2X_4^*)\cos \kappa\right](l_\Delta^{(0)} n_d^{(0)}-l_d^{(0)}
\Delta^{(0)})
\right\} t^\prime \cos \kappa+\nonumber\\
&&\!\!\!\!\!\!\!+\left\{ \left[2{\rm Re}\left(X_2 X_3^*{\rm e}^{{\rm i}\phi}
\right)
-(X_1X_3^*+X_4X_2^*)\cos \kappa\right](n_c^{(0)}-\goth{n}_{\goth{d}}^{(0)})+
\left[2{\rm Re}\left(X_4 X_1^*{\rm e}^{{\rm i}\phi}\right)
-(X_3X_1^*+X_2X_4^*)\cos \kappa\right](n_d^{(0)}-\goth{n}_{\goth{d}}^{(0)})+
\right. \nonumber \\
&&\!\!\!\!\!\!\!\left.+
\left[2{\rm Re}\left((X_2 X_1^*+X_4X_3^*){\rm e}^{{\rm i}\phi}\right)
-(|X_1|^2+|X_2|^2+|X_3|^2+|X_4|^2)\cos \kappa\right]
\Delta^{(0)}
\right\} V_0+\nonumber \\
&&\!\!\!\!\!\!\!+\left\{ \left(X_0^2-|X_1|^2+|X_2|^2+|X_3-X_0|^2+|X_4|^2\right)
\cos\kappa-2{\rm Re}\left(X_0 X_4{\rm e}^{{\rm i}\phi}\right)+\left[|X_1|^2-
|X_2|^2-|X_3-X_0|^2+|X_4-X_5|^2\right]\times \right. \nonumber \\
&&\!\!\!\!\!\!\!\times n_d^{(0)} \cos \kappa+\left[(-X_0^2+|X_1|^2
+|X_2|^2+|X_3|^2-|X_4|^2+|X_5|^2)\cos \kappa +2{\rm Re}\left((X_0 X_4-X_5X_3^*)
{\rm e}^{{\rm i}\phi}\right)\right]n_c^{(0)}+\nonumber \\
&&\!\!\!\!\!\!\!\left. +2{\rm Re}\left[-(X_0X_2+X_5X_1^*){\rm e}^{{\rm i}\phi}+
(-X_0X_1-2X_1X_3^*+2X_2X_4^*-X_2X_5^*)\cos \kappa\right]\Delta^{(0)}\right\}
V_2\cdot m\,,
\label{eq:commwithdH}
\end{eqnarray}
\end{widetext}
where we omitted the (somewhat cumbersome) $V_1$ term. Carrying out the 
algebra, we arrive at 
Eq. (\ref{eq:delowT}).

Turning now to the {\it high-temperature,} phase-disordered case of 
Sec. \ref{sec:high-T}, we notice
that the contribution of $\delta {\cal H}$ to $\delta E$, 
Eq. (\ref{eq:Efluct2}) vanishes upon averaging over phase fluctuations 
[note that every term in 
Eq. (\ref{eq:commwithdH}) contains at 
least one of $\cos \kappa$ or $\exp({\rm i} \phi)$]. The 
contribution of ${\cal H}$ is derived in a similar way and 
$\langle \Psi |\hat{S}^\dagger \left[{\cal H}, 
\hat{S}\right] |\Psi \rangle_{T',\varphi_0}$
is still given by the r.\ h.\ s. of  
Eq. (\ref{eq:commwithH}), where however one has to replace 
the Fermi-distribution averages $l_{c,\Delta}^{(0)}$ [Eq. (\ref{eq:lcdelta0})] 
with the quantities
\begin{eqnarray}
l_c&=&\frac{1}{2} \sum_a\langle c_0^\dagger c_a \rangle_{T'}=l_c^{(0)}+
\delta l_c \,,
\label{eq:lcT}\\
l_\Delta&=&\frac{1}{2}{\rm e}^{i \varphi_0}\sum_a\langle d_0^\dagger c_a 
\rangle_{T'}=l_\Delta^{(0)}+\delta l_\Delta\,, 
\label{eq:ldeltaT}
\end{eqnarray}
which are averaged over the thermal fluctuations of $\beta$ on the 
neighbouring sites 
$a$. In a direct 
analogy to the derivation of Eq. (\ref{eq:Efluct2}) in Sec. \ref{sec:osdm}, 
we find
\begin{eqnarray}
\delta l_c &=& \frac{1}{2} \sum_a \langle c_0^\dagger S_a^\dagger[c_a,S_a]
\rangle_{T'}\,,\,\,
\label{eq:dlc0}\\
\delta l_\Delta&=&\frac{1}{2}{\rm e}^{i \varphi_0}\sum_a
\langle d_0^\dagger \hat{S}_a^\dagger[c_a,\hat{S}_a]\rangle_{T'}\,,
\label{eq:dldelta0}
\end{eqnarray}  
where $\hat{S}_a$ is the operator of an on-site perturbation acting on 
site $a$. It is given by the same
expression (\ref{eq:Sgen}--\ref{eq:X5}), but all the fermion operators 
and angles now carry a 
site index $a$ [instead of index  0, suppressed in 
Eqs. (\ref{eq:Sgen}--\ref{eq:X5})]. 
Upon carrying out the calculation in Eqs. (\ref{eq:dlc0}-\ref{eq:dldelta0}), it
is convenient to exchange the site indices, $0 \leftrightarrow a$, 
so that the averaging is again 
carried out over the on-site thermal fluctuations at site 0, 
in our notation $\langle ... \rangle_T$.
We find
\begin{widetext}
\begin{eqnarray}
\delta l_c &=&\left\langle l_\Delta^{(0)} X_0X_1+  l_c^{(0)}X_0(X_3-X_0)+  
\left(l_\Delta^{(0)} \Delta^{(0)} - l_c^{(0)}n_d^{(0)}\right)
\left[|X_1|^2+|X_4|^2-X_4^*X_5+X_0X_3-X_0^2  \right]+\right.
\nonumber \\
&&+\left. \left(l_\Delta^{(0)} n_c^{(0)} - l_c^{(0)}\Delta^{(0)}\right)
\left[ (X_3^*-X_0)X_1+X_2^* (X_4-X_5) \right] \right\rangle_T\,, 
\label{eq:dlcTg} \\
\delta l_\Delta &=&\left\langle l_d^{(0)} X_0X_1+  l_\Delta^{(0)}X_0(X_3-X_0)+  
\left(l_d^{(0)} \Delta^{(0)} - l_\Delta^{(0)}n_d^{(0)}\right)
\left[|X_1|^2+|X_4|^2-X_4^*X_5+X_0X_3-X_0^2  \right]+\right.
\nonumber \\
&&+\left. \left(l_d^{(0)} n_c^{(0)} - l_\Delta^{(0)}\Delta^{(0)}\right)
\left[ (X_3^*-X_0)X_1+X_2^* (X_4-X_5) \right] \right\rangle_T\,, 
\label{eq:dldeltaTg}
\end{eqnarray}
which at $\gamma_i=0$ yields Eqs. (\ref{eq:dlcT}--\ref{eq:dldeltaT}) 
(note that both phases $\varphi_0$ and $\phi$ cancel out). 

For the energy cost 
of a fluctuation, Eq. (\ref{eq:commwithH}) with 
$l_{c,\Delta}^{(0)} \rightarrow l_{c,\Delta}$ , 
at $\gamma_i=0$ we find Eq. (\ref{eq:dET}) [note that the value 
of $\goth{n}_\goth{d}^{(0)}$ in Eq. (\ref{eq:dET}) is given by the 
Hartree--Fock expression (\ref{eq:doubleHF})]. If we allow
for non-zero $\gamma_i$ at site 0 {\it only}, there arises an additional term,
\begin{eqnarray}
\delta_\gamma E&=&4\left(l_\Delta n_c^{(0)} -l_c\Delta^{(0)}\right) 
\sin \frac{\gamma_3}{2} \left[\cos 
\frac{\beta}{2} \sin \frac{\beta^{(0)}}{2}\sin(\gamma_1+\gamma_2+ 
\frac{\gamma_3}{2}) -
\sin 
\frac{\beta}{2} \cos \frac{\beta^{(0)}}{2}
\sin(\gamma_1-\gamma_2+ \frac{\gamma_3}{2})\right]+
\nonumber \\  
&&+4\left(l_c n_d^{(0)} -l_\Delta \Delta^{(0)}\right) 
\sin \frac{\gamma_3}{2} \left[\cos 
\frac{\beta}{2} \cos \frac{\beta^{(0)}}{2}\sin
(\gamma_1+\gamma_2+ \frac{\gamma_3}{2}) +
\sin 
\frac{\beta}{2} \sin \frac{\beta^{(0)}}{2}\sin
(\gamma_1-\gamma_2+ \frac{\gamma_3}{2})\right]+ 
\nonumber \\
&&+4l_\Delta \left[\cos 
\frac{\beta}{2} \sin \frac{\beta^{(0)}}{2}\sin^2 
\left(\frac{\gamma_1+\gamma_2}{2}\right)-
\sin 
\frac{\beta}{2} \cos \frac{\beta^{(0)}}{2}\sin^2 
\left(\frac{\gamma_1-\gamma_2}{2}\right)\right]+
\nonumber\\
&&+4l_c \left[\cos 
\frac{\beta}{2} \cos \frac{\beta^{(0)}}{2}\sin^2 
\left(\frac{\gamma_1+\gamma_2}{2}\right)+
\sin 
\frac{\beta}{2} \sin \frac{\beta^{(0)}}{2}\sin^2 
\left(\frac{\gamma_1-\gamma_2}{2}\right)\right]\,,
\label{eq:dETgamma}
\end{eqnarray}
which should be added to $\delta E(\beta)$, Eq. (\ref{eq:dET}). 
Note that the angles $\theta_i$ are 
still assumed frozen at
$\theta_i=\theta^{(0)}_i$.
\end{widetext}


\begin{thebibliography}{999}
\bibitem{extfield} The effect of $V_{0,1}$ terms in Eq. (\ref{eq:pert}) is 
similar to that of an external field at the Curie point, as instead of a 
second-order phase transition (at $V_1=V_0=0$) one finds a well-defined 
crossover. See Sec. \ref{sec:low-T} for details.
\bibitem{selfcons} Additional self-consistency conditions may (and do, see 
Sec. \ref{sec:high-T}) arise, related
to the intersite average values, which in turn affect the energy cost of a 
single-site fluctuation.
\bibitem{diffsites} This means, in particular, that one can 
simultaneously and independently apply operator $\hat{S}$ on different sites.
\bibitem{HFcalc} This corresponds to omitting the second terms on the r.\ h.\
s.\ of Eqs. (\ref{eq:mfe0}). The $T$-dependence of the resultant solution is 
due solely to the
presence of the Fermi distribution functions in Eqs. 
(\ref{eq:Delta}--\ref{eq:nd}).
\bibitem{secular} Eqs. (7-10) of Ref. \onlinecite{prb12} for the 
$\delta {\cal H}=0$ case should be solved for $F_i$, and the result 
substituted into Eqs. (11-14). At $\vec{q} \rightarrow 0$, these are reduced
to a system of two homogeneous linear equations for the quantities $A_{a,b}$,
and the l.\ h.\ s. of our present 
Eq. (\ref{eq:secularHF})  [valid also for $n \neq 1$] is the determinant of that system.
\bibitem{substit} Including ${\goth F}$ leads to a substitution 
$\Delta \rightarrow \Delta +{\goth F}/(2U)$ everywhere in the expressions for
$\Delta_{\vec{k}}$ and $n^{c,d}_{\vec{k}}$ [Eqs. (\ref{eq:Delta}--\ref{eq:nd})], 
which enter the mean-field 
equations.
\bibitem{smallcorr} The associated small parameter is the strength of
the perturbation $\delta{\cal H}$ [Eq. (\ref{eq:pert})] or, equivalently,
$T_{cr}/T_* \ll 1$.
\bibitem{finiteq} The fact that this occurs at a finite momentum $q_0$ is 
presumably due to the specific band and crystal structure of the compound, 
and is not important for our qualitative discussion.
\end{thebibliography}
\end{document}